\documentclass[12pt]{iopart}

\usepackage{iopams}
\usepackage[utf8]{inputenc}
\usepackage[english]{babel}

\usepackage{etoolbox}

\expandafter\let\csname equation*\endcsname\relax
\expandafter\let\csname endequation*\endcsname\relax 

\usepackage{amsfonts}
\usepackage{amssymb}
\usepackage{amsmath}
\usepackage{amsthm}
\usepackage{color}
\usepackage{mathrsfs}
\usepackage{relsize}
\usepackage{bm}
\usepackage{hyperref}
\usepackage{soul}
\usepackage{lineno}
\usepackage{lmodern}
\usepackage[T1]{fontenc}
\usepackage{tikz-cd}
\usepackage{graphicx}
\usepackage{ragged2e}
\usepackage{relsize}
\usepackage[all]{xy}
\usepackage[mathscr]{euscript}

\usepackage{verbatim} 

\providecommand{\fibbun}[2]{\pi_{#2\! #1}}

\def\keywords#1{\vspace{10pt}
     \begin{indented}
     \item[]\rm Keywords: #1\par
     \end{indented}}

\def\AMS#1{\vspace{10pt}
     \begin{indented}
     \item[]\rm AMS classification scheme numbers: #1\par
     \end{indented}}

\begin{document}

\title[A review on geometric formulations for classical field theory]{A review on geometric formulations for classical field theory: the Bonzom-Livine 
model for gravity}

\author{Jasel Berra--Montiel$^{1,2}$, Alberto Molgado$^{1,2}$ and Ángel Rodríguez--López$^{1}$}

\address{$^{1}$ Facultad de Ciencias, Universidad Autonoma de San Luis 
Potosi \\
Campus Pedregal, 
Av.~Parque Chapultepec 1610,
Col.~Privadas del Pedregal,
San Luis Potosi, SLP, 78217, Mexico}
\address{$^{2}$ Dual CP Institute of High Energy Physics, Colima, Col, 28045, Mexico}

\eads{\mailto{jasel.berra@uaslp.mx}, \mailto{alberto.molgado@uaslp.mx}, \mailto{angelrodriguez@fc.uaslp.mx}}

\begin{abstract}
Motivated by the study of physical models associated with General Relativity, we review
some finite-dimensional, geometric and covariant formulations  that allow us to characterize in a simple manner the symmetries for classical field theory
by implementing an appropriate fibre-bundle structure, either at the 
Lagrangian, the multisymplectic or the polysymplectic levels.  
In particular, we are able to formulate Noether's theorems by means of the covariant momentum maps and to systematically introduce a covariant Poisson-Hamiltonian framework.   
Also, by focusing on the space plus time decomposition for a 
generic classical field theory and its relation to these geometric formulations, we are able to successfully recover the gauge content and the true local degrees of freedom for the theory. In order to illustrate the relevance of these geometric frameworks, we center our attention to the analysis of a model for $3$-dimensional theory of General Relativity that involves an arbitrary Immirzi-like parameter. At the Lagrangian level, we reproduce the field equations of the system which for this model turn out to be equivalent to the vanishing torsion condition and the Einstein equations.  We also concentrate on the analysis of the gauge symmetries of the system in order to obtain the Lagrangian covariant momentum map associated with the theory and, consequently, its corresponding Noether currents.  Next, within the multisymplectic approach, we aim our attention to describing how the gauge symmetries of the model yield covariant canonical transformations on the covariant multimomenta phase-space, thus giving rise to the existence of a covariant momentum map. Besides, we analyze the physical system under consideration within the De Donder-Weyl canonical theory implemented at 
the polysymplectic level, thus 
establishing a relation from the covariant momentum map to the conserved currents of the theory within this covariant Hamiltonian approach. Finally, after performing the space plus time decomposition of the space-time manifold, and taking as a starting point the multisymplectic formulation, we are able to recover both the extended Hamiltonian and the gauge structure content that characterize the gravity model of our interest within the instantaneous Dirac-Hamiltonian formulation.
\end{abstract}


\AMS{70S15, 70S10, 83C05, 53D20.}

\section{Introduction}

A useful method for studying classical field theory is the instantaneous Dirac-Hamiltonian approach as it is able to reveal the true local degrees of freedom and gauge symmetries for a given classical field theory \cite{QGS}. However, as it is well known, in order to identify a temporal direction, the instantaneous Dirac-Hamiltonian formulation for classical field theory starts by performing a foliation of the space-time manifold into Cauchy surfaces, thus concealing the true covariant nature of the theory under analysis. Hence, when implementing 
a suitable canonical quantization scheme, the instantaneous Dirac-Hamiltonian formalism gives rise to non-covariant approaches for quantum field theory \cite{Forger1}. By contrast, based on the De Donder-Weyl canonical theory \cite{DDT, WT}, the multisymplectic approach provides a finite-dimensional, geometric and covariant Hamiltonian-like formulation for classical field theory \cite{GIMMSY1, Crampin, Gotay1, DeLeon1}. Roughly speaking, after identifying the fields of the theory as sections of an appropriate fibre-bundle, the multisymplectic formalism starts by introducing the covariant multimomenta phase-space, namely, a finite-dimensional manifold locally constructed by associating to each field variable of the system a set of multimomenta (or polymomenta) variables, which are noting but a covariant extension of the standard instantaneous momenta variables implemented within the instantaneous Dirac-Hamiltonian formulation \cite{Forger1}. Particularly, the covariant multimomenta phase-space is endowed with a multisymplectic $(n+1)$-form which is considered as fundamental as it allows not only to obtain the correct field equations but also to describe the symmetries of a given classical field theory. In this regard, we have that within the multisymplectic framework the symmetries of a classical field theory give rise to covariant canonical transformations, that is, transformations on the covariant multimomenta phase-space that preserve the multisymplectic $(n+1)$-form. Thus, for the case of infinitesimal covariant canonical transformations generated by the action of the symmetry group of the theory on the covariant multimomenta phase-space, we can construct the so-called covariant momentum map which allows not only to extend Noether's theorems to the De Donder-Weyl Hamiltonian formulation in a natural way but also to induce the momentum map that characterizes the system from the instantaneous Dirac-Hamiltonian perspective~\cite{GIMMSY1, DeLeon1, GIMMSY2}. In fact, for classical field theories with localizable symmetries, the latter issue is particularly important since, in light of the second Noether theorem, one may see that the admissible space of Cauchy data for the evolution equations of the system is determined by the zero level set of the momentum map. Hence, according to \cite{GIMMSY1, GIMMSY2, Fischer}, for this kind of classical field theories, the vanishing of the momentum map gives rise to the complete set of first-class constraints of the theory, in Dirac's terminology~\cite{QGS}. In that sense, taking as a starting point the multisymplectic approach, we have a natural way to recover the constrained structure for a given  
singular Lagrangian system.

It is important to mention that, on the covariant multimomenta phase-space, one may construct a Poisson-like bracket for a certain set of differential forms known as Poisson-forms \cite{Romer1, Forger3, Forger4}. However, since there is no
known product between Poisson-forms yet such that the Poisson-like bracket satisfies Leibniz's rule, the set of Poisson-forms defines a Lie superalgebra but not a Poisson superalgebra. From the point of view of physics, such problem results particularly unsatisfactory since the introduction of a well-defined Poisson bracket within the multisymplectic formalism may be considered as the first step to implement either a pre-canonical quantization scheme or a deformation quantization associated with the De Donder-Weyl canonical theory \cite{CPVD, IKQFTPV, kana1, IKHEQFT, IKPSWF, kana2}. Bearing this in mind,  the polysymplectic framework for classical field theory introduces a canonical $(n+1)$-form on a subspace of the covariant multimomenta phase-space referred to as the polymomenta phase-space.  Such canonical form, known as the polysymplectic form, encodes the relevant physical data of a given classical field theory in order to construct a well-defined Poisson-Gerstenhaber bracket for a set of appropriately prescribed differential 
Hamiltonian forms, thus allowing us to analyze an arbitrary classical field theory in a covariant Poisson-Hamiltonian framework \cite{IKCSCHF, IKCSPPS, IKGPA}. Some physically motivated examples for which 
the multisymplectic and polysymplectic formalisms have been applied may be 
encountered in references~\cite{GIMMSY1, DeLeon1, GIMMSY2, IKCSCHF, Angel2, DVVG, Ibort, Roman, DeLeon3, Chirco, IKHFVG, Angel, IKPQYMT, JEA, JAD, kana3}.
Despite their mathematical elegance, from our point of view,  the analysis of the gauge content for a given classical field theory from the perspective of such geometric formulations has been rarely exploited, especially when considering certain highly non-trivial gauge models associated with General Relativity.  Therefore, our main aim 
is to apply those formalisms to the description of the gauge structure associated with the Bonzom-Livine model for gravity, in particular, by capitalizing the involved symmetries through the introduction of the covariant momentum map.

Since the novel work by Ach{\'u}carro and Townsend \cite{Achucarro}, it is widely known that the $3$-dimensional Einstein theory of General Relativity can be reformulated as a Chern-Simons gauge theory. Such reformulation has been extensively studied at the classical and quantum levels in reference~\cite{Witten}, for example, and may be thought of as a 3-dimensional analogue of the gauge theory interpretation of the Hamiltonian constraint equations of $4$-dimensional
gravity developed by Ashtekar \cite{Ashtekar, Ashtekar2}, which in turn is a fundamental theoretical framework for the Loop Quantum Gravity approach \cite{Thiemann}. Indeed, based on Witten's reformulation of $3$-dimensional gravity as a Chern-Simons gauge theory, Bonzom and Livine have introduced an action that corresponds to a formulation of $3$-dimensional Einstein theory of General Relativity that involves an arbitrary Immirzi-like parameter \cite{Bonzom}, which may be regarded as a $3$-dimensional analogue of the Holst action \cite{Holst}. In brief, the 
so-called Immirzi parameter corresponds to a quantization ambiguity that arises within the Loop Quantum Gravity approach, as pointed out in~\cite{IGRCCCG, Immirzi2}. In particular, it has been shown that such ambiguity is related to a one-parameter family of canonical transformations on the phase-space of General Relativity that can not be implemented unitarily at the quantum level, as discussed in detail in~\cite{ILQG}. Bearing this in mind, the Holst action includes the Immirzi parameter within the $4$-dimensional Einstein theory of General Relativity, thus giving rise to a class of theories for $4$-dimensional gravity that are classically indistinguishable but quantum mechanically nonequivalent. In a similar fashion, it has been found that the Immirzi-like parameter inherent to the action proposed by Bonzom and Livine has no effect on the field equations of $3$-dimensional gravity but, however, it modifies the spectra of certain geometric objects in the corresponding quantum theory \cite{Bonzom, Barbosa}. At the classical level, the Bonzom-Livine model for gravity has been extensively studied from different perspectives, including the instantaneous Dirac-Hamiltonian, the Faddeev-Jackiw and the covariant canonical formalisms \cite{Escalante, Basu}, which have confirmed the correspondence between such gravity model and a topological field theory which is invariant under both gauge and diffeomorphism transformations.

Taking into account the previous arguments, our main aim in this paper is to study the Bonzom-Livine model for gravity from the finite-dimensional, geometric and covariant perspectives of the Lagrangian, multisymplectic and polysymplectic formulations for classical field theory. For this purpose, after mathematically introducing the system
of our interest, we start by performing a complete analysis for the Bonzom-Livine model within the so-called geometric-covariant Lagrangian formalism, where the dynamical fields of the theory are thought of as (local) sections of a fibre-bundle over the space-time manifold on which the system is defined. Particularly, we obtain 
the field equations for the model and analyze its gauge symmetries, which allows us to establish the Lagrangian covariant momentum map associated with the extended gauge symmetry group of the theory and, consequently, its corresponding Noether currents. Furthermore, we describe how the symmetries of the system correspond to localizable symmetries. Subsequently, we proceed to study the Bonzom-Livine model for gravity within the multisymplectic framework. In order to do so, we begin by introducing both the covariant multimomenta phase-space and the multisymplectic $(n+1)$-form associated with the system. Then, we focus our attention on describing how the symmetries of the theory give rise to infinitesimal covariant canonical transformations on the covariant multimomenta phase-space, which, analogously to the Lagrangian case, allows us to construct the covariant momentum map associated with the extended gauge symmetry group of the model. Besides, at the polysymplectic level, we start by defining the polymomenta phase-space. Thus, in order to carry out a consistent polysymplectic formulation of the Bonzom-Livine model for gravity, we implement 
the proposal developed in \cite{IKGD} to analyze singular Lagrangian systems within the polysymplectic approach. Following Dirac's terminology, we thus find that on the polymomenta phase-space the theory of our interest is characterized by a set of second-class constraint $(n-1)$-forms which specifically arise when applying the covariant Legendre map. Afterwards, in order to treat such constraint $(n-1)$-forms as strong identities, we construct a Dirac-Poisson bracket that allows us to obtain the correct De Donder-Weyl-Hamilton field equations of the system, and consequently, the vanishing torsion condition and the Einstein equations. As far as we know, this is the first time
within this covariant formulation that the Dirac-Poisson bracket is implemented for a non-trivial physical example 
related to General Relativity.  In addition, we also discuss the way in which the covariant momentum map associated with the extended gauge symmetry group of the model allows us to construct the conserved currents of the system within the De Donder-Weyl canonical theory. Finally,  closely following references \cite{GIMMSY1, DeLeon1, GIMMSY2, Fischer, Angel2, DeLeon2}, we  explicitly perform the space plus time decomposition for the gravity model of our interest at both the Lagrangian and multisymplectic level, which enables us to recover not only the extended Hamiltonian but also the generator of infinitesimal gauge transformations as well as the first- and second-class constraints that characterize the system within the instantaneous Dirac-Hamiltonian  formulation as developed in reference~\cite{Escalante}. 

The rest of the paper is organized as follows: in section \ref{Geom Cov Formalisms}, we briefly review the geometric-covariant Lagrangian, multisymplectic and polysymplectic formulations for classical field theory. Additionally, we introduce a general description of the space plus time decomposition for classical field theory which is fundamental in order to establish a well-defined relation between the multisymplectic approach and the instantaneous Dirac-Hamiltonian formalism. In section \ref{Analysis of BLmodel}, we mathematically define the Bonzom-Livine model for gravity. Subsequently, we proceed to analyze such physical system within the geometric-covariant Lagrangian, multisymplectic and polysymplectic approaches,  paying special attention to the study of both the field equations and the gauge symmetries of the theory. Furthermore, we carry out the space plus time decomposition for the gravity model of our interest at both the Lagrangian and multisymplectic level, which in turn allows us to recover the instantaneous Dirac-Hamiltonian
analysis of the system. Lastly, in section \ref{sec:Conclu}, we present some concluding remarks.

\section{Geometric-covariant formalisms for classical field theory}\label{Geom Cov Formalisms}

In this section, we will start by introducing the appropriate notation 
that will be helpful in order to describe the Lagrangian and multisymplectic formalisms 
as finite-dimensional, geometric and covariant formulations for the 
characterization of classical field theory. Particularly, we will focus our attention on the study of symmetries within these frameworks, issue that will be fundamental to formulate Noether's theorems and to analyze the gauge content of a given classical field theory. Also, we will introduce the polysymplectic approach, where we will concentrate on 
the description of the Poisson-Gerstenhaber bracket and its algebraic properties, which will help us to inspect a classical field theory from a covariant Poisson-Hamiltonian perspective. 
Finally, we will provide a brief description of the process to perform the space plus time decomposition for a generic classical field theory at both the Lagrangian and multisymplectic level, which will eventually allow us to recover the instantaneous Dirac-Hamiltonian analysis of the system under consideration. For this general purpose, we will review the main ideas of the Lagrangian, multisymplectic and polysymplectic formalisms as developed in references~\cite{Forger1, GIMMSY1, Crampin, Gotay1, DeLeon1, GIMMSY2, Fischer, IKCSCHF, IKCSPPS, DeLeon2}. Hence, we would like to encourage the reader to examine those 
references for further technical details. It is important to mention that, for the
sake of simplicity, all of the mathematical objects to be considered below, such as manifolds, fibre-bundles and maps,  are assumed to be of class $C^{\infty}$.

\subsection{Conventions on fibre-bundles}\label{SSNotation}

Before proceeding with our discussion, we will set the notation that we will use 
throughout the rest of the article, however, we will refer the reader 
to~\cite{Fischer, Saunders, Sardanashvily1} for a 
detailed account on geometric aspects of fibre-bundles. To start, let $M$ be a manifold. We introduce the triplet $(E, \fibbun{\,E}{M},M)$ to denote a fibre-bundle, where $M$ stands for the base space, $E$ corresponds to the total space (a fibre manifold), and $\fibbun{\,E}{M}:E\rightarrow M$ represents the projector map. At $p\in M$, the subset $\fibbun{\,E}{M}^{-1}(p)$ of $E$, denoted by $E_{p}$, is called the fibre of $E$ over $p$.  Unless otherwise specified, we introduce $\mathscr{E}_{M}$ to denote the set of sections of $\fibbun{\,E}{M}$, that is, the set of maps $\tilde{\kappa}:M\rightarrow E$ satisfying the condition $\fibbun{\,E}{M}\circ \tilde{\kappa}=\mathrm{Id}_{M}$, where $\mathrm{Id}_{M}:M\rightarrow M$ corresponds to the identity map on $M$. Avoiding cumbersome notation, henceforward, a fibre-bundle $(E, \fibbun{\,E}{M},M)$ will be just referred to as $\fibbun{\,E}{M}$. 

In particular, given the tangent bundle $(TM, \fibbun{~TM}{M}, M)$, 
we introduce $(\Lambda^{k}\,TM, \fibbun{~\Lambda^{k}\,TM}{M},M)$ to be the k-th exterior power of the tangent bundle of $M$, and we define $\mathfrak{X}^{\,k}(M)$ as the set of sections of $ \fibbun{~\Lambda^{k}\,TM}{M}$, namely the collection of $k$-multivector fields on $M$. Thus, for $\xi\in \mathfrak{X}^{\,k}(M)$, we denote by $\xi(p)\in \Lambda^{k}\,T_{p}M$ the $k$-tangent vector at $p\in M$ assigned by $\xi$. Besides, given the cotangent bundle $(T^{*}M,\fibbun{~T^{*}M}{M}, M)$, we introduce $(\Lambda^{k}\,T^{*}M, \fibbun{~\Lambda^{k}\,T^{*}M}{M}, M)$ to be the $k$-th exterior power of the cotangent bundle of $M$, and we identify $\Omega^{\;\!k}(M)$ as the set of sections of $\fibbun{~\Lambda^{k} T^{*}M}{M}$, that is, the family of $k$-forms on $M$. Thence, for $\alpha\in \Omega^{\,k}(M)$, we denote by $\alpha(p)\in \Lambda^{k}\,T^{*}_{p}M$ the $k$-cotangent vector at $p\in M$ assigned by $\alpha$. In addition, we define $\Omega^{\;\!0}(M)$ as the set of functions on $M$.

Bearing this in mind, for $\fibbun{\,E}{M}$ a fibre-bundle,  we define $(VE, \fibbun{\,\,V\!E}{E}, E)$ as the vertical tangent bundle of $\fibbun{\,E}{M}$, where the fibre over $e\in E$ is defined as
\begin{equation}
V_{\,e\,}E:= \big\lbrace \zeta\in T_{\,e\,}E~\big|~ T_{\,e\,}\fibbun{\,E}{M}(\zeta)=0\in T_{\fibbun{\,E}{M}(e)}M\big\rbrace\, ,
\end{equation}
being $T_{\,e\,}\fibbun{\,E}{M}:T_{e}E\rightarrow T_{\fibbun{\,E}{M}(e)}M$ the tangent map of $\fibbun{\,E}{M}$ at $e\in E$. Here, we identify $\mathfrak{X}^{V}\!(E)$ as the set of sections of $\fibbun{\,\,V\!E}{E}$, that is, the collection of vector fields on $E$ tangent to the fibres of $\fibbun{\,E}{M}$ \cite{Saunders}. Further, we define $(\Lambda^{k}_{\,r}\,T^{*}E, \fibbun{~\Lambda^{k}_{\,r}\,T^{*}E}{E}, E)$ as the vector bundle of horizontal $(k;r)$-forms over $E$, where the fibre over $e\in E$ is given by
\begin{equation}
\Lambda^{k}_{\,r}\,T^{*}_{\,e\,}E:=\Big\lbrace \omega\, \in \Lambda^{k}\,T^{*}_{\,e\,}E~\Big|~ \zeta_{1}\,\lrcorner \cdots \zeta_{r}\,\lrcorner\, \omega = 0\,,~\forall\, \zeta_{1}, \cdots\!, \zeta_{r} \in V_{\,e}E \Big\rbrace \, .
\end{equation} 
Thus, we introduce $\Omega^{\,k}_{\,r}(E)$ to denote the set of sections of $\fibbun{~\Lambda^{k}_{\,r}\,T^{*}E}{E}$, namely the family of horizontal $(k;r)$-forms on $E$.

Finally, being $(E,\fibbun{\,E}{M},M)$ and $(F,\fibbun{\,F}{N},N)$ a pair of fibre-bundles, we define a bundle morphism from $\fibbun{\,E}{M}$ to $\fibbun{\,F}{N}$ as a pair of maps $(\Phi,\bar{\Phi})$, such that $\Phi:E\rightarrow F$ and $\bar{\Phi}:M\rightarrow N$ satisfy the relation $\fibbun{\,F}{N}\circ \Phi=\bar{\Phi}\circ \fibbun{\,E}{M}$ \cite{Saunders}. Henceforth, we introduce $\Phi_{p}:E_{p}\rightarrow F_{\bar{\Phi}(p)}$ to denote the restriction of $\Phi$ to the subset $E_{p}$. Thus, a map $f:M\rightarrow N$ induces a bundle morphism from $\fibbun{~TM}{M}$ to $\fibbun{~TN}{N}$ denoted by $(Tf,f)$, where $Tf:TM\rightarrow TN$ is called the tangent map of $f$. At $p\in M$, given $v\in T_{p}M$, $T_{p}f(v)\in T_{f(p)}N $ is known as the pushforward of $v$ by $f$ and is defined as $T_{p}f(v)\,[\,g\,]:=v[\,g\circ f\,]$, being $g\in \Omega^{\;\!0}(N)$ a function on $N$ \cite{Fischer}. 

It is important to mention that we will implement the above described basic structures  for any manifold or any fibre-bundle to be considered in the following sections.

\subsection{Lagrangian formalism}\label{Lagsec}

In the present subsection,  we will introduce the basic geometric objects to develop the Lagrangian formulation of classical field theory within the fibre-bundle formalism
following, as close as possible, references \cite{Forger1, GIMMSY1, Crampin, DeLeon1, GIMMSY2, Fischer, Angel2, Saunders, Sardanashvily1}. In particular, we will focus on the emergence of the symmetries for a generic classical field theory  within this geometric-covariant Lagrangian formulation. As previously mentioned, our main motivation to do this is to enunciate the celebrated Noether's theorems, which will be a cornerstone for our study. 

To start, let $X$ be an $n$-dimensional space-time manifold without boundary and locally represented by $(x^{\mu})$, $\mu=0,\dots,n-1 $. Then, given a classical field theory, we define the covariant configuration space of the system as a finite-dimensional fibre-bundle $(Y,\fibbun{\,Y}{X}, X)$, where the classical fields can be identified with $\mathscr{Y}_{X}$, namely the set of sections of $\fibbun{\,Y}{X}$ \cite{GIMMSY1}. Note that, for $x\in X$, the fibre $Y_{x}$ can be regarded as an $m$-dimensional manifold locally represented by $(y^{a})$, $a=1,\dots,m$, which allows to identify $(x^{\mu},y^{a})$ as an adapted coordinate system on $Y$. Hence, a section $\phi\in\mathscr{Y}_{X}$ can be locally represented as $(x^{\mu},\phi^{a}(x^{\mu}))$. 

As discussed in the literature, the natural arena for describing a first order classical field theory within the Lagrangian approach corresponds to the first order jet bundle $(J^{1}Y,\fibbun{\,J^{1}Y}{X}, X)$. In order to define it here, we start by introducing the affine jet bundle $(J^{1}Y,\fibbun{\,J^{1}Y}{Y}, Y)$ first, where the fibre over $y\in Y$ is defined as
\begin{equation}
J^{1}_{y}Y:=\big\lbrace \varkappa \in L(T_{x}X,T_{y}Y)~\big|~T_{\,y\,}\fibbun{\,Y}{X}\circ \varkappa=\mathrm{Id}_{T_{x}X}\big\rbrace\, ,
\end{equation}
being $L(T_{x}X,T_{y}Y)$ the set of all linear mappings $\varkappa:T_{x}X\rightarrow T_{y}Y$ and $x=\fibbun{\,Y}{X}(y)\in X$ the base point associated with $y$\footnote{In fact, $\fibbun{\,J^{1}Y}{Y}$ is an affine bundle since $J^{1}_{y}Y$ is an affine space modeled on the vector space $L\left(T_{x} X, V_{y}Y\right)$, being $V_{y}Y$ the vertical tangent space over $y\in Y$ \cite{Forger1, Crampin, Saunders}.}. In local coordinates an element $\varkappa\in J^{1}_{y}Y$ can be written as
\begin{equation}
\varkappa:=dx^{\mu}\otimes\left(\frac{\partial}{\partial x^{\mu}}+y^{a}_{\mu}\frac{\partial}{\partial y^{a}}\right)\, ,
\end{equation}
which allows to identify $(x^{\mu},y^{a}, y^{a}_{\mu})$ as an adapted coordinate system on $J^{1}Y$. Consequently, the composition map $\fibbun{\,J^{1}Y}{X}:=\fibbun{\,Y}{X}\circ\fibbun{\,J^{1}Y}{Y}$ gives rise to the first order jet bundle  $(J^{1}Y, \fibbun{\,J^{1}Y}{X}, X)$.  In addition, it is not difficult to see that the tangent map of a section $\phi\in \mathscr{Y}_{X}$ at $x\in X$, $T_{x}\phi:T_{x}X\rightarrow T_{\phi(x)}Y$, is an element of $J^{1}_{\phi(x)}Y$, and therefore we can assign to each section $\phi\in \mathscr{Y}_{X}$ a section $j^{1}\phi\in \mathscr{J}^{1}\mathscr{Y}_{X}$ of $\fibbun{\,J^{1}Y}{X}$ known as the first jet prolongation of $\phi$, which can be locally represented as $(x^{\mu},\phi^{a}(x^{\mu}),\phi^{a}_{\nu}(x^{\mu}))$, where we have adopted the notation $\phi^{a}_{\nu}(x^{\mu}):=\partial_{\nu}\:\!\phi^{a}(x^{\mu})$, being $\partial_{\mu}:=\partial/\partial x^{\mu}$ the partial derivative with respect to the base space coordinate $x^{\mu}$.

Bearing this in mind, the action associated with the system can be written as
\begin{equation}\label{APFT}
\mathcal{S}[\phi]:=\int_{ X} ( j^{1}\phi)^{*}\mathcal{L}\, ,
\end{equation} 
where the Lagrangian density of the theory, $\mathcal{L}:J^{1}Y\rightarrow \Lambda^{n}\,T^{*}X$, is locally given by $\mathcal{L}:=L(x^{\mu}, y^{a}, y^{a}_{\mu})\,d^{\,n}x$, with $L:J^{1}Y\rightarrow\mathbb{R}$ denoting the Lagrangian function of the system\footnote{Note that the pullback of the Lagrangian density with a section $j^{1}\phi\in\mathscr{J}^{1}\mathscr{Y}_{X}$ gives rise to an $n$-form on $X$.}. We would like to emphasize that, from our point of view, the main geometric objects of interest within the geometric-covariant Lagrangian formulation are the so-called Poincar\'e-Cartan forms, $\Theta^{(\mathcal{L})}\in\Omega^{\;\!n} (J^1Y)$ and $\Omega^{(\mathcal{L})}\in \Omega^{\;\!n+1}(J^{1}Y)$, which locally read
\begin{subequations}\label{PCforms}
\begin{align}
\Theta^{(\mathcal{L})}\!&:= \frac{\partial L}{\partial y^{a}_{\mu}}dy^{a}\wedge d^{\,n-1}x_{\mu}+\left(L-\frac{\partial L}{\partial y^{a}_{\mu}}y^{a}_{\mu}\right) d^{\,n}x\, ,\label{PCF}\\
\Omega^{(\mathcal{L})}\!&:=
dy^{a}\wedge d\left(\frac{\partial L}{\partial y^{a}_{\mu}}\right)\wedge d^{\,n-1}x_{\mu}-d\left(L-\frac{\partial L}{\partial y^{a}_{\mu}}y^{a}_{\mu}\right)\wedge d^{\,n}x\,,
\end{align}
\end{subequations}
where $d^{\,n-1}x_{\mu}:=\partial_{\mu}\lrcorner\,d^{\,n}x$ denotes the basis for the $\fibbun{\,J^{1}Y}{X}$-horizontal $(n-1;1)$-forms on $J^{1}Y$. One may straightforwardly check that the Poincar{\'e}-Cartan $(n+1)$-form is simply defined through the relation
$\Omega^{(\mathcal{L})}=-d\,\Theta^{(\mathcal{L})}$.

The set of differential forms \eqref{PCforms} is relevant as it contains all the information of the theory under consideration. In fact, it is not difficult to see that, for any section $\phi\in\mathscr{Y}_{X}$, the Poincar{\'e}-Cartan $n$-form satisfies the relation $(j^{1}\phi)^{*}\Theta^{(\mathcal{L})}=(j^{1}\phi)^{*}\mathcal{L}$, which allows us to rewrite the action \eqref{APFT} in terms of the 
Poincaré-Cartan form~(\ref{PCF}) and a section $j^{1}\phi\in \mathscr{J}^{1}\mathscr{Y}_{X}$ (see~\cite{Forger1} for details). Moreover, the Poincar{\'e}-Cartan $(n+1)$-form is necessary in order to write the Euler-Lagrange field equations of the system in an invariant fashion \cite{GIMMSY1, Crampin, DeLeon1}. In other words, we have that, given  a critical point of the action principle \eqref{APFT}, $\phi\in\mathscr{Y}_{X}$, and  an arbitrary vector field on $J^{1}Y$, namely $V\in\mathfrak{X}^{\,1}(J^{1}Y)$, the condition
\begin{equation}\label{ELFE}
(j^{1}\phi)^{*}\left(V\lrcorner\, \Omega^{(\mathcal{L})} \right)=0\, ,
\end{equation}
is completely equivalent to the Euler-Lagrange field equations of the theory, namely,
\begin{equation}\label{Euler-Lagrange equations}
\frac{\partial L}{\partial y^{a}}\big(x^{\mu}, \phi^{a}, \phi^{a}_{\mu}\big)-\frac{\partial}{\partial x^{\mu}}\left(\frac{\partial L}{\partial y^{a}_{\mu}}\big(x^{\mu},\phi^{a},\phi^{a}_{\mu}\big)\right)=0\,. \,
\end{equation}

Now, we are interested in studying the symmetries of the classical field theory \eqref{APFT} within the geometric-covariant Lagrangian formalism, and in particular the action of a Lie group on the covariant configuration space associated with the system. To start, let us consider $\mathcal{G}$ a Lie group and $\mathfrak{g}$ its corresponding Lie algebra. Then, for all $\eta\in\mathcal{G}$, we say that the action of $\eta$ on $\fibbun{\,Y}{X}$ is given by a $\fibbun{\,Y}{X}$-bundle automorphism $(\eta_\mathsmaller{Y}, \eta_\mathsmaller{X})$, where the maps $\eta_\mathsmaller{Y}:Y\rightarrow Y$ and $\eta_\mathsmaller{X}:X\rightarrow X$ satisfy the relation $\fibbun{\,Y}{X}\circ\eta_\mathsmaller{Y}=\eta_\mathsmaller{X}\circ \fibbun{\,Y}{X}$. Locally, these transformations read $\eta_\mathsmaller{X}(x^{\mu}):=\eta^{\mu}_\mathsmaller{X}(x^{\nu})$ and $\eta_\mathsmaller{Y}(x^{\mu},y^{a}):=(\eta^{\mu}_\mathsmaller{X}(x^{\nu}), \eta^{a}_\mathsmaller{Y}(x^{\nu},y^{b}))$. Subsequently, given the infinitesimal generator of $\eta$, $\xi_{\eta}\in\mathfrak{g}$, we introduce $\xi_{\eta}^X\in\mathfrak{X} ^{\,1}\left( X\right)$ and $\xi_{\eta}^{Y}\in\mathfrak{X}^{\,1}\left(Y\right)$ to denote the infinitesimal generators associated with the transformations $\eta_\mathsmaller{X}$ and $\eta_\mathsmaller{Y}$, respectively. Thus, since $\xi^{X}_{\eta}$ and $\xi^{Y}_{\eta}$ must satisfy the condition $T_{\,y\,}\fibbun{\,Y}{X}(\xi^{Y}_{\eta}(y))=\xi^{X}_{\eta}(x)$ for $y\in Y$ and $x=\fibbun{\,Y}{X}(y)\in X$, then we get the local expressions $\xi_{\eta}^X:=\xi^{\mu}(x^{\nu})\partial_{\mu}$ and $\xi_{\eta}^{Y}:=\xi^{\mu}(x^{\nu})\partial_{\mu}+\xi^{a}(x^{\nu},y^{b})\partial_{a}$, where $\partial_{a}:=\partial/\partial y^{a}$ denotes the partial derivative with respect to the fibre coordinate $y^{a}$. As a result, by considering $y\in Y$ and $x=\fibbun{\,Y}{X}(y)\in X$, we have that the action of $\eta$ on $J^{1}_{y}Y$ may be thought of as the map 
$\eta_\mathsmaller{J^{1}_{y}Y}:J^{1}_{y}Y\rightarrow J^{1}_{\eta_{Y}(y)}Y$, which is
explicitly given by $\eta_\mathsmaller{J^{1}_{y}Y}(\varkappa):=T_{\,y\,}\eta_\mathsmaller{Y}\circ \,\varkappa\,\circ {(T_{\,x\,}\eta_{\mathsmaller{X}})}^{-1}, \,\forall \varkappa\in J^{1}_{y}Y$. Here, $T_{\,y\,}\eta_\mathsmaller{Y}:T_{y}Y\rightarrow T_{\eta_{Y}(y)}Y$ corresponds to the tangent map of $\eta_\mathsmaller{Y}$ at $y$, while $(T_{\,x\,}\eta_\mathsmaller{X})^{-1}:T_{\eta_{X}(x)}X\rightarrow T_{x}X$ stands for the tangent maps of $\eta^{-1}_\mathsmaller{X}$ at $\eta_\mathsmaller{X}(x)$. Therefore, the induced action of $\eta$ on $J^{1}Y$, $\eta_\mathsmaller{J^{1}Y}:J^{1}Y\rightarrow J^{1}Y$, locally reads
\begin{equation}\notag
\eta_\mathsmaller{J^{1}Y}(x^{\mu},y^{a},y^{a}_{\mu}):=\left(\eta^{\mu}_\mathsmaller{X}(x^{\nu}), \eta^{a}_\mathsmaller{Y}(x^{\nu},y^{b}), \left(\partial_{\sigma}\eta ^{a}_\mathsmaller{Y}(x^{\nu},y^{b})+\partial_{c}\:\!\eta^{a}_\mathsmaller{Y}(x^{\nu},y^{b})y^{c}_{\sigma}\right)\partial_{\mu}\:\!\big((\eta^{-1}_\mathsmaller{X})^{\sigma}(x^{\nu})\big)\right)\, ,
\end{equation}
where we have introduced $\eta^{-1}_\mathsmaller{X}(x^{\mu})
:=(\eta^{-1}_\mathsmaller{X})^{\mu}(x^{\nu})$ to denote the local representation of $\eta^{-1}_\mathsmaller{X}$.  In consequence, a straightforward calculation shows that the infinitesimal generator $\xi^{J^{1}Y}_{\eta}\in\mathfrak{X}^{\,1}(J^{1}Y)$ associated with the transformation $\eta_\mathsmaller{J^{1}Y}$ is explicitly given by
\begin{equation}\label{PJB}
\xi_{\eta}^{J^{1}Y}\!:=\xi^{\mu}(x^{\nu})\partial_{\mu}+\xi^{a}(x^{\nu},y^{b})\partial_{a}+\left(\partial_{\mu}\xi^{a}(x^{\nu},y^{b})+\partial_{c}\:\!\xi^{a}(x^{\nu},y^{b})y^{c}_{\mu}-\partial_{\mu}\:\!\xi^{\sigma}(x^{\nu})y^{a}_{\sigma}\right)\partial^{\mu}_{a}\, ,
\end{equation}
with $\partial^{\mu}_{a}:=\partial / \partial y^{a}_{\mu}$ denoting the partial derivative with respect to the fibre coordinate $y^{a}_{\mu}$. Indeed, $\xi^{J^{1}Y}_{\eta}\!:=j^{1}\xi^{Y}_{\eta}$ corresponds to the so-called first jet prolongation of the vector field $\xi^{Y}_{\eta}$ (see for instance \cite{GIMMSY1, Saunders, Sardanashvily1}). Henceforward, for all $\eta\in \mathcal{G}$, the pair $\big(\eta_\mathsmaller{J^{1}Y}, \xi^{J^{1}Y}_{\eta}\big)$ consisting of a fibre-preserving transformation $\eta_\mathsmaller{J^{1}Y}:J^{1}Y\rightarrow J^{1}Y$ and its corresponding infinitesimal generator $\xi^{J^{1}Y}_{\eta}\in \mathfrak{X}^{\,1}(J^{1}Y)$ will be just referred to as an infinitesimal transformation on $J^{1}Y$. 

Thus, in the spirit of \cite{GIMMSY1, DeLeon1}, we say that $\mathcal{G}$ is the symmetry group of theory if for all $\eta\in \mathcal{G}$, the associated infinitesimal transformation $\big(\eta_\mathsmaller{{J^{1}Y}}, \xi^{J^{1}Y}_{\eta}\big)$ on $J^{1}Y$ satisfies the condition
\begin{equation}
\label{eq:noethersym}
\mathfrak{L}_{\xi_{\eta}^{J^{1}\!Y}}\Theta^{(\mathcal{L})}=d\alpha^{(\mathcal{L})}_{\eta}\, ,
\end{equation}
where $\mathfrak{L}_{\xi^{J^{1}Y}_{\eta}}$ symbolizes the Lie derivative along the vector field $\xi^{J^{1}Y}_{\eta}\in \mathfrak{X}^{\,1}(J^{1}Y)$, while $\alpha^{(\mathcal{L})}_{\eta}\in \Omega^{\,n-1}_{\,1}(J^{1}Y)$ denotes a $\fibbun{\,J^{1}Y}{X}$-horizontal $(n-1;1)$-form on $J^{1}Y$ locally represented by
\begin{equation}\label{Lagrangian alpha}
\alpha^{(\mathcal{L})}_{\eta}=\alpha_{\eta}^{\nu}(x^{\mu},y^{a})d^{\,n-1}x_{\nu}\, .
\end{equation} 
From now on, for all $\eta\in \mathcal{G}$, the pair $\left(\xi^{J^{1}Y}_{\eta}\!,\alpha^{(\mathcal{L})}_{\eta} \right)$ consisting of a vector field  $\xi^{J^{1}Y}_{\eta}\in \mathfrak{X}^{\,1}(J^{1}Y)$ and a $\fibbun{\,J^{1}Y}{X}$-horizontal $(n-1;1)$-form $\alpha^{(\mathcal{L})}_{\eta}\in \Omega^{\,n-1}_{\,1}(J^{1}Y)$ related through condition \eqref{eq:noethersym} will be referred to as a Noether symmetry \cite{DeLeon1}. 

Bearing this in mind, given the dual pairing $\langle \cdot , \cdot \rangle:\mathfrak{g}^{*}\times \mathfrak{g}\rightarrow \mathbb{R}$, it is possible to see that, the action of $\mathcal{G}$ on $J^{1}Y$ has an associated Lagrangian covariant momentum map \cite{Marsden}, that is, a map $J^{(\mathcal{L})}:J^{1}Y\rightarrow \mathfrak{g}^{*}\otimes\Lambda^{n-1}T^{\,*}\!J^{1}Y$ such that for all $\xi_{\eta}\in \mathfrak{g}$,
\begin{equation}\label{Lagrangian Momentum Map}
d\,J^{\left(\mathcal{L}\right)}\!\left(\xi_{\eta}\right)=\xi_{\eta}^{J^{1}Y}\!\!\lrcorner\,\,\Omega^{(\mathcal{L})}\, ,
\end{equation}
where $J^{(\mathcal{L})}(\xi_{\eta}):=\langle J^{(\mathcal{L})},\xi_{\eta} \rangle\in \Omega^{\,n-1}(J^{1}Y)$ corresponds to the $(n-1)$-form on $J^{1}Y$ explicitly given by
\begin{equation}
\label{eq:LagCurrent}
J^{\left(\mathcal{L}\right)}\left(\xi_{\eta}\right)=\xi_{\eta}^{
J^{1}Y}\!\lrcorner\,\Theta^{(\mathcal{L})}-\alpha ^{(\mathcal{L})}_{\eta}\, .
\end{equation}
In particular, the Lagrangian covariant momentum map is relevant as it allows to construct conserved currents for the solutions of the Euler-Lagrange field equations of the system. In other words, given $\xi_{\eta}\in\mathfrak{g}$ and $\phi\in\mathscr{Y}_{X}$ a critical point of the action principle \eqref{APFT}, the quantity defined as
\begin{equation}\label{NoethCurrDef}
\mathcal{J}^{(\mathcal{L})}\!\left(\xi_{\eta}\right):=(j^{1}\phi)^{*}J^{(\mathcal{L})}\!\left(\xi_{\eta}\right)\, ,
\end{equation} 
corresponds to a conserved current of the theory \cite{DeLeon1}. To see this, it is enough to consider that   a critical point of the action \eqref{APFT},
 $\phi\in\mathscr{Y}_{X}$,  satisfies the condition \eqref{ELFE} and therefore we get that $(j^{1}\phi)^{*}\big(\xi_{\eta}^{J^{1}\!Y}\!\!\lrcorner\,\Omega^{(\mathcal{L})}\big)=d\big((j^{1}\phi)^{*}J^{\left(\mathcal{L}\right)}\!\left(\xi_{\eta}\right)\big)=0$. Concisely, the above result corresponds to the formulation of the first Noether theorem within the geometric-covariant Lagrangian formalism \cite{GIMMSY1, Crampin, DeLeon1, Fischer, Sardanashvily1}.

Further, in order to adapt our geometric formalism to gauge field theories, we will introduce the notion of localizable symmetries. To this end, we will closely follow the definition of this kind of symmetries as presented in \cite{Fischer, Marsden, Lee, Avery}. Thus, a set $\mathcal{C}_\mathsmaller{\mathrm{\,LS}}$ of pairs $\left(\xi^{J^{1}Y}, \alpha^{(\mathcal{L})}\right)$ consisting of a vector field $\xi^{J^{1}Y}\in \mathfrak{X}^{\,1}(J^{1}Y)$ and a $\fibbun{\,J^{1}Y}{X}$-horizontal $(n-1;1)$-form $\alpha^{(\mathcal{L})}\in \Omega^{\,n-1}_{\,1}(J^{1}Y)$ (locally represented as in \eqref{Lagrangian alpha}) is said to be a collection of localizable symmetries if the following conditions are satisfied: $(i)$ $\mathcal{C}_\mathsmaller{\mathrm{LS}}$ forms a vector space. $(ii)$ Each pair $\left(\xi^{J^{1}Y}\!, \alpha^{(\mathcal{L})}\right)\in \mathcal{C}_\mathsmaller{\mathrm{LS}}$ is a Noether symmetry. $(iii)$ For every pair $\left(\xi^{J^{1}Y}\!, \alpha^{(\mathcal{L})}\right)\in \mathcal{C}_\mathsmaller{\mathrm{LS}}$ and any two open sets $U_{1}, U_{2}\subset X$ with disjoint closures there exists a pair $\left(\tilde{\xi}^{J^{1}Y}\!, \tilde{\alpha}^{(\mathcal{L})}\right)\in \mathcal{C}_\mathsmaller{\mathrm{LS}}$ satisfying
\begin{equation}\label{Condition 3}
\begin{aligned}[b]
\left. 
\begin{aligned}
\xi^{J^{1}Y}(\beta)&=\tilde{\xi}^{J^{1}Y}(\beta)\\
\alpha^{(\mathcal{L})}(\beta)&=\tilde{\alpha}^{(\mathcal{L})}(\beta)
\end{aligned}
~\right\} &\,\forall\, \beta\in \fibbun{\,J^{1}Y}{X}^{-1}\left(U_{1}\right)\,,\\
\left. 
\begin{aligned}
\tilde{\xi}^{J^{1}Y}(\beta)&=0~~~\,\,\,\\
\tilde{\alpha}^{(\mathcal{L})}(\beta)&=0~~~\,\,\,
\end{aligned}
~~~~~\right\} &\,\forall\, \beta\in \fibbun{\,J^{1}Y}{X}^{-1}\left(U_{2}\right)\,.
\end{aligned}
\end{equation}
Basically, the above definition states that one may deform a Noether symmetry to zero in the fibres of $J^{1}Y$ over a certain region of the base space $X$. 

According to \cite{GIMMSY1, GIMMSY2, Fischer, Marsden, Lee, Avery}, the presence of localizable symmetries in a classical field theory gives rise to trivial Lagrangian Noether charges. In other words, we know that, given $\Sigma_{t}$ a Cauchy surface of $X$ (a compact oriented $(n-1)$-dimensional submanifold of $X$ without boundary), $\left({\xi}^{J^{1}Y}\!, \alpha^{(\mathcal{L})}\right)\in \mathcal{C}_\mathsmaller{\mathrm{LS}}$ a localizable symmetry and $\phi\in\mathscr{Y}_{X}$ a critical point of the action principle \eqref{APFT}, the associated Lagrangian Noether charge vanishes, namely
\begin{equation}\label{NoetCharVanishing}
\mathcal{Q}^{\left(\mathcal{L}\right)}_{\,\Sigma_{t}}\!\left(\xi\right):=\int_{\Sigma_{t}}(j^{1}\phi\circ i_{t}
)^{*}J^{\left(\mathcal{L}\right)}\!\left(\xi\right)=0\, ,
\end{equation}
where $i_{t}:\Sigma_{t}\rightarrow X$ denotes the inclusion map. In fact, this last corresponds to the second Noether theorem, which states that the vanishing of the Lagrangian Noether charges associated with each of the localizable symmetries of the theory is a necessary condition to extend $j^{1}\phi\circ i_{t}:\Sigma_{t}\rightarrow J^{1}Y$ (a section $j^{1}\phi\in \mathscr{J}^{1}\mathscr{Y}_{X}$ restricted to a Cauchy surface $\Sigma_{t}$) to a solution of the Euler-Lagrange field equations of the system. In consequence, the latter is one of the main sources of constraints on the space of Cauchy data for the evolution equations of the classical field theories with localizable symmetries \cite{Fischer}. 

In the following subsection, we will introduce the multisymplectic framework, which will allow us to study classical field theory in a covariant Hamiltonian-like formulation. In particular, we will focus our attention on describing how the symmetries of a generic classical field theory arise within the multisymplectic
approach.

\subsection{Multisymplectic formalism}
\label{SectionMulti}\label{sec:multi}

Next, we will give a brief description of the multisymplectic formalism, which provides a finite-dimensional, geometric and covariant Hamiltonian-like formulation for classical field theory. As in the previous subsection, we will pay special attention to the study of the symmetries of a given classical field theory within this covariant framework.  These symmetries will allow us to introduce Noether's theorems  within the context of the De Donder-Weyl canonical theory. 

In order to develop a covariant Hamiltonian-like formulation for an arbitrary  classical field theory, we first need to construct a suitable arena where such formulation will take place. To this end, given a covariant configuration space for the classical field theory \eqref{APFT} characterized by the fibre-bundle $(Y,\fibbun{\,Y}{X},X)$,  we define $(Z^{\star}:=\Lambda^{n}_{2}\,T^{*}Y,\fibbun{\,Z^{\star}}{Y}, Y)$ as the covariant multimomenta phase-space associated with the system, namely the vector bundle of horizontal $(n\,;2)$-forms over $Y$. In local coordinates, an element $\Xi\in Z^{\star}_{y}$ can be written as
\begin{equation}
\label{eq:upperxi}
\Xi:=p\,d^{\,n}x+p^{\mu}_{a}dy^{a}\wedge d^{\,n-1}x_{\mu}\, ,
\end{equation}
which allows us to identify $(x^{\mu},y^{a},p,p^{\mu}_{a})$ as an adapted coordinate system on $Z^{\star}$. Additionally, the composition map $\fibbun{\,Z^{\star}}{X}:=\fibbun{\,Y}{X}\circ\fibbun{\,Z^{\star}}{Y}$ yields the bundle structure $(Z^{\star}, \fibbun{\,Z^{\star}}{X}, X)$. Here, in agreement with \cite{IKCSCHF, IKCSPPS}, the fibre coordinates $(p^{\mu}_{a})$ will be referred to as the polymomenta, while the fibre coordinate $p$ will be related with the covariant Hamiltonian of the theory. In fact, it is important to mention that the covariant multimomenta phase-space may be identified with the affine dual jet bundle $(J^{1}Y^{\star},\fibbun{\,J^{1}Y^{\star}}{Y}, Y)$, that is, the affine dual bundle corresponding to the affine jet bundle $\fibbun{\,J^{1}Y}{Y}$ \cite{Forger1} (see also \cite{Saunders, Sardanashvily1} for technical details about dual bundles). This identification may be realized by 
noticing that the spaces $J^{1}Y^{\star}$ and $Z^{\star}$ are canonically isomorphic as vector bundles over $Y$ \cite{GIMMSY1}. As we will see below, the covariant multimomenta phase-space will allow us to introduce a well-defined arena for a covariant Hamiltonian analysis of the classical field theory of our interest. 

In particular, we would like to emphasize that, the vector space $Z^{\star}$ is endowed with a canonical $n$-form $\Theta^{(\mathcal{M})}\in\Omega^{n}(Z^{\star})$, which is locally given by\footnote{Here we must clarify that, even though the local representations of $\Xi$ and $\Theta^{(\mathcal{M})}$ may seem indistinguishable, however, they may not be confused as the former corresponds to an arbitrary element of the fibre of $Z^{\star}$ over a point $y\in Y$, while the latter stands for an $(n+1)$-form on $Z^{\star}$, regarding $Z^{\star}$ as a manifold.}
\begin{equation}
\label{eq:potential}
\Theta^{(\mathcal{M})}:=p\, d^{\,n}x + p^{\mu}_{a}dy^{a}\wedge d^{\,n-1}x_{\mu}\, .
\end{equation}
This canonical $n$-form is also known as the multisymplectic potential since it induces the so-called multisymplectic $(n+1)$-form through the relation $\Omega^{(\mathcal{M})}:=-d\,\Theta^{(\mathcal{M})}\in\Omega^{n+1}(Z^{\star})$, or specifically
\begin{equation}
\label{eq:omega}
\Omega^{(\mathcal{M})}=dy^{a}\wedge dp^{\mu}_{a}\wedge d^{\,n-1}x_{\mu}-dp\wedge d^{\,n}x\, ,
\end{equation}
thus defining the pair $\big(Z^{\star},\Omega^{(\mathcal{M})}\big)$ as a multisymplectic manifold \cite{Forger1, GIMMSY1, Crampin, DeLeon1}. 

Besides, given $L:J^{1}Y\rightarrow \mathbb{R}$ the Lagrangian function that characterizes the classical field theory \eqref{APFT}, we have that the affine jet bundle and the covariant multimomenta phase-space may be related through the covariant Legendre transformation $ \mathbb{F}\mathcal{L}:J^{1}Y\rightarrow Z^{\star}$, namely the bundle map over $Y$ locally defined by 
\begin{equation}\label{CLT}
\mathbb{F}\mathcal{L}(x^{\mu},y^{a},y^{a}_{\mu}):=\left(x^{\mu}, y^{a},\;\!p=L-\frac{\partial L}
{\partial y^{a}_{\mu}}y^{a}_{\mu},\;\!p^{\mu}_{a}=\frac{\partial L}{\partial y^{a}
_{\mu}}\right)\, .
\end{equation} 
Thus, by using the covariant Legendre transformation~\eqref{CLT}, we may induce information of the classical field theory under study from the affine jet bundle to the covariant multimomenta phase-space and vice versa \cite{GIMMSY1, Crampin, DeLeon1}. For instance, it is not difficult to see that, we can obtain the Poincar{\'e}-Cartan forms \eqref{PCforms} by means of the following relations
\begin{equation}
\begin{aligned}\notag
\Theta^{(\mathcal{L})}&=\mathbb{F}\mathcal{L}^{*}\Theta^{(\mathcal{M})}\, ,\\ \Omega^{(\mathcal{L})}&=\mathbb{F}\mathcal{L}^{*}\Theta^{(\mathcal{M})}\, .
\end{aligned}
\end{equation}

Now, it is important to mention that the covariant multimomenta phase-space has as a 
subbundle the vector bundle $(\Lambda^{n}_{1}\,T^{*}Y,\fibbun{\, \Lambda^{n}_{1}T^{*}Y}{Y}, Y)$, that is, the vector bundle of horizontal $(n\,;1)$-forms over $Y$, which in turn allows us to introduce a new vector bundle over $Y$ given by the quotient bundle
\begin{equation}\label{Quotientbundle}
P:=Z^{\star}/\Lambda^{n}_{1}\,T^{*}Y\, .
\end{equation}
In local coordinates an element $\vartheta\in P_{y}$ can be written as
\begin{equation}
\vartheta:=p^{\mu}_{a}\,dy^{a}\wedge d^{\,n-1}x_{\mu}\, ,
\end{equation}
which allows us to identify $(x^{\mu},y^{a},p^{\mu}_{a})$ as an adapted coordinate system on $P$. Consequently, the composition map $\fibbun{\,P}{X}:=\fibbun{\,Y}{X}\circ \fibbun{\,P}{Y}$ gives rise to the so-called polymomenta phase-space $(P,\fibbun{\,P}{X}, X)$ \cite{IKCSCHF, IKCSPPS, IKGPA}. Hence, a section $\varrho\in \mathscr{P}_{X}$ of $\fibbun{\,P}{X}$ covering $\phi=\fibbun{\,P}{Y}(\varrho)\in \mathscr{Y}_{X}$ can be locally represented by $(x^{\mu},\phi^{a}(x^{\mu}),\pi^{\nu}_{a}(x^{\mu}))$. In addition, the definition of quotient bundle \eqref{Quotientbundle} yields the bundle structure $(Z^{\star},\fibbun{\,Z^{\star}}{P},P)$, where a section $h\in\mathscr{Z}^{\star}_{P}$ is said to be a Hamiltonian section if it locally reads
\begin{equation}
h\left(x^{\mu},y^{a},p^{\mu}_{a}\right)=\left(x^{\mu},y^{a},p=-H\left(x^{\mu},y^{a},p^{\mu}_{a}\right), p^{\mu}_{a}\right)\, ,
\end{equation} 
being $H$ the denominated Hamiltonian function associated with $h$ \cite{DeLeon1}.

Transitively, the affine jet bundle and the quotient bundle \eqref{Quotientbundle} 
are related through the covariant Legendre map $\mathbb{F}_{\mathsmaller{\mathrm{DW}}}\mathcal{L}:J^{1}Y\rightarrow P$, namely the bundle map over $Y$ locally given by
\begin{equation}\label{LegMap}
\mathbb{F}_\mathsmaller{\mathrm{DW}}\mathcal{L}(x^{\mu},y^{a},y^{a}_{\mu}):=\left(x^{\mu}, y^{a},\;\!p^{\mu}_{a}=\frac{\partial L}{\partial y^{a}
_{\mu}}\right)\, ,
\end{equation}
which allows us to define the denominated De Donder-Weyl Hamiltonian section $h_{\mathsmaller{\mathrm{DW}}}  \in\mathscr{Z}^{\star}_{P}$, that is, a section of $\fibbun{\,Z^{\star}}{P}$ satisfying the condition $h_{\mathsmaller{\mathrm{DW}}}  \circ\mathbb{F}_\mathsmaller{\mathrm{DW}}\mathcal{L}=\mathbb{F}\mathcal{L}$ and whose Hamiltonian function is identified with the so-called De Donder-Weyl Hamiltonian \cite{Crampin}, specifically
\begin{equation}\label{CanonicDeDondWeylHamDef}
H_\mathsmaller{\mathrm{DW}}(x^{\mu},y^{a},p^{\mu}_{a}):=p^{\mu}_{a}y^{a}_{\mu}-L\left(x^{\mu},y^{a},y^{a}_{\mu}\right)\, .
\end{equation} 

Bearing this in mind, we have that, it is possible to induce differential forms on $P$ by pulling-back differential forms on $Z^{\star}$ with a Hamiltonian section $h\in\mathscr{Z}^{\star}_{P}$. For instance, the De Donder-Weyl forms, $\Theta^{(\mathcal{P})}:=h_{\mathsmaller{\mathrm{DW}}}  ^{*}\Theta^{(\mathcal{M})}\in\Omega^{\,n}(P)$ and $\Omega^{(\mathcal{P})}:=h_{\mathsmaller{\mathrm{DW}}}  ^{*}\Omega^{(\mathcal{M})}\in\Omega^{\,n+1}(P)$, which explicitly read
\begin{subequations}\label{DWForms}
\begin{align}
\Theta^{(\mathcal{P})}&=p^{\mu}_{a}\,dy^{a}\wedge d^{\,n-1}x_{\mu}-H_{\mathsmaller{\mathrm{DW}}}  \,d^{\,n}x\, ,\label{DWCanonForm}\\
\Omega^{(\mathcal{P})}&=\,dy^{a}\wedge dp^{\mu}_{a}\wedge d^{\,n-1}x_{\mu}+dH_{\mathsmaller{\mathrm{DW}}}  \wedge d^{\,n}x\, . \label{DWpolyForm}
\end{align}
\end{subequations}
In a similar fashion to the Poincaré-Cartan forms~\eqref{PCforms} within the Lagrangian formalism, the De Donder-Weyl forms \eqref{DWForms} will be the main geometric objects of interest within the covariant Hamiltonian-like formulation for the classical field theory. In fact, it is not difficult to see that we may recover the Poincar{\'e}-Cartan forms  by means of the following expressions
\begin{equation}\notag
\begin{aligned}
\Theta^{(\mathcal{L})}=\mathbb{F}_\mathsmaller{\mathrm{DW}}\mathcal{L}^{*}\Theta^{(\mathcal{P})}\, ,\\
\Omega^{(\mathcal{L})}=\mathbb{F}_\mathsmaller{\mathrm{DW}}\mathcal{L}^{*}\Omega^{(\mathcal{P})}\, .
\end{aligned}
\end{equation}  

From our point of view, the De Donder-Weyl forms are completely relevant as, 
on the one side, we may rewrite the action principle \eqref{APFT} in terms of the 
$n$-form~\eqref{DWCanonForm} and a section $\varrho\in \mathscr{P}_{X}$ (see for instance \cite{Forger1}) while, on the other side, the De Donder-Weyl 
$(n+1)$-form~\eqref{DWpolyForm} is necessary in order to write the De Donder-Weyl-Hamilton field equations of the system in an invariant fashion \cite{Crampin}. In other words, let us consider $\varrho\in \mathscr{P}_{X}$ a section locally represented by $(x^{\mu}, \phi^{a}(x^{\mu}), \pi^{\nu}_{a}(x^{\mu}))$. Then, we have that for an arbitrary vector field on $P$, 
namely $W\in \mathfrak{X}^{\,1}(P)$, a section $\varrho\in \mathscr{P}_{X}$ satisfying the condition
\begin{equation}\label{DWGeoEqua}
\varrho^{*}\left(W\lrcorner\,\Omega^{(\mathcal{P})}\right)=0\, ,
\end{equation}
corresponds to a solution of the De Donder-Weyl-Hamilton field equations of the theory, namely
\begin{equation}\label{DeDonWeylEquDef}
\begin{aligned}[b]
\partial_{\mu}\phi^{a}&=\frac{\partial H}{\partial p^{\mu}_{a}}\left(x^{\mu},\phi^{a},\pi^{\mu}_{a}\right)\, ,\\
\partial_{\mu}\pi^{\mu}_{a}&=-\frac{\partial H}{\partial y^{a}}\left(x^{\mu},\phi^{a},\pi^{\mu}_{a}\right)\, .
\end{aligned}
\end{equation}

Of course, for a non-singular Lagrangian system, that is, a classical field theory characterized by a Lagrangian function obeying the regularity condition
\begin{equation}\label{Regularity Condition}
\mathrm{det}\left(\frac{\partial^{2}L}{\partial y^{a}_{\mu}\partial y^{b}_{\nu}}\right)\neq 0\, ,
\end{equation}
the Euler-Lagrange field equations \eqref{Euler-Lagrange equations} are completely equivalent to the De Donder-Weyl-Hamilton field equations \eqref{DeDonWeylEquDef}. To see this, we start by emphasizing that for a non-singular Lagrangian system the covariant Legendre map \eqref{LegMap} is a diffeomorphism \cite{DeLeon1}. Therefore, given $V\in\mathfrak{X}^{\,1}\left(J^{1}Y\right)$ and $W:=T\,\mathbb{F}_\mathsmaller{\mathrm{DW}}\mathcal{L}\left(V\right)\in\mathfrak{X}^{\,1}\left(P\right)$ a pair of arbitrary vector fields, a straightforward calculation shows that, if a section $j^{1}\phi\in \mathscr{J}^{1}\mathscr{Y}_{X}$ is a solution of the Euler-Lagrange field equations \eqref{Euler-Lagrange equations}, then the section $\varrho:=\mathbb{F}_\mathsmaller{\mathrm{DW}}\mathcal{L}\circ j^{1}\phi\in\mathscr{P}_{X}$ is a solution of the De Donder-Weyl Hamilton field equations \eqref{DeDonWeylEquDef}, and vice versa.  Specifically
\begin{equation}\notag
\begin{aligned}
\varrho^{*}\left(W\,\lrcorner\,\Omega^{(\mathcal{P})}\right)&= (j^{1}\phi)^{*}\big( T\,\mathbb{F}_\mathsmaller{\mathrm{DW}}\mathcal{L}^{-1}(W)\lrcorner \,\mathbb{F}_\mathsmaller{\mathrm{DW}}\mathcal{L}^{*} \Omega^{(\mathcal{P})}\big)\, ,\\
&=(j^{1}\phi)^{*}\big( V\lrcorner \,\Omega^{(\mathcal{L})}\big)\, ,\\
&= 0\, .
\end{aligned}
\end{equation}
Here, we would like to mention that the covariant Legendre map \eqref{LegMap}, the De Donder-Weyl Hamiltonian \eqref{CanonicDeDondWeylHamDef} and the De Donder-Weyl-Hamilton field equations \eqref{DeDonWeylEquDef} are the pillars of the De Donder-Weyl canonical theory which appeared as early as 1935 in~\cite{DDT, WT}. 

Now, we will analyze the symmetries of the classical field theory \eqref{APFT} within the multisymplectic approach. In particular, we will focus our attention on the action of the symmetry group of the system on the covariant multimomenta phase-space. For this purpose, we will start by introducing the notion of covariant canonical transformations, which will allow us to construct the so-called covariant momentum map and, consequently, the conserved currents for the solutions of the De Donder-Weyl-Hamilton field equations of the theory. 

To start, let $\left( Z^{\star},\fibbun{\,Z^{\star}}{Y},Y\right)$ be the covariant multimomenta phase-space associated with the classical field theory defined by the action~\eqref{APFT}. Then, in accordance with \cite{GIMMSY1, Crampin}, a $\fibbun{\,Z^{\star}}{X}$-bundle automorphism $(\Phi_\mathsmaller{Z^{\star}}, \Phi_\mathsmaller{X})$, namely a pair of maps $\Phi_\mathsmaller{Z^{\star}}: Z^{\star}\rightarrow Z^{\star}$ and $\Phi_\mathsmaller{X}:X\rightarrow X$ satisfying the condition $\fibbun{\,Z^{\star}}{X}\circ\Phi_\mathsmaller{Z^{\star}}=\Phi_\mathsmaller{X}\circ\fibbun{\,Z^{\star}}{X}$, is said to be a covariant canonical transformation if it preserves the multisymplectic $(n+1)$-form \eqref{eq:omega}, specifically $\Phi_{\mathsmaller{Z^{\star}}}^{*}\Omega^{(\mathcal{M})}=\Omega^{(\mathcal{M})}$. Therefore, if $\Phi_\mathsmaller{Z^{\star}}$ is the vector flow associated with a vector field $\xi^{Z^{\star}}_{\Phi}\in\mathfrak{X}^{\,1}(Z^{\star})$, it is clear that
\begin{equation}
\mathfrak{L}_{\xi^{Z^{\star}}_{\Phi}}\Omega^{(\mathcal{M})}=0\, . 
\end{equation} 

Bearing this in mind, we are now in the position to describe how fibre-preserving transformations on the covariant configuration space, for instance, those generated by the symmetry group of the theory, transitively induce covariant canonical transformations on the covariant multimomenta phase-space associated with the system. To see this, let us consider $\mathcal{G}$ the symmetry group of the theory. Then, given $(\eta_\mathsmaller{Y}, \eta_\mathsmaller{X})$ the $\fibbun{\,Y}{X}$-bundle automorphism associated with an element $\eta\in\mathcal{G}$, we introduce $\eta_\mathsmaller{Z^{\star}}:Z^{\star}\rightarrow Z^{\star}$ to be the canonical lift of $\eta_\mathsmaller{Y}$ to $Z^{\star}$. Thus, since $\fibbun{\,Z^{\star}}{Y}$ is a vector bundle of differential forms over $Y$, at $y\in Y$, being $\Xi:\Lambda^{n}\,T_{y}Y\rightarrow \mathbb{R}$ an element of $Z^{\star}_{y}$, we have that, $\eta_\mathsmaller{Z^{\star}_{y}}:Z^{\star}_{y}\rightarrow Z^{\star}_{\eta_{Y}(y)}$ is defined by $\eta_\mathsmaller{Z^{\star}_{y}}(\Xi)(V_{1},\cdots\!,V_{n})=\Xi\,((T_{\,y\,}\eta_\mathsmaller{Y})^{-1}(V_{1}),\cdots\!, (T_{\,y\,}\eta_\mathsmaller{Y})^{-1}(V_{n}))$, $\forall\, V_{1},\cdots\!,V_{n}\in T_{\eta_{Y}(y)}Y$, namely $\eta_\mathsmaller{Z^{\star}_{y}}(\Xi):={\eta_{\mathsmaller{Y}}^{-1}}^{*}\,\Xi$. Therefore, given the infinitesimal generators of the transformations $\eta_\mathsmaller{X}$ and $\eta_\mathsmaller{Y}$, ${\xi}_{\eta}^{X}\in\mathfrak{X}^{1}
\left(X\right)$ and  ${\xi}_{\eta}^{Y}
\in\mathfrak{X}^{\,1}\left(Y\right)$, respectively, we get that the vector field $\xi^{Z^{\star}}_{\eta}\in\mathfrak{X}^{\,1}(Z^{\star})$ on $Z^{\star}$ explicitly defined by
\begin{equation}\label{CLMMPS}
\begin{aligned}[b]
{\xi}_{\eta}^{Z^{\star}}\!&:=\xi^{\mu}(x^{\nu})\partial_{\mu}+\xi^{a}(x^{\nu}, y^{b})\partial_{a}-\left(p\,\partial_{\mu}\xi^{\mu}(x^{\nu})+p^{\mu}_{a}\partial_{\mu}\xi^{a}(x^{\nu},y^{b})\right)\partial_{p}\\
&~~~-\left(p^{\mu}_{c}\partial_{a}\xi^{c}(x^{\nu},y^{b})-p^{\sigma}_{a}\partial_{\sigma}\xi^{\mu}(x^{\nu})+p^{\mu}_{a}\partial_{\sigma}\xi^{\sigma}(x^{\nu})\right)\partial^{a}_{\mu}\, ,
\end{aligned}
\end{equation}
corresponds to the infinitesimal generator of the transformation $\eta_{Z^{\star}}$, where we have introduced the notation $\partial_{p}:=\partial/\partial p$ and $\partial^{a}_{\mu}:=\partial/\partial p^{\mu}_{a}$ to denote the partial derivative with respect to the fibre coordinates $p$ and $p^{\mu}_{a}$, respectively. In fact, it is not difficult to show that the canonical lift preserves the multisymplectic potential \eqref{eq:potential}, namely $\eta_{Z^{\star}}^{*}\Theta^{(\mathcal{M})}=\Theta^{(\mathcal{M})}$, and  hence $\mathfrak{L}_{\xi_{\eta}^{\,Z^{\star}}}\Theta=0$. As a result, we have that, the canonical lifts associated with $\fibbun{\,Y}{X}$-bundle automorphisms give rise to covariant canonical transformations \cite{GIMMSY1}.

Besides, we know that, if $\mathcal{G}$ corresponds to the symmetry group of the theory, then for all $\eta\in\mathcal{G}$ there is a corresponding Noether symmetry $(\xi^{J^{1}Y}_{\eta},\alpha^{(\mathcal{L})}_{\eta})$, namely a pair consisting of a vector filed $\xi^{J^{1}Y}_{\eta}\in\mathfrak{X}^{1}(J^{1}Y)$ and a horizontal $(n-1;1)$-form $\alpha^{(\mathcal{L})}_{\eta}\in\Omega^{n-1}_{1}(J^{1}Y)$ related through condition \eqref{eq:noethersym}, being $\xi^{J^{1}Y}_{\eta}:=j^{1}\xi^{Y}_{\eta}$ the first jet prolongation of $\xi^{Y}_{\eta}\in\mathfrak{X}^{1}(Y)$. Then, for any $\eta\in\mathcal{G}$, we define the so-called $\alpha^{(\mathcal{M})}_{\eta}$-lift of $\xi^{Y}_{\eta}\in \mathfrak{X}^{\,1}(Y)$ to $Z^{\star}$ as the unique vector field $\xi^{\alpha}_{\eta}\in \mathfrak{X}^{\,1}(Z^{\star})$ that projects by means of $T\fibbun{\,Z^{\star}}{Y}:TZ^{\star}\rightarrow TY$ onto $\xi^{Y}_{\eta}$ and also satisfies the condition
\begin{equation}
\mathfrak{L}_{\xi^{\alpha}_{\eta}}\Theta^{(\mathcal{M})}=d\alpha^{(\mathcal{M})}_{\eta}\, ,
\end{equation} 
where $\alpha^{(\mathcal{M})}_{\eta}\in \Omega^{\,n-1}_{1}(Z^{\star})$ denotes a $\fibbun{\,Z^{\star}}{X}$-horizontal $(n-1;1)$-form on $Z^{\star}$ locally represented as 
\begin{equation}
\alpha^{(\mathcal{M})}_{\eta}=\alpha^{\nu}_{\eta}(x^{\mu},y^{a})d^{\,n-1}x_{\nu}\, .
\end{equation}
Thus, the vector field $\xi^{\alpha}_{\eta}$ not only contains the information associated with the canonical lift \eqref{CLMMPS} but also with the $\fibbun{\,Z^{\star}}{X}$-horizontal $(n-1;1)$-form $\alpha^{(\mathcal{M})}_{\eta}$. Here, it is important to emphasize that, in local coordinates, the components of $\alpha^{(\mathcal{L})}_{\eta}\in \Omega^{\,n-1}_{1}(J^{1}Y)$ and $\alpha^{(\mathcal{M})}_{\eta}\in\Omega^{\,n-1}_{1}(Z^{\star})$ are the same, and therefore $\alpha^{(\mathcal{L})}_{\eta}=\mathbb{F}\mathcal{L}^{*}\alpha^{(\mathcal{M})}_{\eta}$. As pointed out in \cite{DeLeon1}, the relevance of the $\alpha^{(\mathcal{M})}$-lifts lies in the fact that, for $\eta\in \mathcal{G}$, the condition $T_{\,\beta\,}\mathbb{F}\mathcal{L}(\xi^{J^{1}Y}_{\eta}(\beta))=\xi^{\alpha}_{\eta}( \mathbb{F}\mathcal{L}(\beta))$ holds, being $T_{\,\beta\,}\mathbb{F}\mathcal{L}:T_{\,\beta\,}J^{1}Y\rightarrow T_{\,\mathbb{F}\mathcal{L}(\beta)\,}Z^{\star}$ the tangent map of $\mathbb{F}\mathcal{L}$ at $\beta\in J^{1}Y$. 

With all this in mind, given $\eta\in\mathcal{G}$, we introduce $\big(\eta^{\alpha}_{Z^{\star}},\xi^{\alpha}_{\eta}\big)$ to represent the action of $\eta$ on $Z^{\star}$, where $\eta^{\alpha}_{Z^{\star}}:Z^{\star}\rightarrow Z^{\star}$ stands for the vector flow associated with the vector field $\xi^{\alpha}_{\eta}\in\mathfrak{X}^{\,1}(Z^{\star})$. Therefore, since the $\alpha^{(\mathcal{M})}$-lifts give rise to covariant canonical transformations, the action of $\mathcal{G}$ on $Z^{\star}$ has an associated covariant momentum map \cite{GIMMSY1}, that is, a map $J^{(\mathcal{M})}:Z^{\star}\rightarrow \mathfrak{g}^{*}\otimes \Lambda^{n-1}T^{\,*}Z^{\star}$ such that for all $\xi_{\eta}\in \mathfrak{g}$,
\begin{equation}\label{CMM}
d\;\!J^{(\mathcal{M})}\!\left(\xi_{\eta}\right)=\xi_{\eta}^{\alpha}\lrcorner\,\Omega^{(\mathcal{M})}\, ,
\end{equation}
where $J^{(\mathcal{M})}\!\left(\xi_{\eta}\right):=\langle J^{(\mathcal{M})}, \xi_{\eta}\rangle\in \Omega^{\,n-1}(Z^{\star})$ corresponds to the $(n-1)$-form on $Z^{\star}$ explicitly given by
\begin{equation}
\label{eq:MultiCurrent}
J^{(\mathcal{M})}\!\left(\xi_{\eta}\right)=\xi^{\alpha}_{\eta}\lrcorner\,\Theta^{(\mathcal{M})}-\alpha^{(\mathcal{M})}_{\eta}\, .
\end{equation}
In fact, for all $\xi_{\eta}\in \mathfrak{g}$, by pulling-back the covariant momentum map with the De Donder-Weyl Hamiltonian section $h_\mathsmaller{\mathrm{DW}}\in \mathscr{Z^{\star}}_{P}$, namely $J^{(\mathcal{P})}\!\left(\xi_{\eta}\right):=h^{*}_\mathsmaller{\mathrm{DW}}J^{(\mathcal{M})}\!\left(\xi_{\eta}\right)\in \Omega^{\,n-1}(P)$, we can construct a conserved current for the solutions of the De Donder-Weyl-Hamilton field equations \cite{Crampin, DeLeon1}. In other words, we have that, for all $\xi_{\eta}\in \mathfrak{g}$ and for any solution of the De Donder-Weyl-Hamilton field equations \eqref{DeDonWeylEquDef}, $\varrho\in \mathscr{P}_{X}$, the quantity defined by
\begin{equation}\label{Hamiltonian Noether currents}
\mathcal{J}^{(\mathcal{P})}(\xi_{\eta}):=\varrho^{*}J^{(\mathcal{P})}(\xi_{\eta})\, ,
\end{equation}
corresponds to a conserved current of the system. To see this, we start by emphasizing that, it is possible to write $J^{(\mathcal{L})}\!\left(\xi_{\eta}\right)=\mathbb{F}\mathcal{L}^{*}J^{(\mathcal{M})}\!\left(\xi_{\eta}\right)$. Furthermore, we know that, for a non-singular Lagrangian system the covariant Legendre map \eqref{LegMap} is a diffeomorphism, and hence if a section $\varrho\in\mathscr{P}_{X}$ is a solution of the De Donder-Weyl-Hamilton field equations \eqref{DeDonWeylEquDef}, then the section $j^{1}\phi:= \mathbb{F}_\mathsmaller{\mathrm{DW}}\mathcal{L}^{-1}\circ\varrho  \in \mathscr{J}^{1}\mathscr{Y}_{X}$ is a solution of the Euler-Lagrange field equations \eqref{Euler-Lagrange equations}. Thus, by considering the relation $h_\mathsmaller{\mathrm{DW}}=\mathbb{F}\mathcal{L}\circ \mathbb{F}_\mathsmaller{\mathrm{DW}}\mathcal{L}^{-1}$, we find that
\begin{equation}\notag
\begin{aligned}
d\big((j^{1}\phi)^{*}J^{(\mathcal{L})}\!\left(\xi_{\eta}\right)\!\big)&=d\big(\varrho^{*}\,(\mathbb{F}\mathcal{L}\circ\mathbb{F}_\mathsmaller{\mathrm{DW}}\mathcal{L}^{-1})^{*} J^{(\mathcal{M})}\!\left(\xi_{\eta}\right)\! \big)\, ,\\
&=d\big(\varrho^{*}h^{*}_\mathsmaller{\mathrm{DW}}J^{(\mathcal{M})}\!\left(\xi_{\eta}\right) \!\big)\, ,\\
&=d\big(\varrho^{*}J^{(\mathcal{P})}\!\left(\xi_{\eta}\right) \!\big)\, ,\\
&=0\, .
\end{aligned}
\end{equation}
The latter in light of the first Noether theorem, which states that, for $j^{1}\phi\in \mathscr{J}^{1}\mathscr{Y}_{X}$ a solution of the Euler-Lagrange field equations \eqref{Euler-Lagrange equations}, quantity \eqref{NoethCurrDef} defines a conserved current of the system.  

Finally, we have that, for localizable symmetries, given $\Sigma_{t}$ a Cauchy surface of $X$ and $\varrho\in\mathscr{P}_{X}$ a solution of the De Donder-Weyl-Hamilton field equations \eqref{DeDonWeylEquDef}, each Hamiltonian Noether charge must vanish \cite{GIMMSY1, GIMMSY2, Fischer}, namely
\begin{equation}
\mathcal{Q}^{\left(\mathcal{P}\right)}_{\,\Sigma_{t}}\!\left(\xi\right):=\int_{\Sigma_{t}}(\varrho\circ i_{t}
)^{*}J^{\left(\mathcal{P}\right)}\!\left(\xi\right)=0\, ,
\end{equation} 
where $i_{t}:\Sigma_{t}\rightarrow X$ denotes the inclusion map. As we will see in subsequent subsections, after performing the space plus time decomposition of the space-time manifold on which a classical field theory is defined, the covariant momentum map may induce the momentum map that characterizes the system under consideration within the instantaneous Dirac-Hamiltonian formulation. Thus, for classical field theories with localizable symmetries, the second Noether theorem establishes that, the vanishing of the momentum map on the space of Cauchy data for the evolution equations of the system will be related to the set of first-class constraints of the theory, as discussed in \cite{GIMMSY1, GIMMSY2, Fischer}. 

In the following subsection, we will give a brief description of the polysymplectic formalism, which endows the polymomenta phase-space with a Poisson-Gerstenhaber bracket that allows to describe the dynamics of a given classical field theory within a covariant Poisson-Hamiltonian framework. 

\subsection{Polysymplectic formalism}\label{PF}

Next, we will introduce the polysymplectic approach for classical field theory, which is a covariant Poisson-Hamiltonian framework. To this end, we will closely follow the description of the geometric and algebraic structures inherent to the aforementioned formalism as presented in \cite{IKQFTPV, IKHEQFT, IKPSWF, IKCSCHF, IKCSPPS, IKGPA, Angel}. Additionally, we will include a summary of the proposal developed in \cite{IKGD} to analyze singular Lagrangian systems within the polysymplectic approach.

To begin with, let $(P,\fibbun{\,P}{X}, X)$ be the polymomenta phase-space associated with the classical field theory described by the action~\eqref{APFT}. Then, given $(x^{\mu}, y^{a}, p^{\mu}_{a})$ an adapted coordinate system on $P$, we define the so-called polysymplectic structure, namely 
\begin{equation}\label{Polysymplectic form}
\Omega^{V}:=\big[~dp^{\mu}_{a}\wedge dy^{a}\wedge\varpi_{\mu}~\mathrm{mod}~\Omega^{\,n+1}_{\,2}(P)~\big]\, ,
\end{equation}
where $\varpi_{\mu}:=\partial_{\mu}\lrcorner\,\varpi$ denotes the basis for the horizontal $(n-1\,;1)$-forms on $P$, while $\varpi:=dx^{0}\wedge\cdots\wedge dx^{n-1}$ stands for the horizontal $(n\,;1)$ volume form on $P$.  From now on, we will consider only horizontal forms with respect to the projector map $\fibbun{\,P}{X}$. Note that, we have implemented an equivalence class of forms in the polysymplectic structure \eqref{Polysymplectic form} as an alternative for the introduction of a connection on the multimomenta phase-space in order to identify $\Omega^{V}\!$ as a vertical part of the negative of the multisymplectic $(n+1)$-form \eqref{eq:omega}. In fact, these two approaches are equivalent as the fundamental structures inherent to the polysymplectic formalism, for instance the Poisson-Gerstenhaber bracket, have been shown to be independent of both the choice of  representatives in the equivalence class and the choice of a connection \cite{CPVD, IKCSPPS, IKGPA}. 

From the physical point of view, the relevance of the polysymplectic form \eqref{Polysymplectic form} lies in the fact that we can appropriately define by means of it both a set of Hamiltonian vector fields and forms and a Poisson-Gerstenhaber bracket on the polymomenta phase-space. To see this, let us consider $\overset{~p}{X}\in \mathfrak{X}^{\,p}(P)$ a vertical $p$-multivector field, that is, a $p$-multivector field such that $\overset{~p}{X}\lrcorner \,\theta=0$, $\forall\theta\in\Omega^{\,p}_{\,1}(P)$. Then, a horizontal $\left(n-p\,;1\right)$-form $\!\overset{~n-p}{F}\!\in \Omega^{\,n-p}_{\,1}(P)$ is said to be a Hamiltonian $(n-p)$-form if there exists a vertical $p$-multivector field $\overset{~p}{X}_{\!F}\in\mathfrak{X}^{\,p}(P)$ satisfying the condition
\begin{equation}\label{HMFD}
\overset{~p}{X}_{\!F}\,\lrcorner\, \Omega^{V}=d^{\,V}\!\overset{~n-p}{F}\, ,
\end{equation}
where we have introduced the vertical exterior derivative $d^{\,V}:\Omega^{\,p}_{\,q}(P)\rightarrow \Omega^{\,p+1}_{\,q+1}(P)$, namely
\begin{equation}
d^{\,V}\theta:=\big[~d\theta~\mathrm{mod}~\Omega^{\,p+1}_{\,q}(P)~\big]\, ,
\end{equation}
being $d:\Omega^{\,p}_{\,q}(P)\rightarrow \Omega^{\,p+1}_{\,q+1}(P)$ the exterior derivative on $P$, while $\theta\in \Omega^{\,p}_{\,q}(P)$ stands for an arbitrary horizontal $(p\,;q)$-form. In consequence, we have that, when $n-p>0$, not every horizontal $(n-p\,;1)$-form is Hamiltonian since equation \eqref{HMFD} imposes a strong integrability condition \cite{Romer1, IKCSPPS}. Henceforward, we denote by $\Omega^{\,p}_\mathrm{H}(P)$ the set of Hamiltonian $p$-forms on the polymomenta phase-space. 

Bearing this in mind, we may introduce the so-called Poisson-Gerstenhaber bracket $\{\![\,\cdot, \cdot\,]\!\}: \Omega^{\,n-p}_\mathrm{H}(P)\times \Omega^{\,n-q}_\mathrm{H}(P)\rightarrow \Omega^{\,n+1-(p+q)}_\mathrm{H}(P)$, and thus given a pair of Hamiltonian forms, $\!\overset{~n-p}{F}\!\!\in\Omega^{\,n-p}_\mathrm{H}(P)$ and $\!\overset{~n-q}{G}\!\!\in\Omega^{\,n-q}_\mathrm{H}(P)$, it is explicitly defined by
\begin{equation}\label{PGD}
\{\![\overset{~n-p}{F}, \!\overset{~n-q}{G}\,]\!\}:=(-1\,)^{\,p}\overset{~p}{X}_{\!F}\,\lrcorner\,\overset{~q}{X}_{\!G}\,\lrcorner\,\Omega^{V}\, ,
\end{equation}
where $\overset{~p}{X}_{\!F}\in\mathfrak{X}^{\,p}(P)$ and $\overset{~q}{X}_{\!G}\in\mathfrak{X}^{\,q}(P)$ denote the Hamiltonian vector fields associated with the Hamiltonian forms $\overset{~n-p}{F}$ and $\overset{~n-q}{G}$, respectively. It is worth noting that, for $n+1\geqslant p+q$, the operation \eqref{PGD} among a pair of Hamiltonian forms induces a new Hamiltonian form, and hence the set of Hamiltonian forms $\Omega^{\,p}_\mathrm{H}(P)$ defines an algebra under the Poisson-Gerstenhaber bracket \eqref{PGD} \cite{CPVD, IKCSCHF, IKCSPPS, IKGPA}. 

In addition, it is possible to show that, given $\!\overset{~n-p}{F}\!\in\Omega^{\,n-p}_\mathrm{H}(P)$, $\!\overset{~n-q}{G}\!\in\Omega^{\,n-q}_\mathrm{H}(P)$ and $\!\overset{~n-r}{H}\!\in\Omega^{\,n-r}_\mathrm{H}(P)$ a set of arbitrary Hamiltonian forms, the Poisson-Gerstenhaber bracket \eqref{PGD} satisfies the following algebraic properties: 
\begin{itemize}
\item[(i)]
Graded-commutation
\begin{equation}
\{\![\!\overset{~n-p}{F},\!\overset{~n-q}{G}]\!\}=-(-1)^{|F||G|}\{\![\!\overset{~n-q}{G},\!\overset{~n-p}{F}]\!\}\, ,
\end{equation}
where $|F|:=p-1$ and $|G|:=q-1$ denote the degrees of the Hamiltonian forms $\!\overset{~n-p}{F}$ and $\!\overset{~n-q}{G}$ under the bracket structure \eqref{PGD}, respectively. 
\item[(ii)] 
Graded Jacobi identity 
\begin{equation}
\{\![\!\overset{~n-p}{F},\! \{\![\!\overset{~n-q}{G},\!\overset{~n-r}{H}]\!\}]\!\}=\{\![\{\![\!\overset{~n-p}{F},\! \overset{~n-q}{G}]\!\},\! \overset{~n-r}{H}]\!\}+(-1)^{|F||G|}\{\![\!\overset{~n-q}{G},\!\{\![\!\overset{~n-p}{F},\! \overset{~n-r}{H}]\!\}]\!\}\,.
\end{equation}
\item[(iii)] 
Graded Leibniz rule
\begin{equation}\label{GLeibnizR}
\{\![\overset{~n-p}{F},\overset{~n-q}{G}\bullet\overset{~n-r}{H}]\!\}=\{\![\overset{~n-p}{F}, \overset{~n-q}{G}]\!\}\bullet\overset{~n-r}{H}+(-1)^{q\,|F|}\overset{~n-q}{G}\bullet\{\![\overset{~n-p}{F}, \overset{~n-r}{H}]\!\}\, ,
\end{equation}
where the co-exterior product, $\bullet: \Omega^{\,n-q}_\mathrm{H}(P)\times \Omega^{\,n-r}_\mathrm{H}(P)\rightarrow \Omega^{\,n-(q+r)}_\mathrm{H}(P)$, is given by
\begin{equation}\label{CPD}
\overset{~n-q}{G}\bullet\overset{~n-r}{H}:=*^{-1}\Big(*\!\overset{~n-q}{G}\wedge *\overset{~n-r}{H}~\Big)\, ,
\end{equation}
being $*:\Omega^{\,n-q}\left(X\right)\rightarrow\Omega^{\,q}\left(X\right)$ the Hodge dual operator defined on the base space $X$. 
\end{itemize}
In other words, since for $n\geqslant q+r$ the operation \eqref{CPD} among a pair of Hamiltonian forms gives rise to a new Hamiltonian form, we have that, the set of Hamiltonian forms $\Omega^{p}_\mathrm{H}(P)$ not only defines an algebra but also a Gerstenhaber algebra with respect to the Poisson-Gerstenhaber bracket \eqref{PGD} and the co-exterior product \eqref{CPD} \cite{IKGPA}.

Here, we would like to emphasize that the subset of Hamiltonian $(n-1)$-forms $\Omega^{\,n-1}_\mathrm{H}(P)$ will play a primordial role within the polysymplectic formalism: on the one hand, the subset of Hamiltonian $(n-1)$-forms closes as a subalgebra under the Poisson-Gerstenhaber bracket \eqref{PGD} while, on the other hand, we know that for symmetries generated by vertical vector fields on the covariant configuration space the local representation of the induced covariant momentum map on the polymomenta phase-space is characterized by a set of Hamiltonian 
$(n-1)$-forms~\cite{Romer1, IKCSPPS}, which will allows us to study such geometric structure in terms of the Poisson-Gerstenhaber bracket. Therefore, from now on, we will focus our attention to the analysis of the subset of Hamiltonian $(n-1)$-forms $\Omega^{\,n-1}_\mathrm{H}(P)$. 

Besides, note that, it is possible to define a set of canonically conjugate variables for the Poisson-Gerstenhaber bracket \eqref{PGD}, specifically
\begin{equation}
\label{CRD}
\{\![\,p^{\mu}_{a}\,\varpi_{\mu}, y^{b}\varpi_{\nu}\,]\!\}=\delta^{b}_{a}\varpi_{\nu}\, ,
\end{equation}
which allows us to express the De Donder-Weyl-Hamilton field equations \eqref{DeDonWeylEquDef} in a covariant Poisson-Hamiltonian fashion. In other words, we have that, given $\varrho\in \mathscr{P}_{X}$ a section locally represented by $(x^{\mu},\phi^{a}(x^{\mu}),\pi^{\nu}_{a}(x^{\mu}))$, the De Donder-Weyl-Hamilton field equations \eqref{DeDonWeylEquDef} can be written as
\begin{equation}
\begin{aligned}[b]
\partial_{\mu}\phi^{a}
&= 
\varrho^{*}\,\{\![\,H\!_\mathsmaller{\mathrm{DW}},y^{a}\varpi_{\mu}\,]\!\}\,, \\
\partial_{\mu}\pi^{\mu}_{a}
&=\varrho^{*}
\,\{\![\,H\!_\mathsmaller{\mathrm{DW}},p^{\mu}_{a}\,\varpi_{\mu}\,]\!\}\, .
\end{aligned}
\end{equation}

In fact, it is possible to write the equation of motion of an arbitrary Hamiltonian $(n-1)$-form in a covariant Poisson-Hamiltonian fashion.  In order to do so, we start by introducing the total co-exterior derivative $\mathbf{d}\bullet:\Omega^{\,p}_\mathrm{H}(P)\rightarrow \Omega^{\,p-(n-1)}(X)$ which, given $\varrho\in \mathscr{P}_{X}$ a solution of the De Donder-Weyl-Hamilton field equations \eqref{DeDonWeylEquDef} and $\overset{~p}{F}\!\in\Omega^{\,p}_{\,1}\left(P\right)$ a Hamiltonian $p$-form, is defined as
\begin{equation}\label{CDdef}
\mathbf{d}\bullet\overset{~p}{F}:=\frac{s_{m}}{(n-p)!} \left(\varrho^{*}\,d\left(\overset{~p}{F}\bullet\, dx^{\mu_{1}}\wedge\cdots\wedge dx^{{\mu}_{\,n-p}}\right)\right)\bullet\varpi_{\mu_{1}\cdots\mu_{\,n-p}}\, ,
\end{equation}
being $\varpi_{\mu_{1}\cdots\mu_{\,n-p}}:=\partial_{\mu_{1}}\lrcorner\cdots\partial_{\mu_{\,n-p}}\lrcorner\,\varpi$ the basis for the horizontal $(p\,;1)$-forms on $P$, while $s_{m}:=\pm1$ stands for the signature of the metric of the base space $X$ \cite{IKHEQFT}. Consequently, the equation of motion of a Hamiltonian $(n-1)$-form  $\!\overset{~n-1}{F}\!\in\Omega^{\,n-1}_\mathrm{H}\left(P\right)$ explicitly reads
\begin{equation}\label{EvouEqution}
\mathbf{d}\bullet\!\overset{~n-1}{F}=-s_{m}(-1)^{n}\varrho^{*}\, \{\![\,H\!_\mathsmaller{\mathrm{DW}},\!\overset{~n-1}{F}\,]\!\}+\mathbf{d}^{H}\bullet \!\overset{~n-1}{F}\, ,
\end{equation}
where the horizontal co-exterior derivative, $\mathbf{d}^{H}\bullet:\Omega^{\,p}_\mathrm{H}(P)\rightarrow \Omega^{\,p-(n-1)}(X)$, is simply given by
\begin{equation}\label{HCDdef}
\mathbf{d}^{H}\bullet\overset{~p}{F}:=\frac{s_{m}}{(n-p)!} \left(\varrho^{*}\,\partial_{\sigma}\left(\overset{~p}{F}\bullet\, dx^{\mu_{1}}\wedge\cdots\wedge dx^{{\mu}_{\,n-p}}\right)dx^{\sigma}\right)\bullet\varpi_{\mu_{1}\cdots\mu_{\,n-p}}\, .
\end{equation}

In light of this, it is possible to see that by means of the subset of Hamiltonian $(n-1)$-forms $\Omega^{\,n-1}_\mathrm{H}(P)$, the De Donder-Weyl Hamiltonian \eqref{CanonicDeDondWeylHamDef}, the Poisson-Gerstenhaber bracket \eqref{PGD}, and the equation of motion \eqref{EvouEqution}, we can study classical field theory in a covariant Poisson-Hamiltonian framework. In particular, from the physical viewpoint this is very convenient as, commonly, a Poisson-Hamiltonian framework may be seen as the first step towards a quantization of the De Donder-Weyl canonical theory \cite{IKQFTPV, IKHEQFT, IKPSWF}. 

Before ending the present subsection, we will briefly introduce the proposal developed in \cite{IKGD} to analyze singular Lagrangian systems within the polysymplectic approach. For this purpose, we begin by mentioning that a singular Lagrangian system may be simply defined as one that does not satisfy the regularity condition \eqref{Regularity Condition}. In this case, the covariant Legendre map \eqref{LegMap} is not a diffeomorphism since it is not invertible, and therefore we have certain conditions emerging from the definition of the polymomenta $p^{\mu}_{a}=\partial L/\partial y^{a}_{\mu}$, namely
\begin{equation}
\label{PCD}
{\mathcal{C}_{P}^{(k)\,\mu}}(y^{a},p^{\nu}_{a})\approx 0\, ,
\end{equation} 
which, using Dirac's terminology \cite{QGS}, will be referred to as primary constraints. Here, the weak equality symbol $\approx$ is implemented to specify that a certain relation is valid only on the constraint surface, that is, the surface on the polymomenta phase-space delimited by the constraints of the system. Henceforward, for convenience, instead of using the primary constraints \eqref{PCD} explicitly, we will consider the associated primary constraint $(n-1)$-forms $\mathcal{C}_{P}^{(k)}:=\mathcal{C}_{P}^{(k)\,\mu}\varpi_{\mu}$\footnote{For simplicity, all the constraints of the theory are assumed to be linearly independent and Hamiltonian.} (where the index $k$ runs over the complete set of primary constraint $(n-1)$-forms). It is worth noting that, even though Dirac's terminology within the polysymplectic framework
stands for a mere abuse of language it will be very useful for our discussion.

Now, in analogy to the Dirac's approach for singular Lagrangian systems \cite{QGS}, we introduce the total De Donder-Weyl Hamiltonian $\widetilde{H}\!_\mathsmaller{\mathrm{DW}}:P\rightarrow \mathbb{R}$, which is given by
\begin{equation}
\label{TDWHD}
\widetilde{H}\!_\mathsmaller{\mathrm{DW}}:=H\!_\mathsmaller{\mathrm{DW}}+u_{(k)}\bullet\, \mathcal{C}_{P}^{(k)}\, ,
\end{equation}
where $u_{(k)}$ stands for a set of Lagrange multiplier $(1\,;1)$-forms enforcing the primary constraint $(n-1)$-forms. Note that, on the constraint surface, the total De Donder-Weyl Hamiltonian \eqref{TDWHD} is equivalent to the De Donder-Weyl Hamiltonian \eqref{CanonicDeDondWeylHamDef}. However, in order to obtain the correct field equations of the theory under study, we will implement within our analysis the Hamiltonian function \eqref{TDWHD}.

Besides, as proposed in~\cite{IKGD}, a basic consistency requirement to study singular Lagrangian systems within the polysymplectic approach is the preservation of each constraint $(n-1)$-form under the co-exterior derivative, that is, the consistency conditions 
\begin{equation}\label{CCD}
\mathbf{d}\bullet\, \mathcal{C}_{P}^{(k)}\approx 0\, ,
\end{equation}
which extends to the polysymplectic framework the analogous concept within Dirac's approach. In particular, it is possible to see that, by using the total De Donder-Weyl Hamiltonian \eqref{TDWHD} and the equation of motion \eqref{EvouEqution}, the consistency conditions \eqref{CCD} can either be trivially satisfied, impose
further restrictions on the Lagrange multipliers $u_{(k)}$, or they may give rise to new relations independent of both the primary constraints \eqref{PCD} and the Lagrange multipliers $u_{(k)}$. In the latter case, following Dirac's terminology, these new relations will be referred to as secondary constraints. Thus, if there are secondary constraints,  writing them as $(n-1)$-forms $\mathcal{C}_{S}^{(k')}:=\mathcal{C}_{S}^{(k')\,\mu}\varpi_{\mu}$ (where the index $k'$ runs over the set of secondary constraint $(n-1)$-forms), we must impose again the consistency conditions, resulting on either the fixing of the Lagrange multipliers $u_{(k)}$ or obtaining new tertiary constraints. In the case we generate further constraints, we must continue applying the consistency conditions  until we either fix all of the Lagrange multipliers $u_{(k)}$ or whenever these conditions are trivially satisfied. Hence, after the process of generating further constraints is finished, we will have a complete set of constraint $(n-1)$-forms  $\{\mathcal{C}^{(l)}\}$ (where the index $l$ runs over the complete set of primary, secondary, tertiary, etc., constraint $(n-1)$-forms) characterizing the singular Lagrangian system of our interest. 

Following the Dirac formalism for constrained systems, we may also introduce the 
analogue notion of first- and second-class Hamiltonian forms within the polysymplectic approach. In this regard, we say that, a Hamiltonian $p$-form $\overset{~p}{F}\in\Omega^{\,p}_\mathrm{H}\left(P\right)$ is a first-class Hamiltonian $p$-form if its Poisson-Gerstenhaber bracket with each constraint $(n-1)$-form of the system weakly vanishes, namely
\begin{equation}
\label{FCD}
\{\![\,\overset{\,p}{F},\mathcal{C}^{(l)}\,]\!\}\approx 0\,.
\end{equation}
If a Hamiltonian $p$-form is not first-class,  then it will be termed a second-class Hamiltonian $p$-form. Thus, in light of the above definition, we can separate the complete set of constraint $(n-1)$-forms $\{\mathcal{C}^{(l)}\}$ into subsets of first- and second-class constraint $(n-1)$-forms, which can be denoted by $\{\mathcal{A}^{(i)}\}$ and $\{\mathcal{B}^{(i)}\}$, respectively (where the index $i$ runs over the subset of either first- or second-class constraint $(n-1)$-forms).

As a result, it is possible to see that, the $(n-1\,;1)$-form valued matrix
\begin{equation}\label{CRSecClasConst}
\mathcal{B}^{\,(\,i\,,\,j\,)}:=\{\![\mathcal{B}^{(i)},\mathcal{B}^{(j)}]\!\}
\end{equation}
does not vanish on the constraint surface\footnote{Here, the $(n-1\,;1)$-form valued matrix $\mathcal{B}^{\,(i,j)}$ is assumed to be of constant rank on the constraint surface.}. The latter because the subset $\{\mathcal{B}^{(i)}\}$ consists only of second-class constraint $(n-1)$-forms. Furthermore, since $\mathcal{B}^{\,(\,i\,,\,j\,)}$ is a non-singular $(n-1;1)$-form valued matrix, we can construct a $(1\,;1)$-form valued matrix ${\mathcal{B}^{-1}}_{(\,i\,,\,j\,)}$ satisfying
\begin{equation}
{\mathcal{B}^{-1}}_{(\,i\,,\,k\,)}\wedge \mathcal{B}^{\,(\,k\,,\,j\,)}=\delta^{j}_{i}\,\varpi\, ,
\end{equation}
where $\delta^{j}_{i}$ denotes the Kronecker delta. In this way, we have that the $(1\,;1)$-form valued matrix ${\mathcal{B}^{-1}}_{(\,i\,,\,j\,)}$ may be thought of  
as the inverse of the $(n-1\,;1)$-form valued matrix $\mathcal{B}^{\,(\,i\,,\,j\,)}$.

Bearing this in mind and maintaining the analogy with the standard Dirac 
formalism for constrained systems, we are able to define a Dirac-Poisson bracket for Hamiltonian $0$- and $(n-1)$-forms on the polymomenta phase-space. To this end, let us consider $F$ a Hamiltonian $0$- or $(n-1)$-form and $G$ a Hamiltonian $(n-1)$-form. Then, according to \cite{IKGD}, a Dirac-Poisson bracket within the polysymplectic approach can be defined as
\begin{equation}\label{DPGBracket}
\{\![\,F,G\,]\!\}_\mathrm{D}:=\{\![\,F,G\,]\!\}-s_{m}\,\{\![\,F,\mathcal{B}^{(i)}\,]\!\}\bullet \big( {\mathcal{B}^{-1}}_{(\,i\,,\,j\,)} \wedge \{\![\,\mathcal{B}^{(j)},G\,]\!\} \big)\, ,
\end{equation}
which, by construction, eliminates second-class constraint $(n-1)$-forms. From our point of view, this issue is relevant as we can use the Dirac-Poisson bracket \eqref{DPGBracket}, instead of the Poisson-Gerstenhaber bracket \eqref{PGD}, in 
order to obtain the correct field equations of the theory while taking all the second-class constraint $(n-1)$-forms of the system as strong identities. Finally, we would like to encourage the reader to inspect reference \cite{IKGD} for further technical details on the construction of the bracket structure \eqref{DPGBracket} and its algebraic properties.

Next, we will describe the explicit manner in which one may recover the instantaneous Dirac-Hamiltonian formulation for a given classical field theory taking as a starting point the multisymplectic approach.

\subsection{Space plus time decomposition for classical field theory}\label{STDCFT}

In the present subsection, closely following references \cite{GIMMSY1, Gotay1, DeLeon1, GIMMSY2, Fischer, Angel2, DeLeon2}, we will briefly present the description of the space plus time decomposition of the geometric-covariant Lagrangian and multisymplectic formulations for classical field theory. For this purpose, we will start by introducing a slicing of the space-time manifold $X$, and subsequently describing the space plus time decomposition of any fibre-bundle over it, which will allow us to define the elemental ingredients of the instantaneous Lagrangian and Dirac-Hamiltonian formulations for classical field theory. Finally, we will relate the multisymplectic framework with the instantaneous Dirac-Hamiltonian approach by invoking a reduction process.

To begin with, let us consider $\Sigma$ a compact $(n-1)$-dimensional manifold without boundary. Then, a slicing of the space-time manifold $X$ is defined as a diffeomorphism between $\mathbb{R}\times \Sigma$ and $X$, namely  $\bar{\Psi}:\mathbb{R}\times \Sigma\rightarrow  X$. Note that, for $t\in \mathbb{R}$, $\bar{\Psi}_{t}:=\bar{\Psi}\left(t,\cdot\right):\Sigma\rightarrow X$ corresponds to an embedding. Here, we denote by $\mathscr{X}:=\{\bar{\Psi}_{t}\left| \right. t\in\mathbb{R}\}$ and $\Sigma_{t}\subset X$ the set of all embeddings of $\Sigma$ into $X$ and the image of $\Sigma$ by $\bar{\Psi}_{t}\in \mathscr{X}$, respectively. In addition, given $\partial_{t}:=\partial/\partial t\in\mathfrak{X}^{\,1}\left(\mathbb{R}\times \Sigma\right)$ the generator of time translations $(t,u)\mapsto (t+s,u)$ on $\mathbb{R}\times \Sigma$, we define $\zeta^{X}:=T\,\bar{\Psi}\left(\partial_{t}\right)\in\mathfrak{X}^{\,1}\left(X\right)$ as the infinitesimal generator of $\bar{\Psi}$, which by definition is everywhere transverse to $\Sigma_{t}$  \cite{GIMMSY2}. In particular, we say that, $(x^{\mu})$ is a coordinate system adapted to $\bar{\Psi}_{t}$ if the Cauchy surface $\Sigma_{t}$ is locally given by a level set of the coordinate $x^{0}$. Thereby, a coordinate system on $\Sigma_{t}$ can be simply denoted by $(x^{i})$, $i=1,\dots,n-1$~\cite{DeLeon2}.

As expected, the space plus time decomposition process induces a similar decomposition for any fibre-bundle over $X$. To illustrate this, let us consider $\left(K, \fibbun{\,K}{X}, X\right)$ a fibre-bundle. Then, given $\bar{\Psi}$ a slicing of $X$, a compatible slicing of $K$ is defined as a fibre-bundle $\left(K_{\Sigma},\fibbun{\,\,K_{\Sigma}}{\,\Sigma},\Sigma \right)$ and a bundle diffeomorphism $\Psi:\mathbb{R}\times K_{\Sigma}\rightarrow K$ such that the following diagram commutes
\begin{equation}
\begin{tikzcd}
{~~\mathbb{R}\times K_{\Sigma}} \arrow[d, xshift=0ex] \arrow[r, yshift=0ex, "\Psi"] 
& [1mm]{~K} \arrow[d, xshift=0.5mm]\\
{\mathbb{R}\times \Sigma} \arrow[r, shorten=1mm, xshift=1mm, "\bar{\Psi}"] & [1mm]{~X}
\end{tikzcd}
\end{equation}
where the vertical arrows stand for fibre-bundle projections \cite{GIMMSY2}. Observe that, for $t\in\mathbb{R}$, $\Psi_{t}:=\Psi\left(t,\cdot\right): K_{\Sigma}\rightarrow K$ defines an embedding. Here, we denote by $K_{t}\subset K$ the image of $K_{\Sigma}$ by $\Psi_{t}$. As in the previous case, given $\partial_{t}:=\partial/\partial t\in\mathfrak{X}^{\,1}(\mathbb{R}\times K_{\Sigma})$, we define $\zeta^{K}:=T\,\Psi(\partial_{t})\in \mathfrak{X}^{\,1}(K)$ as the infinitesimal generator of $\Psi$, which by construction is everywhere transverse to $K_{t}$ and also projects into $\zeta^{X}$ by means of the tangent map $T\fibbun{\,K}{X}:TK\rightarrow TX$. In light of this, the pairs $(\Sigma_{t}, \zeta^{X})$ and $(K_{t},\zeta^{X})$ will be referred to as an infinitesimal and compatible slicing of $\fibbun{\,K}{X}$, which in turn defines a one-parametric group of $\fibbun{\,K}{X}$-bundle automorphisms \cite{GIMMSY2}. Additionally, we introduce $(K_{t}, \fibbun{\,\,K_{t}}{\,\Sigma_{t}},\Sigma_{t})$ to be the restriction of $\fibbun{\,K}{X}$ to the Cauchy surface $\Sigma_{t}$, where $\fibbun{\,\,K_{t}}{\,\Sigma_{t}}:={\fibbun{\,K}{X}}\,\big|_{\,\Sigma_{t}}:K_{t}\rightarrow \Sigma_{t}$ denotes the corresponding projector map.  Thus, given $(k^{a})$ ($a=1,\dots, m'$) a set of fibre-coordinates on $K$, we can identify $(x^{i},k^{a})$ as an adapted coordinate system on $K_{t}$.

Now, let us consider $\mathscr{K}_{t}$ the set of  sections of $\fibbun{\,\,K_{t}}{\,\Sigma_{t}}$, namely the set of sections of $\fibbun{\,K}{X}$ restricted to the Cauchy surface $\Sigma_{t}$. Then, the collection $\mathscr{K}^{\Sigma}$ defined by
\begin{equation}
\mathscr{K}^{\Sigma}:\,=\bigcup_{\bar{\Psi}_{t}\in\, \mathscr{X}}\mathscr{K}_{t},
\end{equation}
gives rise to an infinite-dimensional fibre-bundle $(\mathscr{K}^{\Sigma}, \fibbun{~\mathscr{K}^{\Sigma}}{\,\mathscr{X}},\mathscr{X})$ as $\mathscr{K}_{t}$ (the fibre over $\bar{\Psi}_{t}\in\mathscr{X}$) defines an infinite-dimensional manifold.  At $\bar{\Psi}_{t}\in\mathscr{X}$, a local section of $\fibbun{~\mathscr{K}^{\Sigma}}{\,\mathscr{X}}$ can be understood as the identification of an element of the fibre over $\bar{\Psi}_{t}$, and therefore a section of $\fibbun{\,\,\mathscr{K}^{\Sigma}}{\,\mathscr{X}}$ induces a section of $\fibbun{\,K}{X}$, and conversely \cite{DeLeon2}.

Here, in agreement with \cite{GIMMSY1, Gotay1, GIMMSY2}, $\mathscr{K}_{t}$ is assumed to be an infinite dimensional smooth manifold. At $\kappa\in \mathscr{K}_{t}$, we introduce the set of functions$(\kappa^{a})$ explicitly given by $\kappa^{a}:=k^{a}\circ\kappa$ to denote a coordinate system on $\mathscr{K}_{t}$, and thus these functions  depend on the coordinates on the Cauchy surface $\Sigma_{t}$ and belong to the chosen functional space \cite{DeLeon2}. The tangent and cotangent bundles over $\mathscr{K}_{t}$ can be defined as follows. To begin with, let us consider $(VK_{t},\fibbun{\,\,VK_{t}}{\,K_{t}}, K_{t})$ the restriction of $\fibbun{\,\,VK}{K}$ (the vertical tangent bundle of $\fibbun{\,K}{X}$) to $K_{t}$, where we have implemented $\fibbun{\,\,VK_{t}}{\,K_{t}}:=\fibbun{\,\,VK}{K}\,\big|_{\,K_{t}}:VK_{t}\rightarrow K_{t}$ and $VK_{t}:=\fibbun{\,\,VK}{K}^{-1}(K_{t})$ to represent the projector map and the total space, respectively. Then, the tangent space to $\mathscr{K}_{t}$ at $\kappa\in \mathscr{K}_{t}$ is defined as $T_{\kappa}\mathscr{K}_{t}:=\big\{\dot{\kappa}:\Sigma_{t}\rightarrow VK_{t}\,\big|\, \fibbun{\,\,VK_{t}}{\,K_{t}}\circ \dot{\kappa}=\kappa\big\}$. Note that, in adapted coordinates, an element $\dot{\kappa}\in T_{\kappa}\mathscr{K}_{t}$ can be written as
\begin{equation}
\dot{\kappa}:=\dot{\kappa}^{a}\frac{\partial}{\partial k^{a}}\, ,
\end{equation}
where the functions $\dot{\kappa}^{a}$ depend on the coordinates on the Cauchy surface $\Sigma_{t}$. As usual, the tangent bundle is given by
\begin{equation}
T\,\mathscr{K}_{t}:=\bigcup_{\kappa\,\in\, \mathscr{K}_{t}}T_{\kappa}\mathscr{K}_{t}.
\end{equation}
Consequently, we introduce $(\kappa^{a},\dot{\kappa}^{a})$ to denote a coordinate system on $T\,\mathscr{K}_{t}$, where the functions $\kappa^{a}$ and $\dot{\kappa}^{a}$ belong to the chosen functional space and take values on the Cauchy surface $\Sigma_{t}$.

Besides, let us consider $(V^{*}K_{t},\fibbun{\,\,V^{*}K_{t}}{\,K_{t}}, K_{t})$ the restriction of $ \fibbun{\,\,V^{*}K}{K}$ (the dual bundle of $\fibbun{\,\,VK}{K}$) to $K_{t}$, where we have introduced $\fibbun{\,\,V^{*}K_{t}}{\,K_{t}}:\fibbun{\,\,V^{*}K}{K}\,\big|_{\,K_{t}}:V^{*}K_{t}\rightarrow K_{t}$ and $ V^{*}K_{t}:=\fibbun{\,\,V^{*}K}{K}^{-1}(K_{t})$ to denote the projector map and the total space, respectively. Then, given the tensor product bundle $\pi^{\otimes}_{t}: V^{*}K_{t}\otimes\fibbun{\,\,K_{t}}{\,\Sigma_{t}}^{*}(\Lambda^{n-1}T^{*}\Sigma_{t})\rightarrow K_{t}$, the cotangent space to $\mathscr{K}_{t}$ at $\kappa\in\mathscr{K}_{t}$ is defined as $T^{*}_{\,\kappa\,}\mathscr{K}_{t}:=\big\{\tau:\Sigma_{t}\rightarrow 
V^{*}K_{t}\otimes\fibbun{\,\,K_{t}}{\,\Sigma_{t}}^{*}(\Lambda^{n-1}T^{*}\Sigma_{t})\,\big|\, \pi_{t}^{\otimes}\circ\tau=\kappa\big\}$, where the pullback bundle $\fibbun{\,\,K_{t}}{\,\Sigma_{t}}^{*}(\fibbun{\,\,\Lambda^{n-1}T^{*}\Sigma_{t}}{\,\Sigma_{t}}):\fibbun{\,\,K_{t}}{\,\Sigma_{t}}^{*}(\Lambda^{n-1}T^{*}\Sigma_{t})\rightarrow K_{t}$ may be identified with the subbundle of horizontal $(n-1;\,1)$-forms over $K_{t}$, namely $\fibbun{\,\,\Lambda^{n-1}_{1}T^{*}K_{t}}{\,K_{t}}:\Lambda^{n-1}_{1}T^{*}K_{t}\rightarrow K_{t}$ (see reference \cite{Saunders} for details about tensor products of vector bundles and pullback bundles). Observe that, in adapted coordinates, an element $\tau\in T^{*}_{\,\kappa\,}\mathscr{K}_{t}$ can be expressed as
\begin{equation}
\tau:=\tau_{a}\,dk^{a}\otimes d^{\,n-1}x_{0}\, ,
\end{equation}
where the functions $\tau_{a}$ depend on the coordinates on the Cauchy surface $\Sigma_{t}$. Naturally, the cotangent bundle is defined by
\begin{equation}
T^{*}\,\mathscr{K}_{t}:=\bigcup_{\kappa\,\in\, \mathscr{K}_{t}}T^{*}_{\,\kappa\,}\mathscr{K}_{t}.
\end{equation}
Analogously to the previous case, we identify $(\kappa^{a},\tau_{a})$ as a coordinate system on $T^{*}\mathscr{K}_{t}$, where the functions $\kappa^{a}$ and $\tau_{a}$ belong to the chosen functional space and take values on the Cauchy surface $\Sigma_{t}$. Hence, the natural pairing between elements $\dot{\kappa}\in T_\kappa \mathscr{K}_t$ and $ \tau \in T^{*}_{\,\kappa\,} \mathscr{K}_t$ is locally given by integration, specifically
\begin{equation}
\langle \dot{\kappa}\,, \tau\rangle:=\int_{\Sigma_{t}}\dot{\kappa}\,\lrcorner\,\,\tau\, .
\end{equation}

Finally, we would like to emphasize that, it is possible to induce differential forms on $\mathscr{K}_{t}$ from those defined on $K$ \cite{Gotay1, DeLeon1, Fischer, DeLeon2}. In other words, we have that for all  $(q+n-1)$-form on $K$, $\alpha\in\Omega^{\,q+n-1}(K)$, there is an associated $q$-form $\alpha_{t}\in\Omega^{\,q}(\mathscr{K}_{t})$ on $\mathscr{K}_{t}$ whose dual pairing with a set of $q$ tangent vectors $V_{k}\in T_{\sigma}\mathscr{K}_{t}$ at $\sigma\in\mathscr{K}_{t}$ explicitly reads
\begin{equation}\label{IFSS}
\alpha_{t}\left(\sigma\right)\left(V_{1},\cdots\!,V_{q}\right):=\int_{\Sigma_{t}}\sigma^{*}\left(V_{q}\lrcorner\,\cdots V_{1}\lrcorner\,\alpha\right)\, ,
\end{equation}
where the contraction is taken along the image of each of the tangent vectors $V_{k}$, $k=1,\dots, q$. 

As discussed below, the above space plus time decomposition of fibre-bundles will allow us to relate the finite- and infinite-dimensional formulations for classical field theory. To illustrate this, let $(\Sigma_{t},\zeta^{X})$ and $(Y_{t}, \zeta^{Y})$ be a $\mathcal{G}$-slicing of the covariant configuration space $\fibbun{\,Y}{X}$, namely an infinitesimal and compatible slicing of $\fibbun{\,Y}{X}$ such that 
for some $\xi_{\eta}\in\mathfrak{g}$, the vector field $\zeta^{Y}\in\mathfrak{X}^{\,1}(Y)$ contains the information of $\xi^{Y}_{\eta}\in\mathfrak{X}^{1}(Y)$, and also its corresponding first jet prolongation, $\zeta^{J^{1}Y}:=j^{1}\zeta^{Y}\in\mathfrak{X}^{\,1}(J^{1}Y)$,  defines an infinitesimal symmetry of the theory. As pointed out in \cite{Gotay1}, the vector field $\zeta^{Y}\in\mathfrak{X}^{\,1}(Y)$ may be thought of as an evolution direction on $Y$. Consequently, we introduce $(Y_{t}, \fibbun{\,\,Y_{t}}{\,\Sigma_{t}}, \Sigma_{t})$ to be the restriction of $\fibbun{\,Y}{X}$ to the Cauchy surface $\Sigma_{t}$, and we denote by $(x^{i},y^{a})$ an adapted coordinate system on $Y_{t}$. Furthermore, we define $\mathscr{Y}_{t}$ as the set of sections of $\fibbun{\,\,Y_{t}}{\,\Sigma_{t}}$, that is, the set of sections of $\fibbun{\,Y}{X}$ restricted to the Cauchy surface $\Sigma_{t}$. Observe that, for all $\varphi\in \mathscr{Y}_{t}$ a section  of $\fibbun{\,\,Y_{t}}{\,\Sigma_{t}}$ there is a section $\phi\in \mathscr{Y}_{X}$ of $\fibbun{\,Y}{X}$ such that $\varphi:=\phi\circ i_{t}$, where $i_{t}:\Sigma_{t}\rightarrow X$ denotes the inclusion map. Besides, given $\mathscr{Y}_{t}$ the $t$-instantaneous configuration space of the classical field theory \eqref{APFT}, we identify $T\,\mathscr{Y}_{t}$ as the $t$-instantaneous space of velocities of the system where the instantaneous Lagrangian formulation of the theory will take place. Here, we introduce $(\varphi^{\,a},\dot{\varphi}^{\,a})$ to denote a coordinate system on $T\,\mathscr{Y}_{t}$, where the functions $\varphi^{\,a}$ and $\dot{\varphi}^{\,a}$ depend on the coordinates on the Cauchy surface $\Sigma_{t}$ and belong to the chosen functional space. In particular, by considering $\phi\in\mathscr{Y}_{X}$ a section such that $\varphi:=\phi\circ i_{t}\in \mathscr{Y}_{t}$, the temporal derivative of the field variables is given by
\begin{equation}\label{TEFV}
\dot{\varphi}^{a}:=(\,\mathfrak{L}_{\,\zeta^{Y}}\phi)^{a}\,\big|_{\Sigma_{t}}=\big(\,T\phi\circ\zeta^{X}-\zeta^{Y}\!\circ \phi\,\big)^{a}\,\big|_{\Sigma_{t}}\, ,
\end{equation} 
where $\mathfrak{L}_{\,\zeta^{Y}}\phi$ denotes the Lie derivative of the section $\phi$ along the vector field $\zeta^{Y}$, while $T\phi:TX\rightarrow TY$ represents tangent map of $\phi$ \cite{GIMMSY2}. Note that, $\mathfrak{L}_{\,\zeta^{Y}}\phi\,\big|_{\Sigma_{t}}\!\!\in T_{\varphi}\mathscr{Y}_{t}$ corresponds to a vertical vector field on the image of $\varphi$, while $(\mathfrak{L}_{\,\zeta^{Y}}\phi)^{a}\,\big|_{\Sigma_{t}}$ stands for the components of such a vector field. 

Now, let us consider $((J^{1}Y)_{t}, \fibbun{\,\,(J^{1}Y)_{t}}{\,Y_{t}}, Y_{t})$ the restriction of $\fibbun{\,J^{1}Y}{Y}$ to $Y_{t}$, where we have implemented $\fibbun{\,\,(J^{1}Y)_{t}}{\,Y_{t}}:={\fibbun{\,J^{1}Y}{Y}}\big|_{Y_{t}}:(J^{1}Y)_{t}\rightarrow Y_{t}$ and $(J^{1}Y)_{t}:=\fibbun{\,J^{1}Y}{Y}^{-1}(Y_{t})$ to denote the projector map and the total space, respectively. Then, given $(x^{i},y^{a},y^{a}_{\mu})$ an adapted coordinate system on $(J^{1}Y)_{t}$, we introduce an affine bundle map $\beta_{\zeta
^{Y}}:\left(J^{1}Y\right)_{t}\rightarrow J^{1}\!\left(Y_{t}\right)\times VY_{t}$ over $Y_{t}$, which is locally defined by $\beta_{\zeta^{Y}}\left(x^{i},y^{a},y^{a}_{\mu}\right):=\left(x^{i}, y^{a}, y^{a}_{i}, \dot{y}^{a}\right)$, where $(J^{1}\!\left(Y_{t}\right), \fibbun{\,\,J^{1}\left(Y_{t}\right)}{\,Y_{t}},Y_{t})$ denotes the affine jet bundle of $\fibbun{\,\,Y_{t}}{\,\Sigma_{t}}$\footnote{In adapted coordinates, $\bar{\varkappa}\in J^{1}_{\bar{y}}(Y_{t})$ an element of the fibre over $\bar{y}\in Y_{t}$ can be locally written as $\displaystyle\bar{\varkappa}:=dx^{i}\otimes \left(\frac{\partial}{\partial x^{i}}+y^{a}_{i}\frac{\partial}{\partial y^{a}}\right)$, which allows us to identify $(x^{i},y^{a},y^{a}_{i})$ as an adapted coordinate system on $J^{1}(Y_{t})$.}. In fact, the latter is relevant as the so-called jet decomposition map $\beta_{\zeta^{Y}}$ induces an isomorphism between $\left(\mathscr{J}^{1}\mathscr{Y}\right)_{t}$ (the set of sections of $\fibbun{\,J^{1}Y}{X}$ restricted to the Cauchy surface $\Sigma_{t}$) and $T\,\mathscr{Y}_{t}$, as discussed in \cite{Gotay1, GIMMSY2}.

Observe that, being $j^{1}\phi\in \mathscr{J}^{1}\mathscr{Y}_{X}$ the first jet prolongation of $\phi\in\mathscr{Y}_{X}$, we can write $\beta_{\zeta^{Y}}\left(j^{1}\phi\circ i_{t}\right)=\left(j^{1}\varphi,\dot{\varphi}\right)$, where $j^{1}\varphi$, the first jet prolongation of $\varphi:=\phi\circ i_{t}\in\mathscr{Y}_{t}$, corresponds to a section of $\fibbun{\,\,J^{1}\left(Y_{t}\right)}{\,\Sigma_{t}}:J^{1}(Y_{t})\rightarrow \Sigma_{t}$. As a result, we can obtain the instantaneous Lagrangian density of the theory, $\mathcal{L}_{\,t,\,\zeta^{Y}}:J^{1}(Y_{t})\times VY_{t}\rightarrow \Lambda^{n-1}T^{*}\Sigma_{t}$, by means of $\mathcal{L}_{\,t,\,\zeta^{Y}}\left(j^{1}\varphi, \dot{\varphi}\right):=\left(j^{1}\phi\circ i_{t}\right)^{*}\left(\zeta^{X}\!\lrcorner\,\mathcal{L}\right)$ and, consequently, the corresponding instantaneous Lagrangian functional $L_{\,t,\,\zeta^{Y}}:T\,\mathscr{Y}_{t}\rightarrow\mathbb{R}$, which in adapted coordinates explicitly reads
\begin{equation}\label{ILCFT}
L_{\,t,\,\zeta^{Y}}\left(\varphi,\dot{\varphi}\right):=\int_{\Sigma_{t}}L\left(j^{1}\varphi, \dot{\varphi}\right)\zeta^{0}d^{\,n-1}x_{0}\, ,
\end{equation}
where $L$ represents the covariant Lagrangian function of the system, $\zeta^{0}$ corresponds to the component of the generator $\zeta^{X}$ along $\partial_{0}$ and  $d^{\,n-1}x_0$ denotes the $(n-1)$-volume form on the Cauchy surface $\Sigma_t$. In light of this, the instantaneous Legendre transformation is defined as a bundle map $\mathbb{F}L_{\,t,\,\zeta^{Y}}:T\,\mathscr{Y}_{t}\rightarrow T^{*}\mathscr{Y}_{t}$ over $\mathscr{Y}_{t}$ locally given by
\begin{equation}\label{Insantaneous Leg Transformation}
\mathbb{F}L_{\,t,\,\zeta^{Y}}\left(\varphi^{\,a},\dot{\varphi}^{\,a}\right):=\Big(\varphi^{\,a},\pi_{\,a}:=\frac{\partial L}{\partial \dot{\varphi}^{\,a}}\Big)\, ,
\end{equation}
which allows us to identify the $t$-primary constraint set of the theory $\mathscr{P}_{\,t,\,\zeta^{Y}}\subset T^{*}\mathscr{Y}_{t}$, that is, the submanifold on the $t$-instantaneous phase-space of the system, $T^{*}\mathscr{Y}_{t}$, characterized by the image of the instantaneous Legendre transformation \eqref{Insantaneous Leg Transformation}. As usual, in adapted coordinates, the non-degenerate symplectic structure $\omega_{t}\in \Omega^{\,2}\left(T^{*}\mathscr{Y}_{t}\right)$ on $T^{*}\mathscr{Y}_{t}$ is simply defined as
\begin{equation}
\omega_{t}\left(\varphi,\pi\right)=\int_{\Sigma_{t}}d\varphi^{\,a}\wedge d\pi_{\,a}\otimes d^{\,n-1}x_{0}\, .
\end{equation}

Now, we will perform the space plus time decomposition for the multimomenta phase-space, and consequently introduce a reduction process to relate the multisymplectic and the instantaneous Dirac-Hamiltonian formalisms. For this purpose, let us consider $(Z^{\star}_{t}, \fibbun{\,\,Z^{\star}_{t}}{\,Y_{t}},Y_{t})$ the restriction of $\fibbun{\,Z^{\star}}{Y}$ to $Y_{t}$, where we have introduced $\fibbun{\,\,Z^{\star}_{t}}{\,Y_{t}}:\fibbun{\,Z^{\star}}{Y}\,\big|_{Y_{t}}\!:Z^{\star}_{t}\rightarrow Y_{t}$ and $Z_{t}:=\fibbun{\,Z^{\star}}{Y}^{-1}(Y_{t})$ to denote the projector map and the total space, respectively. Furthermore, we define $\mathscr{Z}^{\star}_{t}$ as the set of sections of $\fibbun{\,\,Z^{\star}_{t}}{\,\Sigma_{t}}:=\fibbun{\,\,Y_{t}}{\,\Sigma_{t}}\circ\fibbun{\,\,Z^{\star}_{t}}{\,Y_{t}}$, namely the set of sections of $\fibbun{\,Z^{\star}}{X}$ restricted to the Cauchy surface $\Sigma_{t}$, which in fact is a presymplectic manifold. The latter since the multisymplectic structure $\Omega^{(\mathcal{M})}\in \Omega^{\,n+1}(Z^{\star})$ on $Z^{\star}$ induces by means of \eqref{IFSS} a presymplectic structure $\Omega_{t}\in \Omega^{\,2}(\mathscr{Z}^{\star}_{t})$ on $\mathscr{Z}^{\star}_{t}$, that is, a closed $2$-form with a non-trivial kernel \cite{Gotay1}. Thus, in order to identify $T^{*}\mathscr{Y}_{t}$ with the quotient of $\mathscr{Z}^{\star}_{t}$ by the kernel of the presymplectic $2$-form $\Omega_{t}$, we introduce a vector bundle map $R_{t}: \mathscr{Z}^{\star}_{t}\rightarrow T^{*}\mathscr{Y}_{t}$ over $\mathscr{Y}_{t}$ defined by
\begin{equation}\label{Bundle Map R}
\begin{aligned}
\left\langle\, R_{t}\left(\sigma\right),V \,\right\rangle:=\int_{\Sigma_{t}}\varphi^{*}\left(V\lrcorner\,\,\sigma\right)\, ,
\end{aligned}
\end{equation}
where $\sigma\in \mathscr{Z}^{\star}_{t}$ stands for a section such that $\varphi:=\fibbun{\,Z}{Y}\circ\sigma\in\mathscr{Y}_{t}$, while $V\in T_{\varphi}\mathscr{Y}_{t}$ denotes a tangent vector to $\mathscr{Y}_{t}$ at $\varphi$\footnote{Of course, the contraction in \eqref{Bundle Map R} is taken along the images of $\sigma$ and $V$.}. Note that, in adapted coordinates, we can write $R_{t}\left(\sigma\right)=\left(p^{0}_{a}\circ\sigma\right)dy^{a}\otimes d^{\,n-1}x_{0}$, which implies that $\mathrm{ker}\,R_{t}:=\big\{ \sigma\in\mathscr{Z}^{\star}_{t}\,\big|\, p^{0}_{a}\circ\sigma=0\big\}$. According to \cite{Gotay1}, given the non-degenerate symplectic structure on $T^{*}\mathscr{Y}_{t}$, $\omega_{t}\in \Omega^{\,2}\left(T^{*}\mathscr{Y}_{t}\right)$, the relation $R^{*}_{t}\,\omega_{t}=\Omega_{t}$ holds, and hence $\mathrm{ker}\,TR_{t}=\mathrm{ker}\,\Omega_{t}$, where $TR_{t}:T\mathscr{Z}^{\star}_{t}\rightarrow T\,T^{*}\mathscr{Y}_{t}$ represents the tangent map of $R_{t}$. In fact, the latter is relevant as the induced quotient map $\mathscr{Z}^{\star}_{t}/\,\mathrm{ker}\,TR_{t}\rightarrow T^{*}\mathscr{Y}_{t}$ defines a symplectic diffeomorphism, as discussed in \cite{GIMMSY2}. Additionally, we identify $\mathscr{N}_{t}:=\big\{\sigma\in\mathscr{Z}^{\star}_{t}\,\big|\,\sigma=\mathbb{F}\mathcal{L}\circ j^{1}\phi\circ i_{t}\big\}$ as the subset of $\mathscr{Z}^{\star}_{t}$ that projects by means of $R_{t}$ onto the $t$-primary constraint set of the theory $\mathscr{P}_{\,t,\,\zeta^{Y}}$. Indeed, this arises from the fact that for $\sigma\in\mathscr{N}_{t}$ it is possible to write
\begin{equation}
R_{t}\left(\sigma\right)=\frac{\partial L}{\partial y^{a}_{0}}\!\left(j^{1}\varphi, \dot{\varphi}\right) dy^{a}\otimes d^{n-1}x_{0}\, \,
\end{equation}
which corresponds to an element of $\mathscr{P}_{\,t,\,\zeta^{Y}}$. Henceforward, a section $\sigma\in\mathscr{N}_{t}$ that projects by means of $R_{t}$ onto $\left(\varphi,\pi\right)\in\mathscr{P}_{\,t,\,\zeta^{Y}}$ will be referred to as an holonomic lift of $\left(\varphi,\pi\right)$ \cite{Gotay1}.

Bearing this in mind, we are in the position to study the action of $\mathcal{G}$ on $\mathscr{Z}^{\star}_{t}$ and $T^{*}\mathscr{Y}_{t}$. In order to do so, we start by remembering that, $\mathcal{G}$ acts on $Z^{\star}$ by covariant canonical transformations, and therefore there is an associated covariant momentum map $J^{(\mathcal{M})}:Z^{\star}\rightarrow \mathfrak{g}^{*}\otimes \Lambda^{n-1}T^{\,*}Z^{\star}$. In particular, the latter is important as it induces on $\mathscr{Z}^{\star}_{t}$ the so-called energy-momentum map $E_{t}:\mathscr{Z}^{\star}_{t}\rightarrow \mathfrak{g}^{*}$, namely 
\begin{equation}
\label{eq:energy}
\langle E_{t}\left(\sigma\right),\xi_{\eta}\rangle   :=\int_{\Sigma_{t}}\sigma^{*}J^{(\mathcal{M})}\!\left(\xi_{\eta}\right)\, ,
\end{equation}
where $\sigma\in \mathscr{Z}^{\star}_{t}$ and $\xi_{\eta}\in\mathfrak{g}$. Now, we introduce $\mathcal{G}_{t}:=\big\{\eta\in\mathcal{G}\,\big|\, \eta_\mathsmaller{X}(\Sigma_{t})=\Sigma_{t}\big\}$ to denote the subset of $\mathcal{G}$ that acts on $\Sigma_{t}$ by diffeomorphisms. In addition, we define $\mathfrak{g}_{\,t}\subset \mathfrak{g}$ as the Lie algebra of $\mathcal{G}_{t}$. Observe that, for all $\eta\in\mathcal{G}_{t}$, the map $\eta_{t}:=\eta_\mathsmaller{X}\big|_{\Sigma_{t}}$ 
may be identified as an element of the group of diffeomorphisms on $\Sigma_t$, $\mathrm{Diff}\left(\Sigma_{t}\right)$. Hereinafter, given $\xi_{\eta}\in\mathfrak{g}_{\,t}$ the infinitesimal generator of $\eta\in \mathcal{G}_{t}$, we introduce $\xi_{\eta_{t}}^{X}\in\mathfrak{X}^{\,1}(X)$ to denote the infinitesimal generator of $\eta_{t}\in\mathrm{Diff}\left(\Sigma_{t}\right)$, which is a vector field tangent to the Cauchy surface $\Sigma_{t}$, and therefore, in adapted coordinates, it must satisfy the condition $\xi^{0}=0$ on $\Sigma_{t}$, being $\xi^{0}$ the component of such a vector field along $\partial_{0}$. In fact, as discussed in \cite{GIMMSY2}, the action of $\mathcal{G}_{t}$ on $\mathscr{Z}^{\star}_{t}$ preserves the presymplectic structure $\Omega_{t}$, and in consequence the restriction of the energy-momentum map $E_{t}$ to the subspace $\mathfrak{g}_{\,t}$ gives rise to the corresponding momentum map, specifically $\mathcal{J}_{t}:=E_{t}\,\big|_{\mathfrak{g}_{\,t}}\!:\mathscr{Z}^{\star}_{t}\rightarrow \mathfrak{g}_{\,t}^{*}$. Furthermore, since $\left(T^{*}\mathscr{Y}_{t},\omega_{t}\right)$ can be thought of as the quotient of the presymplectic space $\left(\mathscr{Z}^{\star}_{t}, \Omega_{t}\right)$ by the kernel of $TR_{t}$, we have that, the momentum map $\mathscr{J}_{t}:T^{*}\mathscr{Y}_{t}\rightarrow \mathfrak{g}_{\,t}^{*}$ associated with the action of $\mathcal{G}_{t}$ on $T^{*}\mathscr{Y}_{t}$ is given by
\begin{equation}\label{PMM}
\left\langle \mathscr{J}_{t}(\varphi,\pi),\xi_{\eta}\right\rangle:=\left\langle \mathcal{J}_{t}\left(\sigma\right),\xi_{\eta}\right\rangle\, ,
\end{equation}
where $\xi_{\eta}\in\mathfrak{g}_{\,t}$, $\left(\varphi,\pi\right)\in T^{*}\mathscr{Y}_{t}$ and $\sigma\in R^{-1}_{t}\{(\varphi,\pi)\}\subset \mathscr{Z}^{\star}_{t}$.

Besides, since $\zeta^{J^{1}Y}:=j^{1}\zeta^{Y}$ (the first jet prolongation of $\zeta^{Y}$) and its corresponding vector flow define an infinitesimal symmetry of the system, we know that the $\alpha^{(\mathcal{M})}$-lift of $\zeta^{Y}$ to $Z^{\star}$, $\zeta^{Z^{\star}}\in\mathfrak{X}^{\,1}(Z^{\star})$, satisfies the relation $\mathfrak{L}_{\zeta^{Z^{\star}}}\Omega^{(\mathcal{M})}=0$, and hence there is an $(n-1)$-form $H^{\mathcal{(M)}}_{\zeta^{Z^{\star}}}\in\Omega^{\,n-1}(Z^{\star})$ on $Z^{\star}$ such that $ \zeta^{Z^{\star}}\!\!\lrcorner\,\,\Omega^{(\mathcal{M})}=d\,H^{(\mathcal{M})}_{\zeta^{Z^{\star}}}$, which  in turn induces a functional $H_{t\,,\,\zeta^{Y}}:\mathscr{P}_{\,t,\,\zeta^{Y}}\rightarrow \mathbb{R}$ on the $t$-primary constraint set by means of
\begin{equation}\label{IHT}
H_{\,t,\,\zeta^{Y}}(\varphi,\pi):=-\int_{\Sigma_{t}}\sigma^{*}H^{(\mathcal{M})}_{\zeta^{Z^{\star}}}\, ,
\end{equation}
being $\left(\varphi,\pi\right)\in\mathscr{P}_{t}$ and $\sigma\in R^{-1}_{t}\{(\varphi,\pi)\}\cap \mathscr{N}_{t}$. In fact, it is possible to show that  the above functional is nothing but the instantaneous Hamiltonian of the theory, as discussed in detail in references \cite{Gotay1, GIMMSY2, DeLeon2}. 

Finally, we would like to emphasize that, for a classical field theory with localizable symmetries, the second Noether theorem extended to the multisymplectic approach establishes that, the vanishing of the momentum map \eqref{PMM} gives rise to the set of first-class constraints that characterizes the system within the instantaneous Dirac-Hamiltonian formulation \cite{GIMMSY1, GIMMSY2, Fischer}. In other words, we have that, for a classical field theory with localizable symmetries, the admissible space of Cauchy data for the evolution equations of the system is determined by the zero level set of the momentum map \eqref{PMM}, namely
\begin{equation}
\mathscr{J}_{t}^{-1}(0):=\big\{(\varphi,\pi)\in T^{*}\mathscr{Y}_{t}\,\big|\, \langle \mathscr{J}_{t}(\varphi,\pi),\xi_{\eta}\rangle=0\,,\, \forall\,\xi_{\eta}\in\mathfrak{g}_{\,t}\big\}\, .
\end{equation}
Bearing this in mind, both the momentum map \eqref{PMM} and the functional \eqref{IHT} will be fundamental for our discussion in the subsequent section.   

\section{Geometric-covariant analysis of the Bonzom-Livine model for gravity}\label{Analysis of BLmodel}

In the present section, we will study the Bonzom-Livine model for gravity within the geometric-covariant Lagrangian, multisymplectic and polysymplectic formulations for classical field theory.  For this purpose, we will start by introducing a mathematical description of the physical system of our interest, the Bonzom-Livine model for gravity, which consists of a formulation for $3$-dimensional gravity that includes an arbitrary Immirzi-like parameter. Here, we will discuss how the different formalisms introduced in section \ref{Geom Cov Formalisms} allow us to obtain a more general understanding of the features of the theory at the classical level. In this regard, we will not only derive the field equations of the model but we will also analyze its gauge symmetries. Additionally, after performing the space plus time decomposition of the space-time manifold on which the system is defined, we will describe the manner in which it is possible to recover the instantaneous Dirac-Hamiltonian analysis of the theory by considering the multisymplectic formulation  as a starting point.

\subsection{Bonzom-Livine model for gravity}

As mentioned before, the Bonzom-Livine model for gravity corresponds to a formulation of $3$-dimensional Einstein theory of General Relativity that includes an arbitrary Immirzi-like parameter \cite{Bonzom}.  From the point of view of physics, the aforementioned gravity model is particularly interesting since, in close analogy to the role played by the Immirzi parameter in Holst action~\cite{Holst}, the Immirzi-like parameter inherent to the Bonzom-Livine model for gravity not only modifies the symplectic structure of $3$-dimensional gravity but it also alters the spectra of certain geometric objects in the corresponding quantum theory \cite{Bonzom, Barbosa, Escalante}. Next, we will give a mathematical description of the model of our interest.

To begin with, let $X$ be a $3$-dimensional Riemannian manifold without boundary\footnote{From now on, we will set the dimension of the space-time manifold to $n=3$ and also we will fix the signature of the metric to $s_{m}=1$.}. Then, given $\mathfrak{t}$ a Lie algebra (or a vector subspace of a Lie algebra), we identify $\Omega^{\,p}(X, \mathfrak{t}):=\Omega^{\,p}(X)\otimes\,\mathfrak{t}$ as the set of $\mathfrak{t}$-valued $p$-form on $X$. In addition, let us consider $\{T_{a}\}$ a basis for $\mathfrak{t}\,$ and $\tilde{\beta}:\mathfrak{t}\otimes\mathfrak{t}\rightarrow \mathbb{R}$ a symmetric bilinear form. Thus, for any pair of $\mathfrak{t}$-valued forms on $X$, $\mu=\mu^{a}\otimes T_{a}\in\Omega^{\,p}(X, \mathfrak{t})$ and $\lambda=\lambda^{a}\otimes T_{a}\in\Omega^{\,q}(X, \mathfrak{t})$, we define 
the following operations $[\mu,\lambda]:=(\mu^{a}\wedge \lambda^{b})\otimes [T_{a}, T_{b}]$, $d\mu:=d\mu^{a}\otimes T_{a}$ and $\tilde{\beta}(\mu\wedge\lambda):=\mu^{a}\wedge\lambda^{b}\,\tilde{\beta}(T_{a}, T_{b})$, where $[\,\cdot\,,\,\cdot\,]$ denotes the Lie bracket. Now, let us consider the Lie group $\mathcal{H}:=\mathrm{SO(3)}$ and $\mathfrak{h}:=\mathfrak{so}(3)$ its corresponding Lie algebra. Thus, by following the Palatini formalism, we introduce the $\mathfrak{h}$-connection $\omega$ and the co-frame field $e$ to be  the fundamental fields of the $3$-dimensional Einstein theory of General Relativity \cite{Carlip}. As we will see below, the Bonzom-Livine model for gravity arises from the fact that, by combining the $\mathfrak{h}$-connection  and the co-frame field $e$ into a single connection, it is possible to reformulate $3$-dimensional gravity with cosmological constant $\Lambda$ as a Chern-Simons theory with
gauge group $\mathcal{G}$, which may be identified with $\mathrm{SO}(4)$, $\mathrm{ISO}(3)$ or $\mathrm{SO}(3,1)$ depending on whether $\Lambda$ is correspondingly positive, zero or negative, as discussed in \cite{Witten}.

To illustrate this, we start by mentioning that, on the one hand, given $\{J_{a}, P_{a}\}$ a basis of $\mathfrak{g}$ (the Lie algebra of $\mathcal{G}$), we can write
\begin{equation}\label{BL Lie algebra}
[J_{a},J_{b}]={\epsilon_{ab}}^{c}J_{c}\,, \quad 
[J_{a},P_{b}]={\epsilon_{ab}}^{c}P_{c}\,, \quad 
[P_{a},P_{b}]=\Lambda\,{\epsilon_{ab}}^{c}J_{c} \,,
\end{equation}
where $\epsilon_{abc}$ stands for the $3$-dimensional Levi-Civita alternating symbol ($a, b, c =1,2,3$).  As pointed out in \cite{Bonzom, Barbosa}, it is possible to define two symmetric bilinear forms $\beta_{\,i}:\mathfrak{g}\otimes \mathfrak{g}\rightarrow \mathbb{R}$ on the Lie algebra \eqref{BL Lie algebra} ($i=1,2$), which are given by 
\begin{equation}
\begin{aligned}[b]
&\beta_{1}(J_{a},P_{a})=\delta_{ab}\, ,\quad \beta_{1}(J_{a},J_{a})=0\, , \quad~~ \beta_{1}(P_{a},P_{a})=0\, ,\\
&\beta_{2}(J_{a},P_{a})=0\, ,\quad~~ \beta_{2}(J_{a},J_{a})=\delta_{ab}\, , \quad \beta_{2}(P_{a},P_{a})=\Lambda\, \delta_{ab}\, ,
\end{aligned}
\end{equation}
being $\delta_{ab}$ the Kronecker delta. Note that, the bilinear form $\beta_{1}$ is non-degenerated for all $\Lambda$, while $\beta_{2}$ is non-degenerated only for $\Lambda\neq 0$. On the other hand, we have that, by regarding $\mathcal{H}$ as a closed subgroup of $\mathcal{G}$, we can perform the symmetric splitting $\mathfrak{g}=\mathfrak{h}\oplus \mathfrak{p}$, where the supplement space $\mathfrak{p}\subset\mathfrak{g}$ (a vector subspace of $\mathfrak{g}$) satisfies the commutation relations $[\mathfrak{h},\mathfrak{p}]\subset \mathfrak{p}$ and $[\mathfrak{p},\mathfrak{p}]\subset \mathfrak{h}$ (see references \cite{Sardanashvily1, Wise} for details about symmetric splittings of Lie algebras). Here, the Lie algebra $\mathfrak{h}$ (as subalgebra of $\mathfrak{g}$) and the supplement space $\mathfrak{p}$ are spanned by $J_{a}$ and $P_{a}$, respectively.

In light of this, given the bilinear form $\displaystyle\beta_\mathrm{\,BL}:=\beta_{1}+\big(\gamma\sqrt{|\Lambda|}\,\big)^{-1}\beta_{2}$ and thinking of the $\mathfrak{h}$-connection $\omega\in \Omega^{\,1}(X,\mathfrak{h})$ and the co-frame field $e\in \Omega^{\,1}(X,\mathfrak{p})$ as parts of a Cartan connection $A:=\omega+e\in\Omega^{\,1}(X,\mathfrak{g})$ \cite{Wise}, the Bonzom-Livine model for gravity (as a Chern-Simons theory with gauge group $\mathcal{G}$) is defined by 
\begin{equation}
\mathcal{S}_\mathrm{\,BL}[A]:=\int_{X}\beta_\mathrm{\,BL}\big(A\wedge dA+\frac{1}{3}A\wedge [A,A]\,\big)\,  ,
\end{equation}
where $\gamma$ represents an arbitrary parameter \cite{Bonzom, Barbosa}. Observe that, in terms of the $\mathfrak{h}$-connection $\omega$ and the co-frame field $e$, it is possible to write
\begin{equation}\label{BLmodel}
\begin{aligned}[b]
\mathcal{S}_\mathrm{\,BL}[\,e,\omega\,]:=\int_{X}\,&\left[\,2\,e^{a}\wedge F_{a}\,[\omega]+\frac{\Lambda}{3}\,\epsilon_{abc}\,e^{a}\wedge e^{b}\wedge e^{c}\right.\\
&+\left. \frac{1}{\gamma\,\sqrt{|\Lambda|}}\left(\omega^{a}\wedge d\,\omega_{a}+\frac{1}{3}{\epsilon_{abc}}\,\omega ^{a}\wedge\omega^{b}\wedge\omega^{c}+s\,|\Lambda|\,e^{a}\wedge d_{\omega}\,e_{a} \right)  \right]\, ,
\end{aligned}
\end{equation}  
where $s=-1,0,1$ stands for the sign of the cosmological constant $\Lambda$, while $\displaystyle F[\omega]:=d\omega+\frac{1}{2}[\omega,\omega]$ and $d_{\omega}:=d+[\omega, \cdot\,]$ denote the curvature $\mathfrak{h}$-valued $2$-form and the covariant exterior derivative associated with the $\mathfrak{h}$-connection $\omega$, respectively. As we will see in the subsequent subsections, the action principle \eqref{BLmodel} describes the $3$-dimensional Einstein theory of General Relativity with $\gamma$ playing an analogous role to that of the Immirzi parameter in 4-dimensional gravity.  Hence, we will refer to $\gamma$ as the Immirzi-like parameter. 

Finally, we would like to emphasize that, when considering the close relationship of the Bonzom-Livine model for gravity with the Chern-Simons gauge theory, it is not surprising that, the action principle \eqref{BLmodel} defines a topological field theory which is invariant under both gauge transformations and diffeomorphisms. However, the symmetries of the system are not all independent since the space-time diffeomorphisms can be generated through a combination of the gauge transformations associated with the so-called $\mathcal{H}$-gauge and translational symmetries of the theory, which are also strongly related to the first-class constraints that arise within the instantaneous Dirac-Hamiltonian analysis of the model \cite{Bonzom, Escalante}. Therefore, in what follows, we will focus our attention to the study of the gauge symmetries of the system. In this regard, we have that, on the one hand, the gauge transformations associated with the $\mathcal{H}$-gauge symmetry of the theory read
\begin{equation}\label{SU(2)-gauge sym}
\begin{aligned}[b]
e\rightarrow\, e_{\theta}&:=e+[e,\theta]\, ,\\
\omega\rightarrow\,{\omega_{\theta}}\!&:=\omega+d_{\omega}\theta\, ,
\end{aligned}
\end{equation}
where $\theta\in \Omega^{\,0}\left(X,\mathfrak{h}\right)$ denotes an arbitrary $\mathfrak{h}$-valued function on $X$. On the other hand, the gauge transformations associated with the translational symmetry of the model are given by
\begin{equation}\label{Topological sym}
\begin{aligned}[b]
e\rightarrow e_{\chi}&:=e+d_{\omega}\chi\, ,\\
\omega\rightarrow \omega_{\chi}\!&:=\omega+\,[e,\chi]\, ,
\end{aligned}
\end{equation} 
being $\chi\in\Omega^{\,0}\left(X,\mathfrak{p}\right)$ an arbitrary $\mathfrak{p}$-valued function on $X$. As discussed in \cite{Bonzom}, the translational symmetry, also known as topological symmetry \cite{Carlip}, implies that $e$ is a pure gauge field and also it is responsible for the lack of local degrees of freedom of the theory within the instantaneous Dirac-Hamiltonian formulation. Of course, the extended gauge symmetry group of the Bonzom-Livine model for gravity corresponds to $\mathcal{G}$, such that the elements of the Lie subalgebra $\mathfrak{h}$ give rise to the $\mathcal{H}$-gauge symmetry, whereas the elements of the vector subspace $\mathfrak{p}$ induce the translational symmetry. 

Next, we will describe within the geometric-covariant Lagrangian formalism the features of the gauge and diffeomorphism invariant topological field theory \eqref{BLmodel} putting special attention to analyzing the $\mathcal{H}$-gauge and translational symmetries of the system.

\subsection{ Lagrangian analysis}

In this subsection, our main aim is to analyze the Bonzom-Livine model for gravity \eqref{BLmodel} from the point of view of the geometric-covariant Lagrangian formulation for classical field theory. Here, we will not only obtain the field equations of the system but we will also study its gauge symmetries, which will be fundamental in order to construct the associated Noether currents. 


To start, let us consider $Y:=\big(T^{\,*}X\otimes\,\mathfrak{p}\big)\oplus\big(T^{\,*}X\otimes\,\mathfrak{h}\big)$. Then, we define $\big(Y,\fibbun{\,Y}{X}, X\big)$ as the covariant configuration space of the Bonzom-Livine model for gravity. This since the dynamical fields of such physical system can be understood as local sections of $\fibbun{\,Y}{X}$. In particular, given $(x^{\mu})$ a coordinate system on $X$, we identify $(x^{\mu}, a^{a}_{\mu}, b^{a}_{\mu})$ as an adapted coordinate system on $Y$. Thus, a section $\phi\in\mathscr{Y}_{X}$ can be locally represented by $(x^{\mu},e^{a}_{\mu},\omega^{a}_{\mu})$. Additionally, being $(J^{1}Y, \fibbun{\,J^{1}Y}{Y},Y)$ the affine jet bundle over $Y$, we denote by $(x^{\mu}, a^{a}_{\mu}, b^{a}_{\mu}, a^{a}_{\mu\nu}, b^{a}_{\mu\nu})$ an adapted coordinate system on $J^{1}Y$.  Then, a section $j^{1}\phi\in \mathscr{J}^{1}\mathscr{Y}_{X}$ of $\fibbun{\,J^{1}Y}{X}$, the first jet prolongation of a section $\phi\in\mathscr{Y}_{X}$, can be locally expressed as $(x^{\mu},e^{a}_{\mu},\omega^{a}_{\mu},\partial_{\mu}e^{a}_{\nu},\partial_{\mu}\omega^{a}_{\nu})$. In light of this, the action principle \eqref{BLmodel} can be rewritten as
\begin{equation}\notag
\mathcal{S}_\mathrm{\,BL}[e,\omega]=\int_{X}d^{\,n}x\,(j^{1}\phi)^{*}L_{\mathrm{BL}}\, ,
\end{equation}
with the Lagrangian function, $L_{\mathrm{BL}}:J^{1}Y\rightarrow \mathbb{R}$, explicitly given by
\begin{equation}
\begin{aligned}[b]
L_{\mathrm{BL}}&:=\epsilon^{\mu\nu\sigma}\Bigg[a^{a}_{\mu}F_{a\,\nu\sigma}+\frac{\Lambda}{3}{\epsilon_{abc}}\,a^{a}_{\mu}\,a^{b}_{\nu}\,a^{c}_{\sigma}\\
&~~~~~~~~~~+\frac{1}{\gamma\sqrt{|\Lambda|}}\left( b_{a\,\mu}b^{a}_{\nu \sigma}+\frac{1}{3}{\epsilon_{abc}}\,b^{a}_{\mu}\,b^{b}_{\nu}\,b^{c}_{\sigma}+\Lambda\,a_{a\,\mu}D_{\nu}a^{a}_{\sigma}\right) \Bigg]\, ,
\end{aligned}
\end{equation}
where $F^{a}_{\mu\nu}:=b^{a}_{\mu\nu}-b^{a}_{\nu\mu}+{\epsilon^{a}}_{bc}\,b^{b}_{\mu}\,b^{c}_{\nu}$ stands for the components of the curvature $\mathfrak{h}$-valued $2$-form $F$, $D_{\mu}a^{a}_{\nu}:=a^{a}_{\mu\nu}+{\epsilon^{a}}_{bc}\,b^{b}_{\mu}a^{c}_{\nu}$ denotes the local representation of the covariant derivative $d_{\omega}$, 
the symbol $\epsilon^{\mu\nu\sigma}$ stands for the $3$-dimensional Levi-Civita tensor, while $\epsilon_{abc}$ comes from the structure constants of the Lie algebra \eqref{BL Lie algebra}.  Nevertheless, as mentioned in section \ref{Geom Cov Formalisms}, within the geometric-covariant Lagrangian approach, the Lagrangian function is not the main object of interest  but the so-called Poincaré-Cartan forms, $\Theta^{(\mathcal{L})}_{\mathrm{BL}}\in \Omega^{\,n}(J^{1}Y)$ and $\Omega^{(\mathcal{L})}_{\mathrm{BL}}:= -d\Theta^{(\mathcal{L})}_{\mathrm{BL}}\in \Omega^{\,n+1}(J^{1}Y)$, which by construction contain all the dynamical information of the theory. In our case, by using relation \eqref{PCF}, we have that, the Poincar{\'e}-Cartan $n$-form explicitly reads
\begin{equation}
\label{BLPCform}
\begin{aligned}[b]
\Theta^{(\mathcal{L})}_\mathrm{BL}:=\epsilon^{\mu\nu\sigma}&\left[~~ \delta_{ab}\left( \frac{s\sqrt{|\Lambda|}}{\gamma}a^{b}_{\sigma}\,da^{a}_{\nu}+2\left(a^{b}_{\sigma}+\frac{1}{2\gamma\sqrt{|\Lambda|}}\,b^{b}_{\sigma}\right)db^{a}_{\nu}\right)\wedge d^{\,n-1}x_{\mu}\right.\\
&+\left.\epsilon_{abc}\left(a^{a}_{\mu}b^{b}_{\nu}b^{c}_{\sigma}+\frac{\Lambda}{3}a^{a}_{\mu}a^{b}_{\nu}a^{c}_{\sigma}+\frac{1}{\gamma\sqrt{|\Lambda|}}\left(\frac{1}{3}b^{a}_{\mu}b^{b}_{\nu}b^{c}_{\sigma}+\Lambda\,b^{a}_{\mu}a^{b}_{\nu}a^{c}_{\sigma}\right)\right)d^{\,n}x~\right]\, .
\end{aligned}
\end{equation}
As we will see below, the Poincar{\'e}-Cartan forms will allow us to obtain not only the correct field equations of the system but also the conserved currents associated with the $\mathcal{H}$-gauge and translational symmetries of the model.  

For simplicity, in order to obtain the dynamical equations of the Bonzom-Livine model for gravity, let us consider $W\in \mathfrak{X}\left(J^{1}Y\right)$ an arbitrary vector field on $J^{1}Y$ locally represented by 
\begin{equation}
W:={W^{1}}^{a}_{\mu}\frac{\partial}{\partial a^{a}_{\mu}}+{W^{2}}^{a}_{\mu}\frac{\partial}{\partial b^{a}_{\mu}}\, .
\end{equation}
By direct calculation, it is not difficult to see that,
\begin{equation}\label{Identity1}\notag
\begin{aligned}
W\!\lrcorner\,d\,\Theta^{(\mathcal{L})}_{\mathrm{BL}}&=\epsilon^{\mu\nu\sigma}\left[ 2\left(\frac{s\sqrt{|\Lambda|}}{\gamma}\,{W^{1}}^{a}_{\mu}+{W^{2}}^{a}_{\mu}\right)\!\Big(-\delta_{ab}\,da^{b}_{\nu}\wedge d^{\,n-1}x_{\sigma}+\epsilon_{abc}\,b^{b}_{\nu}a^{c}_{\sigma}\,d^{\,n}x\Big)\right.\\
&~~~+\left. \left({W^{1}}^{a}_{\mu}+\frac{1}{\gamma\sqrt{|\Lambda|}}{W^{2}}^{a}_{\mu}\right)\!\Big( -2\,\delta_{ab}\,db^{b}_{\nu}\wedge d^{\,n-1}x_{\sigma}+\epsilon_{abc}\left(b^{b}_{\nu}b^{c}_{\sigma}+\Lambda\,a^{b}_{\nu}a^{c}_{\sigma}\right)d^{\,n}x \Big)\right]\, .
\end{aligned}
\end{equation}
Besides, we know that, given $\phi\in \mathscr{Y}_{X}$ a critical point of the action principle \eqref{BLmodel}, the condition $j^{1}\phi^{*}\left(W\lrcorner\,\,\Omega^{(\mathcal{L})}_{\mathrm{BL}}\right)=0$ must hold, and hence we can write
\begin{equation}
\begin{aligned}\notag
\epsilon^{\mu\nu\sigma}\delta_{ab}\left(2\left(\frac{s\sqrt{|\Lambda|}}{\gamma}\,{W^{1}}^{a}_{\mu}+{W^{2}}^{a}_{\mu}\right)D_{\nu}e^{b}_{\sigma}+\left({W^{1}}^{a}_{\mu}+\frac{1}{\gamma\sqrt{|\Lambda|}}{W^{2}}^{a}_{\mu}\right)\Big(F^{b}_{\nu\sigma}+\Lambda\,{\epsilon^{b}}_{cd}\,e^{c}_{\nu}e^{d}_{\sigma}\Big)\right)=0\, .
\end{aligned}
\end{equation}
In particular, by taking into account that $W$ is an arbitrary vector field on $J^{1}Y$, the above relation gives rise to the following set of equations
\begin{equation}\label{BL Field Equations}
\begin{aligned}[b]
\frac{s\sqrt{|\Lambda|}}{\gamma}D_{[\mu}e^{a}_{\nu]}+\frac{1}{2}\Big(F^{a}_{\mu\nu}+\Lambda\,{\epsilon^{a}}_{bc}\,e^{b}_{\mu}e^{c}_{\nu}\Big)=&\,0\, ,\\
D_{[\mu}e^{a}_{\nu]}+\frac{1}{2\gamma\sqrt{|\Lambda|}}\Big(F^{a}_{\mu\nu}+\Lambda\,{\epsilon^{a}}_{bc}\,e^{b}_{\mu}e^{c}_{\nu}\Big)=&\,0\, ,
\end{aligned}
\end{equation}
which correspond to the field equations of the Bonzom-Livine model for gravity obtained in \cite{Bonzom}. Of course, it is not difficult to show that, for $\gamma^{2}\neq s$, relations \eqref{BL Field Equations} are completely equivalent to the vanishing torsion condition and the Einstein equations~\cite{Witten, Carlip, Wise}, namely
\begin{subequations}
\begin{align}
D_{[\mu}e^{a}_{\nu]}&=0\, ,\\
F^{a}_{\mu\nu}+\Lambda\,{\epsilon^{a}}_{bc}\,e^{b}_{\mu}e^{c}_{\nu}&=0\, .
\end{align}
\end{subequations} 
In other words, despite the $\gamma$ ambiguity inherent to the Bonzom-Livine model for gravity, the action principle \eqref{BLmodel} is able to reproduce the same field equations of the $3$-dimensional Einstein theory of General Relativity, as long as
the condition $\gamma^2\neq s$ holds.  

Now, we will analyze the gauge symmetries associated with the extended gauge symmetry group of the theory at the Lagrangian level. For this purpose, given $\xi_{\theta}\in\mathfrak{h}$ and $\xi_{\chi}\in\mathfrak{p}$, we start by defining $\xi^{Y}_{\theta}\in\mathfrak{X}^{\,1}(Y)$ and $\xi^{Y}_{\chi}\in\mathfrak{X}^{\,1}(Y)$ as the infinitesimal generators of the gauge transformation \eqref{SU(2)-gauge sym} and \eqref{Topological sym}, respectively. In local coordinates these vector fields are explicitly given by
\begin{equation}
\label{Infinitesimal Generators}
\begin{aligned}[b]
\xi^{Y}_{\theta}&:={\epsilon^{a}}_{bc}\,a^{b}_{\mu}\,\theta^{c}\frac{\partial}{\partial a^{a}_{\mu}} + D_{\mu}\theta^{a}\frac{\partial}{\partial b^{a}_{\mu}}\, ,\\
\xi^{Y}_{\chi}&:= D_{\mu}\chi^{a}\frac{\partial}{\partial a^{a}_{\mu}} + \Lambda\,{\epsilon^{a}}_{bc}\,a^{b}_{\mu}\,\chi^{c}\frac{\partial}{\partial b^{a}_{\mu}} \, .
\end{aligned}
\end{equation}
Furthermore, we know that fibre-preserving transformations on the covariant configuration space of a given classical field theory induce fibre-preserving transformations on the corresponding affine jet bundle. In particular, by using formula \eqref{PJB}, it is possible to see that, the first jet prolongations, $\xi^{J^{1}Y}_{\theta}\in\mathfrak{X}^{\,1}(J^{1}Y)$ and $\xi^{J^{1}Y}_{\chi}\in\mathfrak{X}^{\,1}(J^{1}Y)$, associated with the infinitesimal generators \eqref{Infinitesimal Generators} can be locally written as
\begin{subequations}
\begin{align}
\xi^{J^{1}Y}_{\theta}\!&:={\epsilon^{a}}_{bc}\,a^{b}_{\mu}\,\theta^{c}\frac{\partial}{\partial a^{a}_{\mu}} + D_{\mu}\theta^{a}\frac{\partial}{\partial b^{a}_{\mu}}+{\epsilon^{a}}_{bc}\left(a^{b}_{\nu}\partial_{\mu}\theta^{c}+a^{b}_{\mu\nu}\theta^{c}\right)\frac{\partial}{\partial a^{a}_{\mu\nu}}\notag\\
&~~~~+\left(D_{\nu}\left(\partial_{\mu}\theta^{a}\right)+{\epsilon^{a}}_{bc}\,b^{b}_{\mu\nu}\theta^{c}\right)\frac{\partial}{\partial b^{a}_{\mu\nu}}\, ,\label{Pro of the SU(2)-gauge sym}\\
\xi^{J^{1}Y}_{\chi}\!&:= D_{\mu}\chi^{a}\frac{\partial}{\partial a^{a}_{\mu}}+\Lambda\,{\epsilon^{a}}_{bc}\,a^{b}_{\mu}\,\chi^{c}\frac{\partial}{\partial b^{a}_{\mu}}+\left(D_{\nu}\left(\partial_{\mu}\chi^{a}\right)+{\epsilon^{a}}_{bc}\,b^{b}_{\mu\nu}\chi^{c}\right)\frac{\partial}{\partial a^{a}_{\mu\nu}}\notag\\
&~~~~+\Lambda\,{\epsilon^{a}}_{bc}\left(a^{b}_{\nu}\partial_{\mu}\chi^{c}+a^{b}_{\mu\nu}\chi^{c}\right)\frac{\partial}{\partial b^{a}_{\mu\nu}}\, .
\end{align}
\end{subequations} 

Bearing this in mind, we are in the position to study the action of the gauge symmetries of the Bonzom-Livine model for gravity on its associated Poincar{\'e}-Cartan $n$-form \eqref{BLPCform}. To begin with, let us consider the case of the $\mathcal{H}$-gauge symmetry. By direct calculation, we have that, for all $\xi_{\theta}\in\mathfrak{h}$, this gauge symmetry preserves the Poincar{\'e}-Cartan $n$-form up to an exact form, namely
\begin{equation}\label{Action of the SU(2) sym}
\mathcal{L}_{\xi^{J^{1}Y}_{\theta}}\Theta^{(\mathcal{L})}_{\mathrm{BL}}=d\alpha^{(\mathcal{L})}_{\theta}\, ,
\end{equation}
where $\alpha^{(\mathcal{L})}_{\theta}\in \Omega^{\,n-1}_{1}(J^{1}Y)$ represents a $\fibbun{\,J^{1}Y}{X}$-horizontal $(n-1\,;1)$-form on $J^{1}Y$ explicitly given by
\begin{equation}\label{SU(2)-gauge alpha}
\alpha_{\theta}^{(\mathcal{L})}:=\frac{1}{\gamma\sqrt{|\Lambda|}}\,\epsilon^{\mu\nu\sigma}\,\delta_{ab}\,\partial_{\sigma}\theta^{a}b^{b}_{\nu}d^{\,n-1}x_{\mu}\, .
\end{equation}
In a similar manner, a straightforward calculation shows that, for all $\xi_{\chi}\in\mathfrak{p}$, the translational symmetry of the theory also preserves the Poincar{\'e}-Cartan $n$-form up to an exact form, specifically
\begin{equation}\label{Action of the Topological sym}
\mathcal{L}_{\xi^{J^{1}Y}_{\chi}}\Theta^{(\mathcal{L})}_{\mathrm{BL}}=d \alpha^{(\mathcal{L})}_{\chi}\, ,
\end{equation}
being $\alpha^{(\mathcal{L})}_{\chi}\in \Omega^{\,n-1}_{1}(J^{1}Y)$ a $\fibbun{\,J^{1}Y}{X}$-horizontal $(n-1\,;1)$-form on $J^{1}Y$ locally represented by
\begin{equation}\label{Topological alpha}
\alpha_{\chi}^{(\mathcal{L})}:=\epsilon^{\mu\nu\sigma}\left( \epsilon_{abc}\,\chi^{a}\left(b^{b}_{\nu}b^{c}_{\sigma}-\Lambda\, a^{b}_{\nu}a^{c}_{\sigma}\right)+\delta_{ab}\,\partial_{\sigma}\chi^{a}\left(\frac{s\sqrt{|\Lambda|}}{\gamma}\,a^{b}_{\nu}+2\,b^{b}_{\nu}\right) \right)d^{\,n-1}x_{\mu}\, .
\end{equation}

Here, we would like to emphasize that, on the one hand, the $\fibbun{\,J^{1}Y}{X}$-horizontal $(n-1\,;1)$-forms \eqref{SU(2)-gauge alpha} and \eqref{Topological alpha} are of the form \eqref{Lagrangian alpha}, and hence we can apply the techniques described in section \ref{Geom Cov Formalisms} to study the gauge symmetries of the system. On the other hand, we know that, the $\mathcal{H}$-gauge and translational symmetries of the Bonzom-Livine model for gravity are localizable symmetries. To illustrate this, let us consider the case of the $\mathcal{H}$-gauge symmetry since the same arguments are valid for the translational symmetry. To start, note that, for all $\theta\in \Omega^{\,0}(X,\mathfrak{h})$, both the vector field \eqref{Pro of the SU(2)-gauge sym} and the $\fibbun{\,J^{1}Y}{X}$-horizontal $(n-1\,;1)$-form \eqref{Topological alpha} are linear in the components of $\theta$ and their partial derivatives, and therefore the collection $\mathcal{C}^{\,\mathcal{H}}_{\mathsmaller{\mathrm{LS}}}$ of pairs $\big(\xi^{J^{1}Y}_{\theta},\alpha^{(\mathcal{L})}_{\theta}\big)$ is a vector space. In addition, we known that each pair $\big(\xi^{J^{1}Y}_{\theta},\alpha^{(\mathcal{L})}_{\theta}\big)\in \mathcal{C}^{\mathcal{H}}_{\mathsmaller{\mathrm{LS}}}$  corresponds to a Noether symmetry since for any $\theta\in \Omega^{\,0}(X,\mathfrak{h})$ the relation \eqref{Action of the SU(2) sym} holds. Now, for an arbitrary element $\theta\in \Omega^{\,0}(X,\mathfrak{h})$, let us consider $U_{1}, U_{2}\subset X$ two open sets with disjoint closures and $\tilde{\theta}\in\Omega^{\,0}(X,\mathfrak{h})$ an element such that $\tilde{\theta}=\theta$ on $U_{1}$ but $\tilde{\theta}=0$ on $U_{2}$. Then, it is clear that, the pairs $\big(\xi^{J^{1}Y}_{\theta},\alpha^{(\mathcal{L})}_{\theta}\big), \,\big(\xi^{J^{1}Y}_{\tilde{\theta}},\alpha^{(\mathcal{L})}_{\tilde{\theta}}\big)\in\mathcal{C}^{\mathcal{H}}_{\mathsmaller{\mathrm{LS}}}$ satisfy condition \eqref{Condition 3}, and proceeding in the same way for any pair of open sets with disjoint closures $U_{1}, U_{2}\subset X$, it is possible to see that, the $\mathcal{H}$-gauge symmetry of the gravity model \eqref{BLmodel} satisfies the properties of a localizable symmetry. 

Hence, the action of $\mathcal{G}$ on $J^{1}Y$ has an associated Lagrangian covariant momentum map, namely $J^{(\mathcal{L})}:J^{1}Y\rightarrow \mathfrak{g}^{*}\otimes \Lambda^{n-1}T^{\,*}J^{1}Y$. In particular, by taking into account the splitting $\mathfrak{g}=\mathfrak{h}\oplus \mathfrak{p}$, we can  obtain the local representation of such a map, that is, $J^{(\mathcal{L})}(\xi)\in\Omega^{\,n-1}(J^{1}Y)$ for an arbitrary $\xi\in\mathfrak{g}$, as follows. To start, let us consider $\xi_{\theta}\in \mathfrak{h}$ and $\xi_{\chi}\in \mathfrak{p}$. Then, by considering~\eqref{eq:LagCurrent}, and in light of relations \eqref{Action of the SU(2) sym} and \eqref{Action of the Topological sym}, it is possible to write
\begin{equation}\label{BL Lagrangian covariant momentum maps}
\begin{aligned}[b]
J^{(\mathcal{L})}\!\left(\xi_{\theta}\right)\!:\!&=\epsilon^{\mu\nu\sigma}\Bigg[2\,\Big(\delta_{ab}\,\partial_{\nu}\theta^{a}a^{b}_{\sigma}-\epsilon_{abc}\,\theta^{a}b^{b}_{\nu}a^{c}_{\sigma}\Big)\\
&~~~~~~~~~+\frac{1}{\gamma\sqrt{|\Lambda|}}\bigg(2\,\delta_{ab}\,\partial_{\nu}\theta^{b}b^{b}_{\sigma}-\epsilon_{abc}\,\theta^{a}\Big(b^{b}_{\nu}b^{c}_{\sigma}+\Lambda\,a^{b}_{\nu}a^{c}_{\sigma}\Big)\bigg)\Bigg]d^{\,n-1}x_{\mu}\, ,\\
J^{(\mathcal{L})}\!\left(\xi_{\chi}\right)\!:\!&=\epsilon^{\mu\nu\sigma}\Bigg[\frac{2\,s\sqrt{|\Lambda|}}{\gamma}\bigg(\delta_{ab}\,\partial_{\nu}\chi^{a}a^{b}_{\sigma}-\epsilon_{abc}\,\chi^{a}b^{b}_{\nu}a^{c}_{\sigma}\bigg)\\
&~~~~~~~~~+2\,\delta_{ab}\,\partial_{\nu}\chi^{b}b^{b}_{\sigma}-\epsilon_{abc}\,\chi^{a}\Big(b^{b}_{\nu}b^{c}_{\sigma}+\Lambda\,a^{b}_{\nu}a^{c}_{\sigma}\Big)\Bigg]d^{\,n-1}x_{\mu}\, .
\end{aligned}
\end{equation}
Of course, this is important as we can obtain the Noether currents of a given classical field theory by means of \eqref{NoethCurrDef}, namely by pulling-back the Lagrangian covariant momentum map with a solution of the Euler-Lagrange field equations of the system. In our case, we have that, for all $\xi_{\theta}\in \mathfrak{h}$, $\xi_{\chi}\in \mathfrak{p}$ and $\phi\in\mathscr{Y}_{X}$ a solution of the field equations \eqref{BL Field Equations}, the $(n-1)$-forms on $X$ explicitly defined by 
\begin{equation}\label{BL Noether currents}
\begin{aligned}[b]
\mathcal{J}^{(\mathcal{L})}\!\left(\xi_{\theta}\right)\!:\!&=\epsilon^{\mu\nu\sigma}\Bigg[2\,\Big(\delta_{ab}\,\partial_{\nu}\theta^{a}e^{b}_{\sigma}-\epsilon_{abc}\,\theta^{a}\omega^{b}_{\nu}e^{c}_{\sigma}\Big)\\
&~~~~~~~~~+\frac{1}{\gamma\sqrt{|\Lambda|}}\bigg(2\,\delta_{ab}\,\partial_{\nu}\theta^{b}\omega^{b}_{\sigma}-\epsilon_{abc}\,\theta^{a}\Big(\omega^{b}_{\nu}\omega^{c}_{\sigma}+\Lambda\,e^{b}_{\nu}e^{c}_{\sigma}\Big)\bigg)\Bigg]d^{\,n-1}x_{\mu}\, ,\\
\mathcal{J}^{(\mathcal{L})}\!\left(\xi_{\chi}\right)\!:\!&=\epsilon^{\mu\nu\sigma}\Bigg[\frac{2\,s\sqrt{|\Lambda|}}{\gamma}\bigg(\delta_{ab}\,\partial_{\nu}\chi^{a}e^{b}_{\sigma}-\epsilon_{abc}\,\chi^{a}\omega^{b}_{\nu}e^{c}_{\sigma}\bigg)\\
&~~~~~~~~~+2\,\delta_{ab}\,\partial_{\nu}\chi^{b}\omega^{b}_{\sigma}-\epsilon_{abc}\,\chi^{a}\Big(\omega^{b}_{\nu}\omega^{c}_{\sigma}+\Lambda\,e^{b}_{\nu}e^{c}_{\sigma}\Big)\Bigg]d^{\,n-1}x_{\mu}\, ,
\end{aligned}
\end{equation}
are nothing but the Noether currents associated with the $\mathcal{H}$-gauge and translational symmetries of the Bonzom-Livine model for gravity, respectively. Observe that, by imposing the field equations \eqref{BL Field Equations}, the integral of each of the Noether currents \eqref{BL Noether currents} over a Cauchy surface $\Sigma_{t}$ vanishes, that is, relation \eqref{NoetCharVanishing} holds. Naturally, the latter is consistent with the fact that the $\mathcal{H}$-gauge and translational symmetries of the theory are localizable symmetries, and therefore the second Noether theorem establishes that, for all $\phi\in\mathscr{Y}_{X}$ a solution of the Euler-Lagrange field equation of the theory, the Lagrangian Noether charges of the system must vanish.

In the following subsection, we will analyze the gravity model \eqref{BLmodel} within the multisymplectic approach paying special attention to the study of the gauge symmetries of the theory, which will be fundamental for our discussion.

\subsection{Multisymplectic analysis}

In the present subsection, we will perform the multisymplectic analysis of the Bonzom-Livine model for gravity. In particular, we will focus our attention on studying the action of the extended gauge symmetry group of the theory on its corresponding covariant multimomenta phase-space. Here, we will describe how the $\mathcal{H}$-gauge and translational symmetries of the system give rise to covariant canonical transformations, which in turn will allow us to obtain the covariant momentum map associated with the extended gauge symmetry group of the model.

To start, let us consider $(Z^{\star}, \fibbun{\,Z^{\star}}{Y}, Y)$ the covariant multimomenta phase-space associated with the theory. Then, as described in subsection~\ref{SectionMulti}, given $(x^{\mu}, a^{a}_{\mu}, b^{a}_{\mu})$ an adapted coordinate system on $Y$, we introduce $(x^{\mu}, a^{a}_{\mu}, b^{a}_{\mu}, p, p^{\,\mu\nu}_{a}, \bar{p}^{\,\mu\nu}_{a})$ to denote an adapted coordinate system on $Z^{\star}$. Thus, the canonical and multisymplectic forms of the Bonzom-Livine model for gravity, $\Theta^{(\mathcal{M})}_{\mathrm{BL}}\in\Omega^{\,n}(Z^{\star})$ and $\Omega^{(\mathcal{M})}_{\mathrm{BL}}\in\Omega^{\,n+1}(Z^{\star})$, can be locally written as 
\begin{subequations}
\begin{align}
\label{BL Multisymplectic Potential}
\Theta^{(\mathcal{M})}_{\mathrm{BL}}&:=p^{\,\mu\nu}_{a}da^{a}_{\nu}\wedge d^{\,n-1}x_{\mu}+\bar{p}^{\,\mu\nu}_{a}db^{a}_{\nu}\wedge d^{\,n-1}x_{\mu}+p\,d^{\,n}x\, ,\\
\label{BL Multisymplectic Form}
\Omega^{(\mathcal{M})}_{\mathrm{BL}}&:=da^{a}_{\nu}\wedge dp^{\,\mu\nu}_{a}\wedge d^{\,n-1}x_{\mu}+db^{a}_{\nu}\wedge d\bar{p}^{\,\mu\nu}_{a}\wedge d^{\,n-1}x_{\mu}-dp\wedge d^{\,n}x\, .
\end{align}
\end{subequations}

Now, in order to study the $\mathcal{H}$-gauge and translational symmetries of the theory at the multisymplectic level, given $\xi_{\theta}\in\mathfrak{h}$ and $\xi_{\chi}\in\mathfrak{p}$, we start by identifying $\xi^{\alpha}_{\theta}\in\mathfrak{X}^{\,1}(Z^{\star})$ and $\xi^{\alpha}_{\chi}\in\mathfrak{X}^{\,1}(Z^{\star})$, that is, the $\alpha^{(\mathcal{M})}$-lifts of the vector fields \eqref{Infinitesimal Generators} to $Z^{\star}$, as the infinitesimal generators of the gauge transformations \eqref{SU(2)-gauge sym} and \eqref{Topological sym} on the covariant multimomenta phase-space, respectively. In local coordinates these vector field explicitly read
\begin{subequations}\label{BL alpha-lifts}
\begin{align}
\xi^{\alpha}_{\theta}&:=\,{\epsilon^{a}}_{bc}\,a^{b}_{\mu}\,\theta^{c}\frac{\partial}{\partial a^{a}_{\mu}}+D_{\mu}\theta^{a}\frac{\partial}{\partial b^{a}_{\mu}}-\left({\epsilon_{ab}}^{c}\theta^{b}\bar{p}^{\,\mu\nu}_{c}-\frac{1}{\gamma\,\sqrt{|\Lambda|}}\,\epsilon^{\mu\nu\sigma}\,\delta_{ab}\,\partial_{\sigma}\theta^{b} \right)\frac{\partial}{\partial \bar{p}^{\,\mu\nu}_{a}}\notag\\
&~~~~\,-{\epsilon_{ab}}^{c}\theta^{b}p^{\,\mu\nu}_{c}\frac{\partial}{\partial p^{\,\mu\nu}_{a}}-\Big({\epsilon^{a}}_{bc}\,a^{b}_{\nu}\,\partial_{\mu}\theta^{c}p^{\,\mu\nu}_{a}+D_{\nu}(\partial_{\mu}\theta^{a})\bar{p}^{\,\mu\nu}_{a}\Big)\frac{\partial}{\partial p}\, ,\\
\xi^{\alpha}_{\chi}&:=\,D_{\mu}\chi^{a}\frac{\partial}{\partial a^{a}_{\mu}}+\Lambda\,{\epsilon^{a}}_{bc}\,a^{b}_{\mu}\,\chi^{c}\frac{\partial}{\partial b^{a}_{\mu}}-\Big({\epsilon_{ab}}^{c}\chi^{b}p^{\,\mu\nu}_{c}-2\,\epsilon^{\mu\nu\sigma}\,\delta_{ab}\,D_{\sigma}\chi^{b}\Big)\frac{\partial}{\partial \bar{p}^{\,\mu\nu}_{a}}\notag\\
&~~~~\,-\Lambda\left({\epsilon_{ab}}^{c}\chi^{b}\bar{p}^{\,\mu\nu}_{c}-\epsilon^{\mu\nu\sigma}\Bigg(\frac{1}{\gamma\,\sqrt{|\Lambda|}}\,\delta_{ab}\,\partial_{\sigma}\chi^{b}+2\,\epsilon_{abc}\,\chi^{b}a^{c}_{\sigma}\Bigg)\right)\frac{\partial}{\partial p^{\,\mu\nu}_{a}}\notag\\
&~~~~\,-\Bigg(D_{\nu}(\partial_{\mu}\chi^{a})p^{\,\mu\nu}_{a}+\Lambda\,{\epsilon^{a}}_{bc}\,a^{b}_{\nu}\,\partial_{\mu}\chi^{c}\,\bar{p}^{\,\mu\nu}_{a}-\epsilon^{\mu\nu\sigma}\,\epsilon_{abc}\,\partial_{\mu}\chi^{a}\Big(b^{b}_{\nu}b^{c}_{\sigma}-\Lambda\,a^{b}_{\nu}a^{c}_{\sigma}\Big)\Bigg)\frac{\partial}{\partial p}\, .
\end{align}
\end{subequations}

With this in mind, it is not difficult to see that, for all $\xi_{\theta}\in\mathfrak{h}$, the $\mathcal{H}$-gauge symmetry of the model acts on $Z^{\star}$ through covariant canonical transformations, specifically
\begin{equation}
\label{Action of SU(2)-alpha-sym}
\mathcal{L}_{\xi^{\alpha}_{\theta}}\Theta^{(\mathcal{M})}_\mathrm{BL}=d\alpha^{(\mathcal{M})}_{\theta}\, ,
\end{equation}
being $\alpha^{(\mathcal{M})}_{\theta}\in \Omega^{\,n-1}_{1}(Z^{\star})$ a $\fibbun{\,Z^{\star}}{X}$-horizontal $(n-1\,;1)$-form on $Z^{\star}$ locally given by
\begin{equation}\label{BL  multisymplectic SU(2)-alpha}
\alpha^{(\mathcal{M})}_{\theta}:=\frac{1}{\gamma\sqrt{|\Lambda|}}\,\epsilon^{\mu\nu\sigma}\,\delta_{ab}\,\partial_{\sigma}\theta^{a}b^{b}_{\nu}\,d^{\,n-1}x_{\mu}\, .
\end{equation}
In a similar manner, we have that, for all $\xi_{\chi}\in\mathfrak{p}$, the translational symmetry of the system also acts on $Z^{\star}$ through covariant canonical transformations since the following relation holds
\begin{equation}\label{Action of topological-alpha-sym}
\mathcal{L}_{\xi^{\alpha}_{\chi}}\Theta^{(\mathcal{M})}_\mathrm{BL}=d\alpha^{(\mathcal{M})}_{\chi}\, ,
\end{equation}
where $\alpha^{(\mathcal{M})}_{\chi}\in\Omega^{\,n-1}_{1}(Z^{\star})$ denotes a $\fibbun{\,Z^{\star}}{X}$-horizontal $(n-1\,;1)$-form on $Z^{\star}$ locally represented as
\begin{equation}\label{BL multisymplectic topological alpha}
\alpha^{(\mathcal{M})}_{\chi}:=\epsilon^{\mu\nu\sigma}\left( \epsilon_{abc}\,\chi^{a}\left(b^{b}_{\nu}b^{c}_{\sigma}-\Lambda\, a^{b}_{\nu}a^{c}_{\sigma}\right)+\delta_{ab}\,\partial_{\sigma}\chi^{a}\left(\frac{s\sqrt{|\Lambda|}}{\gamma}\,a^{b}_{\nu}+2\,b^{b}_{\nu}\right) \right)d^{\,n-1}x_{\mu}\, .
\end{equation}

Therefore, the action of $\mathcal{G}$ on $Z^{\star}$ has an associated covariant momentum map, namely, $J^{(\mathcal{M})}:Z^{\star}\rightarrow \mathfrak{g}^{*}\otimes\Lambda^{n-1}T^{*}Z^{\star}$. In fact, as in the Lagrangian case, by considering the splitting $\mathfrak{g}=\mathfrak{h}\oplus\mathfrak{p}$, we can express the local representation of such a map, that is, $J^{(\mathcal{M})}(\xi)\in\Omega^{\,n-1}(Z^{\star})$ for an arbitrary $\xi\in\mathfrak{g}$, as follows. To begin with, let us consider  $\xi_{\theta}\in\mathfrak{h}$ and $\xi_{\chi}\in\mathfrak{p}$. Then, by taking into account \eqref{eq:MultiCurrent}, and in light of relations \eqref{Action of SU(2)-alpha-sym} and \eqref{Action of topological-alpha-sym}, it is possible to write
\begin{equation}\label{BL Covariant Momentum Maps}
\begin{aligned}[b]
J^{(\mathcal{M})}\left(\xi_{\theta}\right):=&\Bigg[D_{\nu}\theta^{a}\bar{p}^{\,\mu\nu}_{a}+{\epsilon^{a}}_{bc}\,a^{b}_{\nu}\,\theta^{c}p^{\,\mu\nu}_{a}-\frac{1}{\gamma\,\sqrt{|\Lambda|}}\,\epsilon^{\mu\nu\sigma}\,\delta_{ab}\,\partial_{\sigma}\theta^{a}\,b^{b}_{\nu}\Bigg]d^{\,n-1}x_{\mu}\, ,\\
J^{(\mathcal{M})}\left(\xi_{\chi}\right):=&\Bigg[D_{\nu}\chi^{a}p^{\,\mu\nu}_{a}+\Lambda\,{\epsilon^{a}}_{bc}\,a^{b}_{\nu}\,\chi^{c}\bar{p}^{\,\mu\nu}_{a}\\
&-\epsilon^{\mu\nu\sigma}\Bigg(\epsilon_{abc}\,\chi^{a}\Big(b^{b}_{\nu}b^{c}_{\sigma}-\Lambda\,a^{b}_{\nu}a^{c}_{\sigma}\Big)+\delta_{ab}\,\partial_{\sigma}\chi^{a}\Bigg(\frac{s\,\sqrt{|\Lambda|}}{\gamma}\,a^{b}_{\nu}+2\,b^{b}_{\nu}\Bigg)\Bigg)\Bigg]d^{\,n-1}x_{\mu}\, .
\end{aligned}
\end{equation}
Note that the Lagrangian covariant momentum map \eqref{BL Lagrangian covariant momentum maps} can be recovered by pulling-back the covariant momentum map \eqref{BL Covariant Momentum Maps} with the covariant Legendre transformation \eqref{CLT}, which establishes a link between both geometric structures. Finally,  we would like to mention that, as we will see in the subsequent subsections, the covariant momentum map associated with the extended gauge symmetry group of the model will allow us not only to construct conserved currents for the solutions of the De Donder-Weyl-Hamilton field equations of the system but it will also be fundamental to obtain, on the space of Cauchy data, the first-class constrained structure that characterizes the theory within the instantaneous Dirac-Hamiltonian formulation. 

Next, we will carry out the polysymplectic analysis of the gravity model \eqref{BLmodel}, where we will emphasize the relevance of the Immirzi-like parameter inherent to the system within the De Donder-Weyl Hamiltonian formulation.

\subsection{Polysymplectic analysis}

In this subsection, we will study the Bonzom-Livine model for gravity from the point of view of the polysymplectic formalism. To this end, since the gravity model of our interest corresponds to a singular Lagrangian system, we will make use of the algorithm to study this kind of systems within the polysymplectic approach. Here, we will not only obtain the correct De Donder-Weyl-Hamilton field equations of the theory but we will also describe how the covariant momentum map associated with the extended gauge symmetry group of the system gives rise to the conserved currents of the model within the De Donder-Weyl canonical theory. 

To begin with, let us consider the quotient bundle $\left(P,\fibbun{\,P}{Y},Y\right)$, as described in subsection \ref{PF}. Then, given $(x^{\mu}, a^{a}_{\mu}, b^{a}_{\mu})$ an adapted coordinate system on $Y$, we denote by $\left(x^{\mu},a^{a}_{\mu},b^{a}_{\mu},p^{\,\mu\nu}_{a},\bar{p}^{\,\mu\nu}_{a}\right)$ an adapted coordinate system on $P$. Thus, a section $\rho\in\mathscr{P}_{X}$ of $\fibbun{\,P}{X}$ (the polymomenta phase-space of the theory) can be locally represented as $(x^{\mu}, e^{a}_{\mu}, \omega^{a}_{\mu},\pi^{\mu\nu}_{a}, \bar{\pi}^{\mu\nu}_{a})$. Now, in order to describe the gravity model \eqref{BLmodel} in a covariant Hamiltonian-like formulation, we proceed to apply the covariant Legendre map \eqref{LegMap}, which allows us to write
\begin{equation}\label{BL Polymomenta}
\begin{aligned}[b]
p^{\,\mu\nu}_{a}&:=\frac{\partial L_{\mathrm{BL}}}{\partial a^{a}_{\mu\nu}}=\frac{s\sqrt{|\Lambda|}}{\gamma}\epsilon^{\mu\nu\sigma}\,\delta_{ab}\,a^{b}_{\sigma}\,,\\
\bar{p}^{\,\mu\nu}_{a}&:=\frac{\partial L_{\mathrm{BL}}}{\partial b^{a}_{\mu\nu}}=2\,\epsilon^{\mu\nu\sigma}\,\delta_{ab}\left(a^{b}_{\sigma}+\frac{1}{2\,\gamma\sqrt{|\Lambda|}}\,b^{b}_{\sigma}\right)\, .
\end{aligned}
\end{equation}

Note that, on the one hand, the set of relations \eqref{BL Polymomenta} gives rise to the primary constraints of the Bonzom-Livine model for gravity  which properly identified as $(n-1\,;1)$-forms explicitly read
\begin{subequations}\label{PrimConst}
\begin{align}
{C^{(1)}}^{\nu}_{a}&:=\left(p^{\,\mu\nu}_{a}-\frac{s\sqrt{|\Lambda|}}{\gamma}\epsilon^{\mu\nu\sigma}\,\delta_{ab}\,a^{b}_{\sigma}\right)\varpi_{\mu}\approx 0\, ,\label{PrimConst1} \\
{C^{(2)}}^{\nu}_{a}&:=\left(\bar{p}^{\,\mu\nu}_{a}-2\,\epsilon^{\mu\nu\sigma}\delta_{ab}\left(a^{b}_{\sigma}+\frac{1}{2\,\gamma\sqrt{|\Lambda|}}\,b^{b}_{\sigma}\right)\right)\varpi_{\mu}\approx 0\, .\label{PrimConst2}
\end{align}
\end{subequations}
On the other hand, we have that, on the primary constraint surface $P^{\mathrm{PCS}}_{\mathrm{BL}}\subset P$, namely the surface on $P$ characterized by the vanishing of the primary constraint $(n-1)$-forms \eqref{PrimConst}, the polymomenta variables of the system are anti-symmetric in the space-time indices, and hence on $P^{\mathrm{PCS}}_{\mathrm{BL}}$ we are able to choose a more suitable set of canonically conjugate variables for which the commutation relations under the Poisson-Gerstenhaber bracket are given by 
\begin{equation}\label{BL Conjugate Variables}
\begin{aligned}[b]
\{\![\,p^{\,\mu\nu}_{a}\varpi_{\mu},a^{b}_{[\sigma}\varpi_{\rho]}\,]\!\}&=\delta^{b}_{a}\delta^{\nu}_{[\sigma}\varpi_{\rho]}\, ,\\
\{\![\,\bar{p}^{\,\mu\nu}_{a} \varpi_{\mu},b^{b}_{[\sigma} \varpi_{\rho]}\,]\!\}&=\delta^{b}_{a}\delta^{\nu}_{[\sigma} \varpi_{\rho]}\, .
\end{aligned}
\end{equation}
In particular, this choice of canonically conjugate variables for the bracket structure associated with the physical model under analysis is related to the fact that the dynamical information of a gauge theory whose fields are Lie algebra-valued 1-forms is contained in the anti-symmetric part of the polymomenta variables, as discussed in \cite{DeLeon3, Sardanashvily1}. Thus, the set of canonically conjugate variables introduced in \eqref{BL Conjugate Variables} will allow us to consistently describe the system of our interest within the polysymplectic approach, as we will see throughout this subsection.   

Bearing this in mind, it is not difficult to see that, on the primary constraint surface $P^\mathrm{PCS}_\mathrm{\,BL}$,  the De Donder-Weyl Hamiltonian associated with the Bonzom-Livine model for gravity can be written as
\begin{equation}\label{BL Canonic Hamiltonian}
H_\mathsmaller{\mathrm{DW}}:=-\epsilon^{\mu\nu\sigma}\epsilon_{abc}\left(a^{a}_{\mu}b^{b}_{\nu}b^{c}_{\sigma}+\frac{\Lambda}{3}a^{a}_{\mu}a^{b}_{\nu}a^{c}_{\sigma}+\frac{1}{\gamma\sqrt{|\Lambda|}}\left(\frac{1}{3}b^{a}_{\mu}b^{b}_{\nu}b^{c}_{\sigma}+\Lambda\, a^{a}_{\mu}b^{b}_{\nu}a^{c}_{\sigma}\right)\right)\, .
\end{equation}
However, since the model under consideration corresponds to a singular Lagrangian system, we know that the De Donder-Weyl-Hamilton field equations generated by the above De Donder-Weyl Hamiltonian may not be equivalent to the Euler-Lagrange field equations associated with the theory, as it follows from the Dirac prescription for constrained
systems~\cite{QGS}.

Then, in order to perform a consistent polysymplectic formulation of the gravity model \eqref{BLmodel}, we will implement the algorithm to study singular Lagrangian systems within the polysymplectic approach. To do so, we start by introducing the total De Donder-Weyl Hamiltonian of the system, specifically
\begin{equation}
\tilde{H}_\mathsmaller{\mathrm{DW}}:=H_\mathsmaller{\mathrm{DW}}+{\lambda_{(i)}}^{a}_{\nu}\bullet {C^{(i)}}^{\nu}_{a},
\end{equation}
where ${\lambda_{(i)}}_{\nu}^{a}:={\lambda_{(i)}}^{a}_{\mu\nu}dx^{\mu}\in \Omega^{\,1}_{\,1}\left(P\right)$ denotes a set of Lagrange multiplier $(1\,;1)$-forms enforcing the primary constraint $(n-1)$-forms \eqref{PrimConst}.  We will refer to 
$\tilde{H}_\mathsmaller{\mathrm{DW}}$ as the total De Donder-Weyl Hamiltonian. Consequently, to determine if the set of constraint $(n-1)$-forms of the theory is complete, we need to impose the consistency conditions \eqref{CCD} on each of the primary constraint $(n-1)$-forms of the model. In this regard, it is not difficult to see that the consistency condition applied to the primary constraint $(n-1)$-form \eqref{PrimConst1} allows us to obtain the following relation
\begin{equation}\label{Consistency Con 1}
\epsilon^{\nu\sigma\rho}\left( \epsilon_{abc}\left(\frac{\,s\sqrt{|\Lambda|}}{\gamma}a^{b}_{\sigma}b^{c}_{\rho}+\frac{1}{2}\Big( b^{b}_{\sigma}b^{c}_{\rho}+\Lambda\, a^{b}_{\sigma}a^{c}_{\rho}\Big)\right)
+\delta_{ab}\left( \frac{s\sqrt{|\Lambda|}}{\gamma}\,{\lambda_{(1)}}^{b}_{\sigma\rho} +{\lambda_{(2)}}^{b}_{\sigma\rho}\right)\right)\approx 0\, ,
\end{equation}
while a straightforward calculation shows that the consistency condition associated with the primary constraint $(n-1)$-form \eqref{PrimConst2} gives rise to the following identity
\begin{equation}\label{Consistency Con 2}
\epsilon^{\nu\sigma\rho}\left( \epsilon_{abc}\left(a^{b}_{\sigma}b^{c}_{\rho}+\frac{1}{2\gamma\sqrt{|\Lambda|}}\Big(b^{b}_{\sigma}b^{c}_{\rho}+\Lambda\, a^{b}_{\sigma}a^{c}_{\rho}\Big)\right)+\delta_{ab}\left({\lambda_{(1)}}^{b}_{\sigma\rho} + \frac{1}{\gamma\sqrt{|\Lambda|}}\,
{\lambda_{(2)}}^{b}_{\sigma\rho}\right)\right)\approx 0\, .
\end{equation}
Observe that, relations \eqref{Consistency Con 1} and \eqref{Consistency Con 2} impose restrictions on the Lagrange multiplier $(1\,;1)$-forms, which implies that there are no secondary constraints, and therefore the set of constraint $(n-1)$-forms \eqref{PrimConst} that characterizes the Bonzom-Livine model for gravity within the polysymplectic approach is complete. Indeed, for $\gamma^{2}\neq s$, the consistency conditions allow us to explicitly fix the components of the Lagrange multiplier $(1\,;1)$-forms, namely
\begin{equation}
\begin{aligned}[b]
{\lambda_{(1)}}^{a}_{[\mu\nu]}&=-{\epsilon^{a}}_{bc}\,b^{b}_{[\mu}a^{c}_{\nu]}\, ,\\
{\lambda_{(2)}}^{a}_{[\mu\nu]}&=-\frac{1}{2}{\epsilon^{a}}_{bc}\left(b^{b}_{\mu}b^{c}_{\nu}+\Lambda\,a^{b}_{\mu}a^{c}_{\nu}\right)\, .
\end{aligned}
\end{equation}

Now, we are in the position to classify the set of constraint $(n-1)$-forms of the theory. For this purpose, we have to compute the commutation relations of the constraint $(n-1)$-forms \eqref{PrimConst} under the Poisson-Gerstenhaber bracket, that is, the $(n-1\,;1)$-form valued matrix ${C^{(i\,,\,j)}}^{\mu\nu}_{ab}:=\{\![\,{C^{(i)}}^{\mu}_{a},{C ^{(j)}}^{\nu}_{b}\,]\!\}$, which explicitly reads
\begin{equation}\label{Matrix of Constraints}
{C^{(i\,,\,j)}}^{\mu\nu}_{ab}=2\left(\begin{array}{cc}
\displaystyle\frac{s\,\sqrt{|\Lambda|}}{\gamma} & 1\\
1& \displaystyle \frac{1}{\gamma\,\sqrt{|\Lambda|}}
\end{array}\right) 
\delta_{ab}\,\epsilon^{\mu\nu\sigma}\,\varpi_{\sigma}\, .
\end{equation} 
Then, since the above $(n-1\,;1)$-form valued matrix does not vanish on the constraint surface $P^{\mathrm{PCS}}_{\,\mathrm{BL}}$, it is clear that, according to definition \eqref{FCD}, the set of constraint $(n-1)$-forms \eqref{PrimConst} is not only complete but it is also of second-class, in Dirac's terminology.

Besides, it is worth noting that we may induce the covariant momentum map associated with the Bonzom-Livine model for gravity to the polymomenta phase-space. To illustrate this, let us consider the splitting $\mathfrak{g}=\mathfrak{h}\oplus \mathfrak{p}$. Then, given $\xi_{\theta}\in\mathfrak{h}$ and $\xi_{\chi}\in\mathfrak{p}$, it is possible to see that, by taking the pullback of the local representation \eqref{BL Covariant Momentum Maps} with the De Donder-Weyl Hamiltonian section $h_{\mathsmaller{\mathrm{DW}}}\in\mathscr{Z}^{\star}_{P}$, we can obtain the local representation of the covariant momentum map on the polymomenta phase-space, namely $J^{(\mathcal{P})}(\xi):=h_{\mathsmaller{\mathrm{DW}}}^{*}J^{(\mathcal{M})}(\xi)\in\Omega^{\,n-1}(P)$ for an arbitrary $\xi\in\mathfrak{g}$, which in terms of the horizontal $(n-1;\,1)$-forms on $P$ can be written as
\begin{equation}\label{BL Poly-Covariant momentum maps}
\begin{aligned}[b]
J^{(\mathcal{P})}\!\left(\xi_{\theta}\right):=&\Bigg[ D_{\nu}\theta^{a}\bar{p}^{\,\mu\nu}_{a}+{\epsilon^{a}}_{bc}\,a^{b}_{\nu}\,\theta^{c}p^{\,\mu\nu}_{a}-\frac{1}{\gamma\,\sqrt{|\Lambda|}}\epsilon^{\mu\nu\sigma}\,\delta_{ab}\,\partial_{\sigma}\theta^{a}\,b^{b}_{\nu}\Bigg] \varpi_{\mu}\, ,\\
J^{(\mathcal{P})}\!\left(\xi_{\chi}\right):=&\Bigg[D_{\nu}\chi^{a}p^{\,\mu\nu}_{a}+\Lambda\,{\epsilon^{a}}_{bc}\,a^{b}_{\nu}\,\chi^{c}\bar{p}^{\,\mu\nu}_{a}\\
&~-\epsilon^{\mu\nu\sigma}\!\Bigg(\epsilon_{abc}\,\chi^{a}\Big(b^{b}_{\nu}b^{c}_{\sigma}-\Lambda\,a^{b}_{\nu}a^{c}_{\sigma}\Big)+\delta_{ab}\,\partial_{\sigma}\chi^{a}\Bigg(\frac{s\,\sqrt{|\Lambda|}}{\gamma}\,a^{b}_{\nu}+2\,b^{b}_{\nu}\Bigg)\Bigg)\Bigg]\varpi_{\mu}\, .
\end{aligned}
\end{equation}
Observe that, for all $\xi_{\theta}\in\mathfrak{h}$ and $\xi_{\chi}\in\mathfrak{p}$, the set of horizontal $(n-1;\,1)$-forms \eqref{BL Poly-Covariant momentum maps} corresponds to a set of Hamiltonian $(n-1)$-forms on $P$. In other words, we have that, for each horizontal $(n-1;\,1)$-form of the set \eqref{BL Poly-Covariant momentum maps} there exists a vertical $1$-multivector field on $P$ satisfying condition \eqref{HMFD}. From our point of view, the latter is particularly relevant since this will allow us to study the induced covariant momentum map \eqref{BL Poly-Covariant momentum maps} by means of the Poisson-Gerstenhaber bracket inherent to the polysymplectic formalism. In particular, a straightforward calculation shows that, for all $\xi_{\theta}\in\mathfrak{h}$ and $\xi_{\chi}\in\mathfrak{p}$, the induced covariant momentum map \eqref{BL Poly-Covariant momentum maps} satisfies the following commutation relations
\begin{equation}
\begin{aligned}[b]
\{\![{C^{(1)}}^{\nu}_{a},J^{(\mathcal{P})}(\xi_{\theta})]\!\}&={\epsilon_{ab}}^{c}\theta^{b}{C^{(1)}}^{\nu}_{c}\, ,\\
\{\![{C^{(2)}}^{\nu}_{a},J^{(\mathcal{P})}(\xi_{\theta})]\!\}&={\epsilon_{ab}}^{c}\theta^{b}{C^{(2)}}^{\nu}_{c}\, ,\\
\{\![{C^{(1)}}^{\nu}_{a},J^{(\mathcal{P})}(\xi_{\chi})]\!\}&=\Lambda\,{\epsilon_{ab}}^{c}\chi^{b}{C^{(2)}}^{\nu}_{c}\, ,\\
\{\![{C^{(2)}}^{\nu}_{a},J^{(\mathcal{P})}(\xi_{\chi})]\!\}&={\epsilon_{ab}}^{c}\chi^{b}{C^{(1)}}^{\nu}_{c}\, ,
\end{aligned}
\end{equation}
which, in light of definition \eqref{FCD}, implies that the local representation of the induced covariant momentum map \eqref{BL Poly-Covariant momentum maps} is characterized by a set of first-class Hamiltonian $(n-1)$-forms on $P$.  In addition, we have that, for all $\xi_{\theta}, \xi_{\tilde{\theta}}\in\mathfrak{h}$ and $\xi_{\chi}, \xi_{\tilde{\chi}}\in\mathfrak{p}$, the following algebraic structure holds
\begin{equation}
\begin{aligned}[b]
\{\![J^{(\mathcal{P})}(\xi_{\theta}), J^{(\mathcal{P})}(\xi_{\tilde{\theta}})]\!\}&=J^{(\mathcal{P})}\big([\xi_{\theta},\xi_{\tilde{\theta}}]\big)-\frac{1}{\gamma\,\sqrt{|\Lambda|}}\,d\,\big(\epsilon^{\mu\nu\sigma}\delta_{ab}\,\theta^{a}\partial_{\mu}\tilde{\theta}^{b}\varpi_{\nu\sigma}\big)
\, ,\\
\{\![J^{(\mathcal{P})}(\xi_{\theta}), J^{(\mathcal{P})}(\xi_{\chi})]\!\}&=J^{(\mathcal{P})}\big([\xi_{\theta},\xi_{\chi}]\big)-d\,\big(\epsilon^{\mu\nu\sigma}\delta_{ab}\,\theta^{a}\partial_{\mu}\chi^{b}\varpi_{\nu\sigma}\big)\, ,\\
\{\![J^{(\mathcal{P})}(\xi_{\chi}), J^{(\mathcal{P})}(\xi_{\tilde{\chi}})]\!\}&=J^{(\mathcal{P})}([\xi_{\chi}, \xi_{\tilde{\chi}}])-\frac{s\sqrt{|\Lambda|}}{\gamma}\,d\,\big(\epsilon^{\mu\nu\sigma}\delta_{ab}\,\chi^{a}\partial_{\mu}\tilde{\chi}^{b}\varpi_{\nu\sigma}\big)\, .
\end{aligned}
\end{equation}
Therefore, it is not difficult to see that, under the Poisson-Gerstenhaber bracket and modulo exact forms, the induced covariant momentum map \eqref{BL Poly-Covariant momentum maps} reproduces the Lie algebra \eqref{BL Lie algebra}. 

Bearing this in mind, our main aim now is to construct the Dirac-Poisson bracket \eqref{DPGBracket} for the Bonzom-Livine model for gravity.  As we will see below, this will allow us to eliminate the second-class constraint $(n-1)$-forms \eqref{PrimConst}, and eventually also to obtain the correct De Donder-Weyl-Hamilton field equations of the theory. To this end, let us consider $F$ a Hamiltonian $0$- or $(n-1)$-form and $G$ a Hamiltonian $(n-1)$-form. Then, in our case, the Dirac-Poisson bracket is defined by
\begin{equation}\label{BL Dirac Bracket}
\{\![\,F,G\,]\!\}_\mathrm{D}:=\{\![\,F,G\,]\!\}- \{\![\,F,{C^{(i)}}^{\mu}_{a}\,]\!\}\bullet\left({{C^{-1}}_{(i\,,\,j)}}^{ab}_{\mu\nu}\,\wedge \{\![{C ^{(j)}}^{\nu}_{b},G\,]\!\} \right)\, ,
\end{equation}  
where ${{C^{-1}}_{(i\,,\,j)}}^{ab}_{\mu\nu}$ stands for the $(1\,;1)$-form valued matrix satisfying the condition
\begin{equation}
{{C^{-1}}_{(i\,,\,k)}}^{ac}_{\mu\lambda}\,\wedge\,{C^{(k\,,\,j)}}^{\lambda\nu}_{cb}=\delta^{j}_{i}\,\delta^{a}_{b}\,\delta^{\nu}_{\mu}\varpi\, .
\end{equation}
Of courser, the Latin indices $i$, $j$ and $k$ run over the complete set of constraint $(n-1)$-forms \eqref{PrimConst}, while the $(n-1\,;1)$-form valued matrix ${C^{(i,j)}}^{\mu\nu}_{ab}$ is given by \eqref{Matrix of Constraints}. Thus, by direct calculation, it is possible to show that, the $(1\,;1)$-form valued matrix ${{C^{-1}}_{(i\,,\,j)}}^{ab}_{\mu\nu}$ explicitly reads
\begin{equation}
{{C^{-1}}_{(i\,,\,j)}}^{ab}_{\mu\nu}:=-\frac{\gamma^{2}}{4\left(s-\gamma^{2}\right)} \left(\begin{array}{cc}
\displaystyle\frac{1}{\gamma\sqrt{|\Lambda|}} & -1\\
-1& \displaystyle\frac{s\sqrt{|\Lambda|}}{\gamma}
\end{array}\right) 
\delta^{ab}\,\epsilon_{\mu\nu\sigma}\,dx^{\sigma}\, .
\end{equation}

Now, once we have constructed the Dirac-Poisson bracket associated with the system, we are in the position to study the Bonzom-Livine model for gravity on the constraint surface $P^{\mathrm{PCS}}_{\mathrm{BL}}$. For this purpose, we start by calculating the commutation relations of the canonically conjugate variables of the theory \eqref{BL Conjugate Variables} under bracket structure \eqref{BL Dirac Bracket}. In this regard, it is possible to write
\begin{equation}
\begin{aligned}[b]
\{\![\,p^{\,\mu\nu}_{a}\varpi_{\mu},a^{b}_{[\sigma}\varpi_{\rho]}\,]\!\}_\mathrm{D}&=\frac{1}{2}\frac{s}{(s-\gamma^{2})}\delta^{b}_{a}\delta^{\nu}_{[\sigma}\varpi_{\rho]}\, ,\\
\{\![\,\bar{p}^{\,\mu\nu}_{a}\varpi_{\mu},b^{b}_{[\sigma}\varpi_{\rho]}\,]\!\}_\mathrm{D}&=\frac{1}{2}\frac{s-2\gamma^{2}}{(s-\gamma^{2})}\delta^{b}_{a}\delta^{\nu}_{[\sigma}\varpi_{\rho]}\, .
\end{aligned}
\end{equation}
Furthermore, a straightforward computation shows that, the commutation relations among the field variable $(n-1)$-forms of the model are given by
\begin{equation}
\begin{aligned}[b]
\{\![a^{a}_{[\nu}\varpi_{\mu]},a^{b}_{[\sigma}\varpi_{\rho]}]\!\}_\mathrm{D}&=-\frac{1}{4\sqrt{|\Lambda|}}\frac{\gamma}{\left(s-\gamma^{2}\right)}\delta ^{ab}\epsilon_{[\nu|\sigma\rho}\varpi_{\mu]}\, ,\\
\{\![\,b^{a}_{[\nu}\varpi_{\mu]},b^{b}_{[\sigma}\varpi_{\rho]}\,]\!\}_\mathrm{D}&=-\frac{s\sqrt{|\Lambda|}}{4}\frac{\gamma}{\left(s-\gamma^{2}\right)}\delta ^{ab}\epsilon_{[\nu|\sigma\rho}\varpi_{\mu]}\, ,\\
\{\![\,b^{a}_{[\nu}\varpi_{\mu]},a^{b}_{[\sigma}\varpi_{\rho]}\,]\!\}_\mathrm{D}&=\frac{1}{4}\frac{\gamma ^{2}}{\left(s-\gamma^{2}\right)}\delta ^{ab}\epsilon_{[\nu|\sigma\rho}\varpi_{\mu]}\, .
\end{aligned}
\end{equation}
In a similar manner, we have that, the commutation relations among the polymomenta variable $(n-1)$-forms of the system read
\begin{equation}
\begin{aligned}[b]
\{\![\,p^{\,\mu\nu}_{a}\varpi_{\mu},p^{\,\rho\sigma}_{b} \varpi_{\rho}\,]\!\}_\mathrm{D}&=-\frac{~s^{2}}{2\,\gamma}\frac{\sqrt{|\Lambda|}}{\left(s-\gamma^{2}\right)}\delta _{ab}\epsilon^{\nu\sigma\rho}\varpi_{\rho}\, ,\\
\{\![\,p^{\,\mu\nu}_{a}\varpi_{\mu},\bar{p}^{\,\rho\sigma}_{b}\varpi_{\rho}\,]\!\}_\mathrm{D}&=-\frac{1}{2}\frac{s}{\left(s-\gamma^{2}\right)}\delta _{ab}\epsilon^{\nu\sigma\rho}\varpi_{\rho}\, ,\\
\{\![\,\bar{p}^{\,\mu\nu}_{a}\varpi_{\mu},\bar{p}^{\,\rho\sigma}_{b}\varpi_{\rho}\,]\!\}_\mathrm{D}&=-\frac{1}{2\,\gamma\sqrt{|\Lambda|}}\frac{s}{\left(s-\gamma^{2}\right)}\delta _{ab}\epsilon^{\nu\sigma\rho}\varpi_{\rho}\, .
\end{aligned}
\end{equation}
Note that, all the above commutation relations explicitly depend on the parameter $\gamma$, which implies that, after introducing the Dirac-Poisson bracket \eqref{BL Dirac Bracket} within the polysymplectic formulation, the Immirzi-like parameter inherent to the Bonzom-Livine model for gravity \eqref{BLmodel} not only modifies the fundamental commutation relations of the theory but it also alters its canonical polysymplectic structure to a non-commutative one.  In fact, this last issue stands as the polysymplectic counterpart to the analogous result obtained through the canonical analysis of the system developed in \cite{Bonzom}.

Next, we will study the De Donder-Weyl-Hamilton field equations associated with the system. To do so, we start by emphasizing that, on the constraint surface $P^\mathrm{PCS}_\mathrm{\,BL}$, the correct field equations of the gravity model \eqref{BLmodel} can be obtained in a covariant Poisson-Hamiltonian framework by means of the set of canonically conjugate  variables \eqref{BL Conjugate Variables}, the De Donder-Weyl Hamiltonian \eqref{BL Canonic Hamiltonian} and the Dirac-Poisson bracket \eqref{BL Dirac Bracket}. In other words, we have that, given $\varrho\in\mathscr{P}_{X}$ a section of $\fibbun{\,P}{X}$, the De Donder-Weyl-Hamilton field equations of the Bonzom-Livine model for gravity are given by
\begin{subequations}\label{BL DWH-equations}
\begin{align}
\partial _{[\mu}e^{a}_{\nu]}\,&=\varrho^{*}\{\![{H}_\mathsmaller{\mathrm{DW}}, a^{a}_{[\nu}\varpi_{\mu]}]\!\}_\mathrm{D}=-{\epsilon^{a}}_{bc}\,\omega^{b}_{[\mu}e^{c}_{\nu]}\, ,\label{DWHequa:a} \\
\partial_{\mu}\pi^{\mu\nu}_{a}&=\varrho^{*}\{\![{H}_\mathsmaller{\mathrm{DW}}, p^{\,\mu\nu}_{a}\varpi_{\mu}]\!\}_\mathrm{D}=\frac{s\sqrt{|\Lambda|}}{\gamma}\epsilon_{abc} \epsilon^{\nu\sigma\rho}\omega ^{b}_{\sigma}e ^{c}_{\rho}\, ,\label{DWHequa:b} \\
\partial_{[\mu}\omega^{a}_{\nu]}&=\varrho^{*}\{\![{H}_\mathsmaller{\mathrm{DW}}, b^{a}_{[\nu}\varpi_{\mu]}]\!\}_\mathrm{D}=-\frac{1}{2}{\epsilon^{a}}_{bc}\left(\omega^{b}_{\mu}\omega^{c}_{\nu}+\Lambda\,e^{b}_{\mu}e^{c}_{\nu}\right)\, ,\label{DWHequa:c} \\
\partial_{\mu}\bar{\pi}^{\mu\nu}_{a}&=\varrho^{*}\{\![{H}_\mathsmaller{\mathrm{DW}}, \bar{p}^{\,\mu\nu}_{a}\varpi_{\mu}]\!\}_\mathrm{D}=\epsilon_{abc}\epsilon^{\nu\sigma\rho}\left(2\,e^{b}_{\sigma}\omega^{c}_{\rho}+\frac{1}{2\gamma}\left(\frac{1}{\sqrt{|\Lambda|}}\omega^{b}_{\sigma}\omega^{c}_{\rho}+s\sqrt{|\Lambda|}e^{b}_{\sigma}e^{c}_{\rho}\right) \right)\, .\label{DWHequa:d} 
\end{align}
\end{subequations}
In particular, it is not difficult to see that, equation \eqref{DWHequa:a} can be explicitly written as 
\begin{equation}\label{Compatibility}
D_{[\mu}e^{a}_{\nu]}=0\, ,
\end{equation}
giving rise to the vanishing torsion condition. In a similar manner, equation \eqref{DWHequa:c} allows us to obtain
\begin{equation}\label{Einstein equations}
F^{a}_{\mu\nu}+\Lambda\,{\epsilon^{a}}_{bc}\,e^{b}_{\mu}e^{c}_{\nu}=0\,,
\end{equation}
which implies that the curvature is constant and, particularly,  given by the cosmological constant. Of course, relations~\eqref{Compatibility} and~\eqref{Einstein equations} are nothing but the Einstein equations, which describe the gravitational field on a $3$-dimensional space-time manifold. A straightforward calculation also shows that,  in light of relations \eqref{Compatibility} and \eqref{Einstein equations}, the De Donder-Weyl-Hamilton field equations \eqref{DWHequa:b} and \eqref{DWHequa:d} are trivially satisfied. Furthermore, observe that both the vanishing torsion condition \eqref{Compatibility} and the Einstein equations \eqref{Einstein equations} were derived in the geometric-covariant Lagrangian analysis of the Bonzom-Livine model for gravity, thus demonstrating the equivalence of the Lagrangian and De Donder-Weyl-Hamilton field equations.

Finally, we will analyze how the induced covariant momentum map \eqref{BL Poly-Covariant momentum maps} allows us to construct conserved currents for the solutions of the field equations of the model within the De Donder-Weyl Hamiltonian formulation. To this end, let us consider $\xi_{\chi}\in\mathfrak{h}$ and $\xi_{\chi}\in\mathfrak{p}$. Then, on the constraint surface $P^{\mathrm{PCS}}_{\,\mathrm{BL}}$ and given $\varrho\in\mathscr{P}_{X}$ a solution of the De Donder-Weyl-Hamilton field equations \eqref{BL DWH-equations}, it is possible to write
\begin{equation}
\begin{aligned}[b]
\mathbf{d}\bullet J^{(\mathcal{P})}(\xi_{\theta})&=\varrho^{*}\{\![ H_\mathsmaller{\mathrm{DW}},  J^{(\mathcal{P})}(\xi_{\theta})]\!\}_\mathrm{D}+\mathbf{d}^{H}\bullet  J^{(\mathcal{P})}(\xi_{\theta})\, ,\\
&=\varrho^{*}\left[\epsilon^{\mu\nu\sigma}\epsilon_{a[bc}\,{\epsilon^{a}}_{d]e}\left(\frac{s\sqrt{|\Lambda|}}{\gamma}a^{b}_{\mu}a^{c}_{\nu}b^{d}_{\sigma}+b^{b}_{\mu}b^{c}_{\nu}a^{d}_{\sigma}+\Lambda\,a^{b}_{\mu}a^{c}_{\nu}a^{d}_{\sigma}+\frac{1}{\gamma\sqrt{|\Lambda|}}b^{b}_{\mu}b^{c}_{\nu}b^{d}_{\sigma}\right)\theta^{e}\right]\,,\\
&=0\, ,
\end{aligned}
\end{equation}
where we have implemented the Jacobi identity for the structure constants of the Lie algebra \eqref{BL Lie algebra}. In a similar manner, the following relation holds
\begin{equation}
\begin{aligned}[b]
\mathbf{d}\bullet J^{(\mathcal{P})}(\xi_{\chi})&=\varrho^{*}\{\![ H_\mathsmaller{\mathrm{DW}},  J^{(\mathcal{P})}(\xi_{\chi})]\!\}_\mathrm{D}+\mathbf{d}^{H}\bullet J^{(\mathcal{P})}(\xi_{\chi})\, ,\\
&=\varrho^{*}\left[\epsilon^{\mu\nu\sigma}\epsilon_{a[bc}\,{\epsilon^{a}}_{d]e}\Bigg(\frac{s\sqrt{|\Lambda|}}{\gamma}\left(b^{b}_{\mu}b^{c}_{\nu}a^{d}_{\sigma}+\Lambda\, a^{b}_{\mu}a^{c}_{\nu}a^{d}_{\sigma}\right)+\Lambda\,a^{b}_{\mu}a^{c}_{\nu}b^{d}_{\sigma}+b^{b}_{\mu}b^{c}_{\nu}b^{d}_{\sigma}\Bigg) \chi^{e}\right]\,,\\
&=0\, ,
\end{aligned}
\end{equation}
which implies that, on the constraint surface $P^{\mathrm{PCS}}_{\,\mathrm{BL}}$, the Hamiltonian conserved currents \eqref{Hamiltonian Noether currents} of the Bonzom-Livine model for gravity can be obtained simply by pulling-back the induced covariant momentum map \eqref{BL Poly-Covariant momentum maps} with $\varrho\in\mathscr{P}_{X}$ 
being a solution of the De Donder-Weyl-Hamilton field equations \eqref{BL DWH-equations}.

Next, we will derive the instantaneous Dirac-Hamiltonian analysis of the gravity model \eqref{BLmodel} by performing the space plus time decomposition of the corresponding multisymplectic formulation.

\subsection{The Bonzom-Livine model for gravity in the space of Cauchy data}

In the present subsection, we will carry out the space plus time decomposition for the Bonzom-Livine model for gravity \eqref{BLmodel}. Here, our main aim is to describe how both the first-class constraints and the extended Hamiltonian of the theory can be obtained by means of certain geometric objects defined on the covariant multimomenta phase-space associated with the system. To this end, we will implement the elementary objects defined in subsection \ref{STDCFT} to study classical field theory within the so-called infinite-dimensional formulations. 

To start, let us consider $\Sigma_{t}$ a Cauchy surface characterized by a level set of the local coordinate $x^{0}$, specifically $x^{0}-t=0$, for some $t\in\mathbb{R}$. Subsequently, we identify $\zeta^{X}:=\partial_{0}\in \mathfrak{X}^{\,1}(X)$ as the infinitesimal generator of the slicing of the space-time manifold $X$. Thus, for some 
$\xi_{\theta}\in\mathfrak{h}$ and $\xi_{\chi}\in\mathfrak{p}$, we introduce $(\Sigma_{t},\zeta^{X})$ and $(Y_{t},\zeta^{Y})$ to be a $\mathcal{G}$-slicing of the covariant configuration space $\fibbun{\,Y}{X}$, where the vector field $\zeta^{Y}\in\mathfrak{X}^{\,1}(Y)$ is locally defined as 
\begin{equation}\label{Slicing of the configuration space}
\zeta^{Y}:=\partial_{0}+\xi^{Y}_{\theta}+\xi^{Y}_{\chi}\, ,
\end{equation}
being $\xi^{Y}_{\theta}\in\mathfrak{X}^{\,1}(Y)$ and $\xi ^{Y}_{\chi}\in\mathfrak{X}^{\,1}(Y)$ the vector fields locally represented by \eqref{Infinitesimal Generators}. From now on, $\zeta^{Y}$ will be understood as a temporal direction on $Y$ for the Bonzom-Livine model for gravity. Consequently, we identify $(Y_{t},\fibbun{\,\,Y_{t}}{\,\Sigma_{t}}, \Sigma_{t})$ as the restriction of $\fibbun{\,Y}{X}$ to the Cauchy surface $\Sigma_{t}$, and we denote by $(x^{i},a^{a}_{\mu},b^{a}_{\nu})$ an adapted coordinate system on $Y_{t}$. Thus, given $\mathscr{Y}_{t}$ the set of sections of $\fibbun{\,\,Y_{t}}{\,\Sigma_{t}}$, we define $T\,\mathscr{Y}_{t}$ as the $t$-instantaneous space of velocities of the theory. Here, by considering $\varphi:=\phi\circ i_{t}\in\mathscr{Y}_{t}$ for some $\phi\in\mathscr{Y}_{X}$, we introduce $(e^{a}_{\mu}, \omega^{a}_{\mu},\dot{e}^{a}_{\mu},\dot{\omega}^{a}_{\mu})$ to denote an adapted coordinate system on $T\,\mathscr{Y}_{t}$, where the temporal derivative of the field variables of the model, according to relation \eqref{TEFV}, is explicitly given by
\begin{equation}\label{BL Temporal derivatives}
\begin{aligned}[b]
&\dot{e}^{a}_{\mu}:=\partial_{0}e^{a}_{\mu}-{\epsilon^{a}}_{bc}\,e^{b}_{\mu}\,\theta^{c}-D_{\mu}\chi^{a}\, ,\\
&\dot{\omega}^{a}_{\mu}:=\partial_{0}\omega^{a}_{\mu}-D_{\mu}\theta^{a}-\Lambda\,{\epsilon^{a}}_{bc}\,e^{b}_{\mu}\,\chi^{c}\, .
\end{aligned}
\end{equation}

Bearing this in mind, we are in the position to perform the space plus time decomposition  for the gravity model \eqref{BLmodel} at the Lagrangian level. For this purpose, we begin by introducing $((J^{1}Y)_{t}, \fibbun{\,\,(J^{1}Y)_{t}}{Y_{t}},\,Y_{t})$ to be the restriction of $\fibbun{\,J^{1}Y}{Y}$ to $Y_{t}$. Then, the jet decomposition map $\beta_{\zeta^{Y}}:(J^{1}Y)_{t}\rightarrow J^{1}(Y_{t})\times VY_{t}$ over $Y_{t}$ locally reads 
\begin{equation}\label{BL decomposition map}
\beta_{\zeta_{Y}}\left(x^{i},a^{a}_{\nu}, b^{a}_{\nu}, a^{a}_{\mu\nu},b^{a}_{\mu\nu}\right):=\left(x^{i},a^{a}_{\nu}, b^{a}_{\nu}, a^{a}_{i\nu},b^{a}_{i\nu}, \dot{a}^{a}_{\nu},\dot{b}^{a}_{\nu}\right)\,.
\end{equation}
Hereinafter, the Latin indices $i$ and $j$ will denote spatial indices ($i,j=1,2$). Hence, by using relations \eqref{ILCFT}, \eqref{BL Temporal derivatives} and \eqref{BL decomposition map}, it is not difficult to see that, the instantaneous Lagrangian functional of the theory, $L^{\mathrm{BL}}_{\,t,\,\zeta^{Y}}:T\,\mathscr{Y}_{t}\rightarrow \mathbb{R}$, can be written as
\begin{equation}
\begin{aligned}\label{BL Instantaneous Lagrangian}
&\,L^{\mathrm{BL}}_{\,t,\,\zeta^{Y}}(e,\omega,\dot{e},\dot{\omega}):=\int_{\Sigma_{t}}\!d^{\,n-1}x_{0}\,\epsilon^{0ij}\,\delta_{ab}\Bigg\lbrace s\frac{\sqrt{|\Lambda|}}{\gamma}\,e^{b}_{j}\Big(\dot{e}^{a}_{i}+{\epsilon^{a}}_{c\;\!d}\;\!e^{c}_{i}\;\!\theta^{d}+D_{i}\chi^{a}\Big)\\
&+\left(2\,e^{b}_{j}+\frac{1}{\gamma\,\sqrt{|\Lambda|}}\,\omega^{b}_{j}\right)\Big(\dot{\omega}^{a}_{i}+\Lambda\;\!{\epsilon^{a}}_{c\;\!d}\;\!e^{c}_{i}\;\!\chi^{d}+D_{i}\theta^{a}\Big)+2\,e^{a}_{0}\left(\frac{s\,\sqrt{|\Lambda|}}{\gamma}\,D_{i}e^{b}_{j}+\frac{1}{2}\Big(F^{b}_{ij}+\Lambda\;\!{\epsilon^{b}}_{c\;\!d}\;\!e^{c}_{i}\;\!e^{d}_{j}\Big)\right)\\
&+2\,\omega^{a}_{0}\left(D_{i}e^{b}_{j} +\frac{1}{2\,\gamma\,\sqrt{|\Lambda|}}\Big(F^{b}_{ij}+\Lambda\;\!{\epsilon^{b}}_{c\;\!d}\;\!e^{c}_{i}\;\!e^{d}_{j}\Big)\right)\Bigg\rbrace \, ,
\end{aligned}
\end{equation}
where we have performed some integration by parts and avoided terms on the boundary of the Cauchy surface $\Sigma_{t}$.

In particular, a straightforward calculation shows that, in light of the instantaneous Legendre transformation \eqref{Insantaneous Leg Transformation}, the instantaneous momenta variables of the system are given by
\begin{equation}\label{BL instantaneous momenta variables}
\begin{aligned}[b]
\pi^{\mu}_{a}:=&\frac{\partial L_{\mathrm{BL}}}{\partial \dot{e}^{a}_{\mu}}\big(e,\omega,\dot{e},\dot{\omega}\big)=\frac{s\,\sqrt{|\Lambda|}}{\gamma}\delta^{\mu}_{i}\epsilon^{0ij}\;\!\delta_{ab}\;\!e^{b}_{j}\, ,\\
\bar{\pi}^{\mu}_{a}:=&\frac{\partial L_{\mathrm{BL}}}{\partial \dot{\omega}^{a}_{\mu}}\big(e,\omega,\dot{e},\dot{\omega}\big)= \delta^{\mu}_{i}\epsilon^{0ij}\;\!\delta_{ab}\left( 2\,e^{b}_{j}+\frac{1}{\gamma\,\sqrt{|\Lambda|}}\,\omega^{b}_{j}\right)\, ,
\end{aligned}
\end{equation}
being $L_\mathrm{BL}(e,\omega,\dot{e},\dot{\omega})$ the Lagrangian function of the model after the space plus time decomposition, that is, the integrand in \eqref{BL Instantaneous Lagrangian}. Note that, the set of relations \eqref{BL instantaneous momenta variables} gives rise to the primary constraint surface of the Bonzom-Livine model for gravity $\mathscr{P}_{\,t,\,\zeta^{Y}}\subset T^{*}\mathscr{Y}_{t}$, specifically
\begin{equation}
\begin{aligned}[b]
\mathscr{P}_{\,t,\,\zeta^{Y}}:=&\big\lbrace (e,\omega,\pi,\bar{\pi})\in T^{*}\mathscr{Y}_{t}\,\big|\,\gamma^{0}_{a}=0\,, \Upsilon^{i}_{a}=0\,,  \bar{\gamma}^{0}_{a}=0\,,\bar{\Upsilon}^{i}_{a}=0 \big\rbrace \, ,
\end{aligned}
\end{equation}
where the primary constraints of the theory are explicitly defined by
\begin{subequations}\label{BL instantaneous primary constraints}
\begin{align}
\gamma^{0}_{a}&:=\pi^{0}_{a}\approx 0\,,\\
\Upsilon^{i}_{a}&:=\pi^{i}_{a}-\frac{s\,\sqrt{|\Lambda|}}{\gamma}\epsilon^{0ij}\;\!\delta_{ab}\,e^{b}_{j}\approx 0\,, \label{PSCconst1}\\
\bar{\gamma}^{0}_{a}&:=\bar{\pi}^{0}_{a}\approx 0\, ,\\
\bar{\Upsilon}^{i}_{a}&:=\bar{\pi}^{i}_{a}-\epsilon^{0ij}\;\!\delta_{ab}\left( 2\,e^{b}_{j}+\frac{1}{\gamma\,\sqrt{|\Lambda|}}\,\omega^{b}_{j}\right)\approx0\, .\label{PSCconst2}
\end{align}
\end{subequations}
Of course, as pointed out in \cite{QGS}, the presence of primary constraints in a classical field theory is related with the fact that the instantaneous Legendre transformation of the system is not invertible. Therefore, the gravity model \eqref{BLmodel} corresponds to a singular Lagrangian system. 

Next, we will implement the space plus time decomposition of the theory of our interest at the multisymplectic level. To do so, we start by introducing $(Z^{\star}_{t}, \fibbun{\,\,Z^{\star}_{t}}{Y_{t}},Y_{t})$ to denote the restriction of $\fibbun{\,Z^{\star}}{Y}$ to $Y_{t}$. In addition, we define $\mathscr{Z}^{\star}_{t}$ as the set of sections of $\fibbun{\,\,Z^{\star}_{t}}{\,\Sigma_{t}}:=\fibbun{\,\,Y_{t}}{\Sigma_{t}}\circ \fibbun{\,\,Z^{\star}_{t}}{Y_{t}}$, which is related with $T^{*}\mathscr{Y}_{t}$, the $t$-instantaneous phase-space of the model, through the vector bundle map $R_{t}:\mathscr{Z}^{\star}_{t}\rightarrow T^{*}\mathscr{Y}_{t}$ over $\mathscr{Y}_{t}$, as discussed in subsection \ref{STDCFT}. Besides, we identify $\zeta^{Z^{\star}}\in \mathfrak{X}^{\,1}(Z^{\star})$ as the $\alpha^{(\mathcal{M})}$-lift of \eqref{Slicing of the configuration space} to $Z^{\star}$, namely
\begin{equation}\label{Slicing of the multimomenta space}
\zeta^{Z^{\star}}:=\partial_{0}+\xi^{\alpha}_{\theta}+\xi^{\alpha}_{\chi}\, ,
\end{equation}
where the vector fields $\xi^{\alpha}_{\theta}\in\mathfrak{X}^{\,1}(Z^{\star})$ and $\xi^{\alpha}_{\chi}\in\mathfrak{X}^{\,1}(Z^{\star})$ are locally given by \eqref{BL alpha-lifts}. As we will see below, the vector field \eqref{Slicing of the multimomenta space} will play an important role within our study. 

As previously mentioned, we are interested in describing how the covariant momentum map associated with the extended gauge symmetry group of the Bonzom-Livine model for gravity allows to obtain the complete set of first-class constraints in Dirac's terminology of the system. To illustrate this, it is important to emphasize that, $\mathcal{G}$ (the extended gauge symmetry group of the theory) acts on $\fibbun{\,Y}{X}$ by means of vertical bundle automorphisms, which implies that, the action of $\mathcal{G}$ on $T^{*}\mathscr{Y}_{t}$ is well-defined, and hence there is an associated momentum map, specifically $\mathscr{J}_{t}:T^{*}\mathscr{Y}_{t}\rightarrow \mathfrak{g}^{*}$. Then, in order to obtain the local representation of such a momentum map, we can proceed as follows. To start, we would like to mention that, in light of the splitting $\mathfrak{g}=\mathfrak{h}\oplus\mathfrak{p}$, any element $\xi_{\eta}\in\mathfrak{g}$ can be written as $\xi_{\eta}:=\xi_{\theta}+\xi_{\chi}$, for some $\xi_{\theta}\in\mathfrak{h}$ and $\xi_{\chi}\in\mathfrak{p}$. Now, let us consider $\sigma\in \mathscr{Z}^{*}_{t}$ a section such that $R_{t}(\sigma)=(e,\omega,\pi,\bar{\pi})\in T^{*}\mathscr{Y}_{t}$. Then, by taking into account relation \eqref{PMM}, we know that, the momentum map associated with the action of $\mathcal{G}$ on $T^{*}\mathscr{Y}_{t}$ is locally given by
\begin{equation}\label{BL projected map}
\begin{aligned}
\big\langle \mathscr{J}_{t}(e,\omega,\pi,\bar{\pi}),\xi_{\eta}\big\rangle:=\int_{\Sigma_{t}}\sigma^{*} J^{(\mathcal{M})}\left(\xi_{\eta}\right)\, ,
\end{aligned}
\end{equation}
where $J^{(\mathcal{M})}\left(\xi_{\eta}\right):= J^{(\mathcal{M})}\left(\xi_{\theta}\right)+J^{(\mathcal{M})}\left(\xi_{\chi}\right)\in \Omega^{\,n-1}(Z^{\star})$. In particular, by using the local representation \eqref{BL Covariant Momentum Maps} and after identifying $\pi^{\nu}_{a}=p^{0\nu}_{a}\circ \sigma$ and $\bar{\pi}^{\nu}_{a}=\bar{p}^{\,0\nu}_{a}\circ \sigma$, we find that
\begin{equation}\notag
\begin{aligned}
\big\langle \mathscr{J}_{t}(e,\omega,\pi,\bar{\pi}),\xi_{\eta}\big\rangle&=\int_{\Sigma_{t}}d^{\,n-1}x_{0}\Bigg\lbrace \big(D_{0}\chi^{a}+{\epsilon^{a}}_{b\;\!c}\;\!e^{b}_{0}\;\!\theta^{c}\big)\pi^{0}_{a}+\big(D_{0}\theta^{a}+\Lambda\,{\epsilon^{a}}_{b\;\!c}\;\!e^{b}_{0}\;\!\chi^{c}\big)\bar{\pi}^{0}_{a}\\
&~~~-\theta^{a}\left(D_{i}\bar{\pi}^{i}_{a}+{\epsilon_{ab}}^{c}e^{b}_{i}\;\!\pi^{i}_{c}+\frac{1}{\gamma\,\sqrt{|\Lambda|}}\epsilon^{0ij}\;\!\delta_{ab}\;\!\partial_{i}\omega^{b}_{j}\right)\\
&~~~-\chi^{a}\left(D_{i}\pi^{i}_{a}+\Lambda\,{\epsilon_{ab}}^{c}e^{b}_{i}\;\!\bar{\pi}^{i}_{c}+\epsilon^{0ij}\;\!\delta_{ab}\left(\frac{s\,\sqrt{|\Lambda|}}{\gamma}\partial_{i}e^{b}_{j}+F^{b}_{ij}-\Lambda\,{\epsilon^{b}}_{c\;\!d}\;\!e^{c}_{i}\;\!e^{d}_{j}\right)\right)\!\Bigg\rbrace\, ,
\end{aligned}
\end{equation}
where we have performed some integral by parts and avoided terms on the boundary of the Cauchy surface $\Sigma_{t}$. Observe that, in terms of the primary constraints of the model \eqref{BL instantaneous primary constraints}, the momentum map \eqref{BL projected map}  explicitly reads
\begin{equation}\notag
\begin{aligned}
\!\!\!\!\!\!&\big\langle \mathscr{J}_{t}(e,\omega,\pi,\bar{\pi}),\xi_{\eta}\big\rangle=\int_{\Sigma_{t}}d^{\,n-1}x_{0}\Bigg\lbrace 
\big(D_{0}\chi^{a}+{\epsilon^{a}}_{b\;\!c}\;\!e^{b}_{0}\;\!\theta^{c}\big)\gamma^{0}_{a}+\big(D_{0}\theta^{a}+\Lambda\,{\epsilon^{a}}_{b\;\!c}\;\!e^{b}_{0}\;\!\chi^{c}\big)\bar{\gamma}^{0}_{a}\\
&~~~~~~~~~~ -\theta^{a}\left(D_{i}\bar{\Upsilon}^{i}_{a}+{\epsilon_{ab}}^{c}e^{b}_{i}\;\!\Upsilon^{i}_{c}+2\,\epsilon^{0ij}\;\!\delta_{ab}\left(D_{i}e^{b}_{j}+\frac{1}{2\gamma\sqrt{|\Lambda|}}\Big(F^{b}_{ij}+\Lambda\,{\epsilon^{b}}_{c\;\!d}\;\!e^{c}_{i}\;\!e^{d}_{j}\Big)\right) \right)\\
&~~~~~~~~~~ -\chi^{a}\left(D_{i}\Upsilon^{i}_{a}+\Lambda\,{\epsilon_{ab}}^{c}e^{b}_{i}\;\!\bar{\Upsilon}^{i}_{c}+2\,\epsilon^{0ij}\;\!\delta_{ab}\left(\frac{s\,\sqrt{|\Lambda|}}{\gamma}D_{i}e^{b}_{j}+\frac{1}{2}\Big(F^{b}_{ij}+\Lambda\,{\epsilon^{b}}_{c\;\!d}\;\!e^{c}_{i}\;\!e^{d}_{j}\Big)\right)\right)\!\Bigg\rbrace\, .
\end{aligned}
\end{equation}
Hence, it is not difficult to see that, after introducing the set of parameters 
\begin{equation}\label{BL gauge parameters}
\begin{aligned}[b]
\lambda^{a}_{0}&:=D_{0}\chi^{a}+{\epsilon^{a}}_{b\;\!c}\;\!e^{b}_{0}\;\!\theta^{c}\, ,\\
\lambda^{a}&:=-\chi^{a}\, ,\\
\bar{\lambda}^{\,a}_{\,0}&:=D_{0}\theta^{a}+\Lambda\,{\epsilon^{a}}_{b\;\!c}\;\!e^{b}_{0}\;\!\chi^{c}\, ,\\
\bar{\lambda}^{a}&:=-\theta^{a}\, ,
\end{aligned}
\end{equation}
it is possible to write
\begin{equation}\notag
\big\langle \mathscr{J}_{t}(e,\omega,\pi,\bar{\pi}),\xi_{\eta}\big\rangle=\int_{\Sigma_{t}}d^{\,n-1}x_{0}\Bigg\lbrace \lambda^{a}_{0}\gamma^{0}_{a}+\bar{\lambda}^{a}_{0}\bar{\gamma}^{0}_{a}+\lambda^{a}\Gamma_{a}+\bar{\lambda}^{a}\bar{\Gamma}_{a}\Bigg\rbrace\, ,
\end{equation}
where $\Gamma_{a}$ and $\bar{\Gamma}_{a}$ are defined by
\begin{equation}\label{Gamma Constraints}
\begin{aligned}[b]
\Gamma_{a}&:=D_{i}\Upsilon^{i}_{a}+\Lambda\,{\epsilon_{ab}}^{c}e^{b}_{i}\;\!\bar{\Upsilon}^{i}_{c}+2\,\epsilon^{0ij}\;\!\delta_{ab}\left(\frac{s\,\sqrt{|\Lambda|}}{\gamma}D_{i}e^{b}_{j}+\frac{1}{2}\Big(F^{b}_{ij}+\Lambda\,{\epsilon^{b}}_{c\;\!d}\;\!e^{c}_{i}\;\!e^{d}_{j}\Big)\right)\, ,\\
\bar{\Gamma}_{a}&:=D_{i}\bar{\Upsilon}^{i}_{a}+{\epsilon_{ab}}^{c}e^{b}_{i}\;\!\Upsilon^{i}_{c}+2\,\epsilon^{0ij}\;\!\delta_{ab}\left(D_{i}e^{b}_{j}+\frac{1}{2\gamma\sqrt{|\Lambda|}}\Big(F^{b}_{ij}+\Lambda\,{\epsilon^{b}}_{c\;\!d}\;\!e^{c}_{i}\;\!e^{d}_{j}\Big)\right)\, .
\end{aligned}
\end{equation}
Here, it is worth noting that, the local representation \eqref{BL projected map} exactly coincides with the generator of infinitesimal gauge transformations of the Bonzom-Livine model for gravity obtained by means of Dirac's algorithm in \cite{Escalante}.

Moreover, it is important to remember that the gauge symmetries of the gravity model \eqref{BLmodel} correspond to localizable symmetries, which implies that, according to \cite{GIMMSY1, GIMMSY2, Fischer}, the admissible space of Cauchy data for the evolution equations of the theory is determined by the zero level set of the momentum map \eqref{BL projected map}, specifically
\begin{equation}
{\mathscr{J}_\mathrm{BL}}_{\,t}^{-1}(0):=\big\{\,(e,\omega,\pi,\bar{\pi})\in T^{*}\mathscr{Y}_{t}~\big|~\langle \mathscr{J}_{t}(e,\omega,\pi\,,\bar{\pi}), \xi_{\eta}\rangle=0\,,\, \forall\, \xi_{\eta}\in\mathfrak{g}\,\big\}\, .
\end{equation}
In this regard, we have that, by using the local representation \eqref{BL projected map} and since parameters \eqref{BL gauge parameters} depend on arbitrary functions on the Cauchy surface $\Sigma_{t}$,  it is possible to write
\begin{equation}\notag
{\mathscr{J}_\mathrm{BL}}_{\,t}^{-1}(0)=\Big\lbrace (e,\omega,\pi,\bar{\pi})\in T^{*}\mathscr{Y}_{t}\,\Big|\, \gamma^{0}_{a}=0\,, \bar{\gamma}^{0}_{a}=0\, ,\Gamma_{a}=0\,, \bar{\Gamma}_{a}=0 \Big\rbrace\, ,
\end{equation}
which is nothing but the surface on the $t$-instantaneous phase-space defined by the complete set of first-class constraints that characterizes the Bonzom-Livine model for gravity within the instantaneous Dirac-Hamiltonian formulation as discussed in \cite{Escalante}. In other words, the vanishing of the momentum map \eqref{BL projected map} yields the complete set of first-class constraints of the gravity model \eqref{BLmodel}, namely
\begin{subequations}\label{BL First Class Constraints}
\begin{align}
\gamma^{0}_{a}&\approx 0\, ,\label{PFCconst1}\\
\bar{\gamma}^{0}_{a}&\approx 0\, ,\label{PFCconst2}\\
\Gamma_{a}&\approx 0\,,\label{BL Gamma FCC1}\\
\bar{\Gamma}_{a}&\approx 0\, . \label{BL Gamma FCC2}
\end{align}
\end{subequations}
Therefore, by following Dirac's terminology, the primary constraints \eqref{PSCconst1} and \eqref{PSCconst2}, which do not belong to set \eqref{BL First Class Constraints}, must be second-class constraints. Furthermore, it is not difficult to see that by considering the second-class primary constraints \eqref{PSCconst1} and \eqref{PSCconst2} as strong identities, the first-class constraints \eqref{BL Gamma FCC1} and \eqref{BL Gamma FCC2} reduce to the following weak identities
\begin{equation}
\begin{aligned}[b]
\Phi_{a}&:=2\,\epsilon^{0ij}\;\!\delta_{ab}\left(\frac{s\,\sqrt{|\Lambda|}}{\gamma}D_{i}e^{b}_{j}+\frac{1}{2}\Big(F^{b}_{ij}+\Lambda\,{\epsilon^{b}}_{c\;\!d}\;\!e^{c}_{i}\;\!e^{d}_{j}\Big)\right)\approx0\,,\\ 
\bar{\Phi}_{a}&:=2\,\epsilon^{0ij}\;\!\delta_{ab}\left(D_{i}e^{b}_{j}+\frac{1}{2\gamma\sqrt{|\Lambda|}}\Big(F^{b}_{ij}+\Lambda\,{\epsilon^{b}}_{c\;\!d}\;\!e^{c}_{i}\;\!e^{d}_{j}\Big)\right)\approx 0\,,
\end{aligned}
\end{equation}
which correspond to the secondary constraints of the theory, that is, the constraints that arise within the instantaneous Dirac-Hamiltonian formulation of the system by imposing the consistency conditions to the first-class primary constraints \eqref{PFCconst1} and \eqref{PFCconst2}, respectively \cite{Escalante}. 

With this in mind, we are in the position to recover the extended Hamiltonian of the theory. For this purpose, we start by emphasizing that, for some $\xi_{\theta}\in\mathfrak{h}$ and $\xi_{\chi}\in\mathfrak{p}$, the vector field \eqref{Slicing of the multimomenta space} satisfies the condition ${\mathfrak{L}_{\zeta^{Z^{\star}}}}\Omega^{(\mathcal{M})}_{\mathrm{BL}}=0$, and hence there is an $(n-1)$-form $H^{(\mathcal{M})}_{\zeta^{Z^{\star}}}:=\zeta^{Z^{\star}}\!\!\lrcorner \,\,\Theta^{(\mathcal{M})}_{\mathrm{BL}}-\alpha^{(\mathcal{M})}_{\theta}-\alpha^{(\mathcal{M})}_{\chi}\in\Omega^{\,n-1}(Z^{\star})$ on $Z^{\star}$ such that $\zeta^{Z^{\star}}\!\!\lrcorner\,\, \Omega^{(\mathcal{M})}_{\mathrm{BL}}=d H^{(\mathcal{M})}_{\zeta^{Z^{\star}}}$, being $\alpha^{(\mathcal{M})}_{\theta}\in\Omega^{\,n-1}(Z^{\star})$ and $\alpha^{(\mathcal{M})}_{\chi}\in\Omega^{\,n-1}(Z^{\star})$ the $\fibbun{\,Z^{\star}}{X}$-horizontal $(n-1;\,1)$-forms on $Z^{\star}$ locally represented by \eqref{BL  multisymplectic SU(2)-alpha} and \eqref{BL multisymplectic topological alpha}, respectively. Now, for some section $\phi\in\mathscr{Y}_{X}$, let us consider $\sigma:=\mathbb{F}\mathcal{L}\circ j^{1}\phi\circ i_{t}\in\mathscr{N}_{t}$ the canonical lift of an element $(e,\omega,\pi,\bar{\pi})\in\mathscr{P}_{\,t,\,\zeta^{Y}}$. Then, according to \cite{Gotay1, GIMMSY2, DeLeon2}, the instantaneous Hamiltonian of the system can be obtained by means of \eqref{IHT}, together with relation $ \zeta^{Z^{\star}}\!\!\lrcorner\,\,\Omega^{(\mathcal{M})}=d\,H^{(\mathcal{M})}_{\zeta^{Z^{\star}}}$. It is important to mention that, contrary to the geometric formulation implemented here, the standard procedure for studying relation \eqref{IHT} consists of imposing the primary constraints of the system as strong identities. However, in our case, by following Dirac's algorithm \cite{QGS}, we will treat such constraints as weak identities, which will eventually allow us to obtain the extended Hamiltonian for the gravity model \eqref{BLmodel}. To illustrate this, we begin by introducing the functional $H^{\mathrm{BL}}_{\,t,\,\zeta^{Y}}:\mathscr{P}_{\,t,\,\zeta^{Y}}\rightarrow \mathbb{R}$, which is given by
\begin{equation}\label{BL Extended Hamiltonian}
\begin{aligned}
H^{\mathrm{BL}}_{\,t,\,\zeta^{Y}}(e,\omega,\pi,\bar{\pi}):=-\int_{\Sigma_{t}}\sigma^{*}H^{(\mathcal{M})}_{\zeta^{Z^{\star}}}\, .
\end{aligned}
\end{equation}
Subsequently, a straightforward calculation shows that after identifying $\pi^{\nu}_{a}=p^{0\nu}_{a}\circ \sigma$ and $\bar{\pi}^{\nu}_{a}=\bar{p}^{\,0\nu}_{a}\circ \sigma$, it is possible to write
\begin{equation}\notag
\begin{aligned}
&H^{\mathrm{BL}}_{\,t,\,\zeta^{Y}}(e,\omega,\pi,\bar{\pi})=\int_{\Sigma_{t}}d^{\,n-1}x_{0}\Bigg\lbrace\dot{e}^{a}_{0}\pi^{0}_{a}+\dot{\omega}^{a}_{0}\bar{\pi}^{0}_{a}+\partial_{0}e^{a}_{i}\left(\pi^{i}_{a}-\frac{s\,\sqrt{|\Lambda|}}{\gamma}\epsilon^{0ij}\;\!\delta_{ab}\;\!e^{b}_{j}\right)\\
&+\partial_{0}\omega^{a}_{i}\left(\bar{\pi}^{i}_{a}-\epsilon^{0ij}\;\!\delta_{ab}\left(2\,e^{b}_{j}+\frac{1}{\gamma\,\sqrt{|\Lambda|}}\,\omega^{b}_{j}\right)\right)-2\,\delta_{ab}\,e^{a}_{0}\left(\frac{s\,\sqrt{|\Lambda|}}{\gamma}D_{i}e^{b}_{j}+\frac{1}{2}\left(F^{b}_{ij}+\Lambda\,{\epsilon^{b}}_{c\;\!d}\;\!e^{c}_{i}\;\!e^{d}_{j}\right)\right)\\
&-2\,\delta_{ab}\,\omega^{a}_{0}\left(D_{i}e^{b}_{j}+\frac{1}{2\gamma\sqrt{|\Lambda|}}\left(F^{b}_{ij}+\Lambda\,{\epsilon^{b}}_{c\;\!d}\;\!e^{c}_{i}\;\!e^{d}_{j}\right)\right)
+\theta^{a}\left(D_{i}\bar{\pi}^{i}_{a}+{\epsilon_{ab}}^{c}e^{b}_{i}\;\!\pi^{i}_{c}+\frac{1}{\gamma\,\sqrt{|\Lambda|}}\epsilon^{0ij}\;\!\delta_{ab}\;\!\partial_{i}\omega^{b}_{j}\right)\\
&+\chi^{a}\left(D_{i}\pi^{i}_{a}+\Lambda\,{\epsilon_{ab}}^{c}e^{b}_{i}\;\!\bar{\pi}^{i}_{c}+\delta_{ab}\left(\frac{s\,\sqrt{|\Lambda|}}{\gamma}\partial_{i}e^{b}_{j}+F^{b}_{ij}-\Lambda\,{\epsilon^{b}}_{c\;\!d}\;\!e^{c}_{i}\;\!e^{d}_{j}\right)\right)\!
\Bigg\rbrace\, ,
\end{aligned}
\end{equation} 
where, once again, we have performed some integrations by parts and avoided terms on the boundary of the Cauchy surface $\Sigma_{t}$. Note that, on the one hand, in terms of the primary constraint of the theory \eqref{BL instantaneous primary constraints}, the functional \eqref{BL Extended Hamiltonian} explicitly reads
\begin{equation}\notag
\begin{aligned}
H^{\mathrm{BL}}_{\,t,\,\zeta^{Y}}(e,\omega,\pi,\bar{\pi})&=\int_{\Sigma_{t}}\!d^{\,n-1}x_{0}\Bigg\lbrace\dot{e}^{a}_{0}\gamma^{0}_{a}+\dot{\omega}^{a}_{0}\bar{\gamma}^{0}_{a}+\partial_{0}e^{a}_{i}\Upsilon^{i}_{a}+\partial_{0}\omega^{a}_{i}\bar{\Upsilon}^{i}_{a}+\chi^{a}\Gamma_{a}+\theta^{a}\bar{\Gamma}_{a}\\
&~~~~~~~~~~~~~~~~~~-2\,\delta_{ab}\,e^{a}_{0}\left(\frac{s\,\sqrt{|\Lambda|}}{\gamma}D_{i}e^{b}_{j}+\frac{1}{2}\left(F^{b}_{ij}+\Lambda\,{\epsilon^{b}}_{c\;\!d}\;\!e^{c}_{i}\;\!e^{d}_{j}\right)\right)\\
&~~~~~~~~~~~~~~~~~~-2\,\delta_{ab}\,\omega^{a}_{0}\left(D_{i}e^{b}_{j}+\frac{1}{2\gamma\sqrt{|\Lambda|}}\left(F^{b}_{ij}+\Lambda\,{\epsilon^{b}}_{c\;\!d}\;\!e^{c}_{i}\;\!e^{d}_{j}\right)\right)
\Bigg\rbrace\, .
\end{aligned}
\end{equation} 
On the other hand, we know that, for $\gamma^{2}\neq s$, the field equations of the system \eqref{BL Field Equations} give rise to the following set of relations
\begin{equation}\label{BL fixed Lagrange multipliers}
\begin{aligned}[b]
\partial_{0}e^{a}_{i}&=D_{i}e^{a}_{0}-{\epsilon^{a}}_{b\;\!c}\;\!\omega^{b}_{0}\;\!e^{c}_{i}\, ,\\
\partial_{0}\omega^{a}_{i}&=\partial_{i}\omega^{a}_{0}-{\epsilon^{a}}_{bc}\left(\omega^{b}_{0}\;\!\omega^{c}_{i}+\Lambda\,e^{b}_{0}\;\!e^{c}_{i}\right)\, ,
\end{aligned}
\end{equation} 
which allows us to identify the fixed Lagrange multipliers that enforce the second-class primary constraints \eqref{PSCconst1} and \eqref{PSCconst2} into the functional \eqref{BL Extended Hamiltonian}, respectively. Hence,
by considering $\phi$ a section that satisfies relations \eqref{BL fixed Lagrange multipliers}, we can write
\begin{equation}\notag
H^{\mathrm{BL}}_{\,t,\,\zeta^{Y}}(e,\omega,\pi,\bar{\pi})=\int_{\Sigma_{t}}d^{\,n-1}x_{0}\Big\lbrace\dot{e}^{a}_{0}\gamma^{0}_{a}+\dot{\omega}^{a}_{0}\bar{\gamma}^{0}_{a}+\chi^{a}\Gamma_{a}+\theta^{a}\bar{\Gamma}_{a}-e^{a}_{0}\Gamma_{a}-\omega^{a}_{0}\bar{\Gamma}_{a}\Big\rbrace\approx 0\,,
\end{equation}
where we have again performed some integration by parts and avoided terms on the boundary of the Cauchy surface $\Sigma_{t}$. Thereby, it is not difficult to see that, the functional \eqref{BL Extended Hamiltonian} is nothing but the extended Hamiltonian that characterizes the Bonzom-Livine model for gravity within the instantaneous Dirac-Hamiltonian formulation. Observe that  the extended Hamiltonian of the system is a linear combination of the first-class constraints \eqref{BL First Class Constraints}, which reflects the general covariance of the classical field theory under consideration. In addition, it is important to highlight that under the gauge fixing $e^{a}_{0}\approx 0$ and $\omega^{a}_{0}\approx 0$ and by considering the primary constraints \eqref{BL instantaneous primary constraints} as strong identities, the first-class constraints \eqref{BL First Class Constraints} and the functional \eqref{BL Extended Hamiltonian} reduce to the first-class constraints and the instantaneous Hamiltonian studied in \cite{Bonzom}, respectively, where the instantaneous Dirac-Hamiltonian formulation of the gravity model \eqref{BLmodel} is developed in an even
smaller $t$-instantaneous reduced phase-space, as discussed in \cite{Escalante}. 

Finally, we can compute the number of local degrees of freedom of the system. In this regard, we start by emphasizing that, on the one hand, the number of linear independent canonical variables of the theory is $N_\mathsmaller{\mathrm{CV}}:=2\cdot18$. On the other hand, we know that, the number of linear independent first-class constraints of the system corresponds to $N_\mathsmaller{\mathrm{FCC}}:=4\cdot3$, while the number of linear independent second-class constraints of the model is $N_\mathsmaller{\mathrm{SCC}}:=2\cdot6$. Therefore, according to \cite{QGS}, the  number of local degrees of freedom of the Bonzom-Livine model for gravity corresponds to $N^{\mathrm{BL}}_{\mathsmaller{\mathrm{DF}}}:=N_\mathsmaller{\mathrm{CV}}-2N_\mathsmaller{\mathrm{FCC}}-N_\mathsmaller{\mathrm{SCC}}=0$, thus implying that the gravity model \eqref{BLmodel} is a topological field theory. 

In summary, by carrying out the space plus time decomposition of the Bonzom-Livine model for gravity at the multisymplectic level, it is possible to recover not only the first- and second-class constrained content of the system but also its extended Hamiltonian. Of course, all the results that we have discussed in this subsection are consistent with those obtained within the instantaneous Dirac-Hamiltonian analysis of the theory.

\section{Conclusions}
\label{sec:Conclu}

In this paper, we have reviewed the Lagrangian, multisymplectic and polysymplectic  geometric and covariant formalisms for the analysis of gauge field theories at the classical level. By judiciously identifying the fields with sections of appropriate
fibre-bundles, one is able to implement a classical field theory in a 
finite-dimensional setup.   As discussed above, such geometric formalisms are particularly  transparent in the manner they characterize the symmetries for classical field theory by analyzing the action of the symmetries on the Poincaré-Cartan forms at 
the Lagrangian level or the 
canonical forms at the multisymplectic level. In particular, this allowed us to  formulate Noether's theorems by means of 
the  covariant versions of the momentum map and also to identified the Noether currents and charges 
in a succinct manner.  Important to mention is the fact that at the multisymplectic level the symmetries of the theory were associated with covariant canonical transformations in a straightforwardly way.  Besides, by considering the polymomenta phase-space,
we were able to introduce a covariant Poisson-Hamiltonian framework based on a well-defined Poisson-Gerstenhaber graded bracket.   
We were also able to undertake the analysis of theories characterized by singular Lagrangian systems through the implementation of a Dirac-Poisson bracket in the polymomenta phase-space, thus recovering the correct 
field equations.  Further, 
by suitably enforcing a space plus time decomposition of the background manifold,
we studied the manner in which one may recover the complete gauge content and the 
true degrees of freedom for a 
classical field theory with localizable symmetries solely by considering the admissible space of Cauchy data for the evolution equations of the system which turned out to be determined by the zero level set of the induced momentum map. In consequence, we consider that the 
geometric formulations reviewed here provide a 
legitimate way to analyze the dynamics and constrained structure for singular Lagrangian 
systems associated with classical field theories.

Motivated by the application of such geometric formalisms to physical models that 
mimic certain aspects of General Relativity, here
we have studied the Bonzom-Livine model for gravity within the geometric-covariant Lagrangian, multisymplectic and polysymplectic formulations for classical field theory. At the Lagrangian level, we have obtained the field equations of the system, which have been shown to be equivalent to the vanishing torsion condition and the Einstein equations, as discussed in \cite{Bonzom}. Besides, we have analyzed the so-called $\mathcal{H}$-gauge and translational symmetries of the model, which not only preserve the Poincar{\'e}-Cartan $n$-form of the theory up to an exact form but also correspond to localizable symmetries. Subsequently, we have computed the local representation of the Lagrangian covariant momentum map associated with the extended gauge symmetry group of the system which allowed us to construct the Noether currents of the model. In addition, for any solution of the field equations of the theory, we have discussed how the Lagrangian Noether charges of the system vanish, which is consistent with the fact that the symmetries of the gravity model of our interest are localizable symmetries. Besides, within the multisymplectic approach we have shown that the $\mathcal{H}$-gauge and translational symmetries of the theory act on the corresponding covariant multimomenta phase-space by infinitesimal covariant canonical transformations, thus demonstrating the existence of a covariant momentum map, which in turn is related with the Lagrangian covariant momentum map of the system by means of the covariant Legendre transformation. Further, at the polysymplectic level, we have found that the Bonzom-Livine model for gravity corresponds to a singular Lagrangian system. As a consequence, in order to carry out a consistent polysymplectic formulation of the gravity model of our interest, we have implemented the algorithm proposed in \cite{IKGD} to study singular Lagrangian systems within the polysymplectic framework. In particular, we have shown that, on the polymomenta phase-space, the Bonzom-Livine model for gravity is characterized by a set of second-class constraint $(n-1)$-forms which arise when implementing the covariant Legendre map. Furthermore, we have described how the covariant momentum map associated with the theory not only induces 
an appropriate set of first-class Hamiltonian $(n-1)$-forms on the polymomenta phase-space but also reproduces under the Poisson-Gerstenhaber bracket, modulo exact forms, the Lie algebra corresponding to the extended gauge symmetry group of the model. Finally, in order to eliminate the second-class constraint $(n-1)$-forms of the system, we have constructed the Dirac-Poisson bracket associated with the theory. In this regard, we have found that under the Dirac-Poisson bracket the fundamental commutation relations of the model explicitly depend on the Immirzi-like parameter inherent to the system, thus modifying the canonical polysymplectic structure of the theory to a non-commutative one, as analogously discussed within the canonical analysis developed 
in~\cite{Bonzom}. Next, we have calculated the correct De Donder-Weyl-Hamilton field equations of the model which on the constraint surface allowed us to obtain the vanishing torsion condition and the Einstein equations, thus demonstrating the equivalence of the Lagrangian and De Donder-Weyl-Hamilton field equations. In addition, we have discussed how the covariant momentum map associated with the extended gauge symmetry group of the system gives rise to the conserved currents of the theory within the De Donder-Weyl Hamiltonian formulation. 

Moreover, we have also performed the space plus time decomposition for the Bonzom-Livine model for gravity at both the Lagrangian and multisymplectic level. By appropriately considering adapted coordinates, we have introduced a slicing of the space-time manifold, which allowed us to foliate it into Cauchy surfaces. Subsequently, we have proceeded to introduce a $\mathcal{G}$-slicing of the covariant configuration space of the theory generated by a vector field that can be thought of as an evolution direction of the system. Thus, after carrying out the space plus time decomposition of the affine jet bundle, we obtained the instantaneous Lagrangian functional of the model which by means of the instantaneous Legendre transformation enabled us to compute not only the instantaneous momenta variables of the theory but also the related primary constraints. Further, as previously mentioned, we have also addressed the space plus time decomposition of the multisymplectic formulation for the model of our interest. In this regard, we have found that the covariant momentum map associated with the extended gauge symmetry group of the system induces a momentum map on the $t$-instantaneous phase-space of the theory. In particular, we have realized that, since the symmetries of the model are localizable symmetries, the zero level set of the induced momentum map corresponds to the surface on the $t$-instantaneous phase-space determined by the complete set of first-class constraints of the system, which arise within the instantaneous Dirac-Hamiltonian analysis of the model as described in \cite{Escalante}. In addition, we have discussed how the $\alpha^{(\mathcal{M})}$-lift of the generator of the $\mathcal{G}$-slicing of the covariant configuration space of the model gives rise to an infinitesimal covariant canonical transformation, thus resulting in the existence of an $(n-1)$-form on the covariant multimomenta phase-space that allowed us to recover the extended Hamiltonian of the theory on the $t$-instantaneous phase-space. 

By taking into account all of these results, we have shown that, the geometric-covariant Lagrangian, multisymplectic and polysymplectic formalisms for classical field theory enable us to describe in a geometric, consistent and elegant way the features of the Bonzom-Livine model for gravity. In particular, we have discussed how the instantaneous Dirac-Hamiltonian analysis of the system can be straightforwardly recovered by performing the space plus time decomposition of its corresponding multisymplectic formulation. Also, we have described how the algorithm to study singular Lagrangian systems within the polysymplectic approach is necessary in order to obtain a consistent polysymplectic formulation of the gravity model of our interest. In this regard, it is worth noting that, such an algorithm has been scarcely explored in the literature, see for instance \cite{IKHFVG, Angel, Kanatchikov1}. However, based on the results of the present work, we can conclude that the proposal developed in \cite{IKGD} to study singular Lagrangian systems within the polysymplecic formalism is completely adequate for the analysis of physically motivated classical field theories. In addition, as far as we know, our study provides the first non-trivial example where the Dirac-Poisson bracket \eqref{DPGBracket} is explicitly implemented in the 
context of a model related with General Relativity. Thus, our work may shed some light on a deeper understanding of such bracket structure, which resulted fundamental in order to reproduce the correct De Donder-Weyl-Hamilton field equations of the theory under consideration. 
Finally, we would like to mention that our results may be relevant at the 
quantum level by considering the so-called pre-canonical quantization approach for the De Donder-Weyl canonical theory~\cite{IKQFTPV, IKHEQFT, IKPSWF,kana2}. Indeed, such quantization program is strongly based on the polysymplectic formalism for classical field theory, and thus our study may constitute a first step towards the quantum analysis of $3$-dimensional gravity with Immirzi-like parameter from the aforementioned quantization scheme. Certainly, a relevant issue 
in this direction will be to implement both the gauge transformations and the
diffeomorphisms invariance at the quantum level.  This will be done elsewhere.

\section*{Acknowledgments}

The authors thank Alexis Tepale-Luna for discussions and collaboration. The authors would also like to acknowledge financial support from CONACYT-Mexico under
the project CB-2017-283838.

\end{document}